\documentclass[final,authoryear,5p,times,twocolumn]{elsarticle}
\usepackage{amsmath}
\usepackage[usenames,dvipsnames]{color}
\usepackage{booktabs}
\usepackage{lineno,hyperref}
\modulolinenumbers[5]

\journal{Journal of Quantitative Spectroscopy \& Radiative Transfer}

\newcommand{\tripoli}{{\sc Tripoli-4}\textsuperscript{ \textregistered}}

\begin{document}

\begin{frontmatter}

\title{Monte Carlo particle transport in random media: the effects of mixing statistics}


\author[label1]{Coline Larmier}
\address[label1]{Den-Service d'Etudes des R\'eacteurs et de Math\'ematiques Appliqu\'ees (SERMA), CEA, Universit\'e Paris-Saclay, 91191 Gif-sur-Yvette, FRANCE.}
\author[label1]{Andrea Zoia\corref{cor1}}
\cortext[cor1]{Corresponding author. Tel. +33 (0)1 69 08 95 44}
\ead{andrea.zoia@cea.fr}
\author[label1]{Fausto Malvagi}
\author[label2]{Eric Dumonteil}
\address[label2]{IRSN, 31 Avenue de la Division Leclerc, 92260 Fontenay aux Roses, FRANCE.}
\author[label1]{Alain Mazzolo}





\begin{abstract}
Particle transport in random media obeying a given mixing statistics is key in several applications in nuclear reactor physics and more generally in diffusion phenomena emerging in optics and life sciences. Exact solutions for the ensemble-averaged physical observables are hardly available, and several approximate models have been thus developed, providing a compromise between the accurate treatment of the disorder-induced spatial correlations and the computational time. In order to validate these models, it is mandatory to resort to reference solutions in benchmark configurations, typically obtained by explicitly generating by Monte Carlo methods several realizations of random media, simulating particle transport in each realization, and finally taking the ensemble averages for the quantities of interest. In this context, intense research efforts have been devoted to Poisson (Markov) mixing statistics, where benchmark solutions have been derived for transport in one-dimensional geometries. In a recent work, we have generalized these solutions to two and three-dimensional configurations, and shown how dimension affects the simulation results. In this paper we will examine the impact of mixing statistics: to this aim, we will compare the reflection and transmission probabilities, as well as the particle flux, for three-dimensional random media obtained by resorting to Poisson, Voronoi and Box stochastic tessellations. For each tessellation, we will furthermore discuss the effects of varying the fragmentation of the stochastic geometry, the material compositions, and the cross sections of the transported particles.
\end{abstract}

\begin{keyword}
Stochastic geometries \sep Benchmark \sep Monte Carlo \sep {\sc Tripoli-4}\textsuperscript{ \textregistered} \sep Poisson \sep Voronoi \sep Box
\end{keyword}

\end{frontmatter}


\section{Introduction}

Linear transport in heterogeneous and random media emerges in several applications in nuclear reactor physics, ranging from the analysis of the effects of the grain size distribution in burnable poisons (especially Gadolinium) and of Pu agglomerates in MOX fuel pellets, the quantification of water density variations in concrete structures and in the moderator fluid during operation (e.g., steam flow in BWRs) or accidents (e.g., local boiling in PWRs), the assessment of the probability of a re-criticality accident in a reactor core after melt-down (corium), or the investigation of neutron diffusion in pebble-bed reactors~\cite{pomraning, larsen, zimmerman}. The spectrum of applications of stochastic media is actually far reaching~\cite{santalo, kendall, torquato, bouchaud}, and concerns also inertial-confinement fusion~\cite{haran}, light propagation through engineered optical materials~\cite{NatureOptical, PREOptical, PREQuenched}, atmospheric radiation transport~\cite{davis, kostinski, clouds}, tracer diffusion in biological tissues~\cite{tuchin}, and radiation trapping in hot atomic vapours~\cite{NatureVapours}, only to name a few.

The stochastic nature of particle transport stems from the materials composing the traversed medium being randomly distributed according to some statistical law: thus, the total cross section, the scattering kernel and the source are in principle random fields. Particle transport theory in random media is therefore aimed at providing a description of the ensemble-averaged angular particle flux $\langle \varphi({\bf r}, {\bf v}) \rangle$ and related functionals. For the sake of simplicity, in the following we will focus on mono-energetic transport in non-fissile media, in stationary (i.e., time-independent) conditions. However, these hypotheses are not restrictive, as described in~\cite{pomraning}.

A widely adopted model of random media is the so-called binary stochastic mixing, where only two immiscible materials are present~\cite{pomraning}. In principle, it is possible to formally write down a set of coupled linear Boltzmann equations describing the evolution of the particle flux in each immiscible phase. Nonetheless, it has been shown that these equations form generally speaking an infinite hierarchy (exact solutions can be exceptionally found, such as for purely absorbing media), so that in most cases it is necessary to truncate the infinite set of equations with some appropriate closure formulas, depending on the underlying mixing statistics. Perhaps the best-known of such closure formulas goes under the name of the Levermore-Pomraning model, initially developed for homogeneous Markov mixing statistics~\cite{pomraning, levermore}. Several generalisations of this model have been later proposed, including higher-order closure schemes~\cite{pomraning, su}. Along the development of deterministic equations for the ensemble-averaged flux, Monte Carlo methods have been also proposed, such as the celebrated Chord Length Sampling~\cite{zimmerman, zimmerman_adams, sutton, donovan}. The common feature of these approaches is that they allow a simpler, albeit approximate, treatment of transport in stochastic mixtures, which might be convenient in practical applications where a trade-off between computational time and precision is needed.

In order to assess the accuracy of the various approximate models it is therefore mandatory to compute reference solutions for linear transport in random media. Such solutions can be obtained in the following way: first, a realization of the medium is sampled from the underlying mixing statistics (a stochastic tessellation model); then, the linear transport equations corresponding to this realization are solved by either deterministic or Monte Carlo methods, and the physical observables of interest are determined; this procedure is repeated many times so as to create a sufficiently large collection of realizations, and ensemble averages are finally taken for the physical observables.

For this purpose, a number of benchmark problems for Markov mixing have been proposed in the literature so far~\cite{benchmark_adams, renewal, brantley_benchmark, brantley_conf, brantley_conf_2, vasques_suite2}. In a previous work~\cite{larmier_benchmark}, we have revisited the benchmark problem originally proposed by Adams, Larsen and Pomraning for transport in binary stochastic media with Markov mixing~\cite{benchmark_adams}, and later extended in~\cite{brantley_benchmark, brantley_conf, brantley_conf_2, vasques_suite2}. In particular, while these authors had exclusively considered $1d$ slab or rod geometries, we have provided reference solutions obtained by Monte Carlo particle transport simulations through $2d$ extruded and $3d$ Markov tessellations, and discussed the effects of dimension on the physical observables.

In this work, we further generalize these findings by probing the impact of the underlying mixing statistics on particle transport. The nature of the microscopic disorder is known to subtly affect the path of the travelling particles, so that the observables will eventually depend on the statistical laws describing the shape and the material compositions of the random media~\cite{torquato, bouchaud, davis, renewal}. This is especially true in the presence of distributed absorbing traps~\cite{bouchaud}. We will consider three different stochastic $3d$ tessellations and compute the ensemble-averaged reflection and transmission probabilities, as well as the particle flux. Two distinct benchmark configurations will be considered, the former including purely scattering materials and voids, and the latter containing scattering and absorbing materials. This paper is organized as follows: in Sec.~\ref{section_models} we will introduce the mixing statistics that we have chosen, namely homogeneous and isotropic Poisson (Markov) tessellations, Poisson-Voronoi tessellations, and Poisson Box tessellations, and we will show how the free parameters governing the mixing statistics can be chosen in order for the resulting stochastic media to be comparable. In Sec.~\ref{size_effects} we will illustrate the statistical features of such tessellations, which is key to understanding the effects on particle transport. In Sec.~\ref{transport_benchmark} we will propose two benchmark problems, provide reference solutions by resorting to the \tripoli{} Monte Carlo code, and discuss how mixing statistics affects ensemble-averaged observables. Conclusions will be finally drawn in Sec.~\ref{conclusions}.

\section{Description of the mixing statistics}
\label{section_models}

In this section, we introduce three mixing statistics leading to random media with distinct features. The subscript or superscript $m$ will denote the class of the stochastic mixing: $m={\cal P}$ for Poisson tessellations, $m={\cal V}$ for Voronoi tessellations, and $m={\cal B}$ for Box tessellations. For each stochastic model, we describe the strategy for the construction of three-dimensional tessellations, spatially restricted to a cubic box of side $L$. Without loss of generality, we assume that the cubes are centered at the origin.

\subsection{Isotropic Poisson tessellations}
\label{poisson}

Markovian mixing is generated by resorting to isotropic Poisson geometries, which form a prototype process of stochastic tessellations: a domain included in a $d$-dimensional space is partitioned by randomly generated $(d-1)$-dimensional hyper-planes drawn from an underlying Poisson process~\cite{santalo}. In order to construct three-dimensional homogeneous and isotropic Poisson tessellations restricted to a cubic box, we use an algorithm recently proposed for finite $d$-dimensional geometries~\cite{serra, mikhailov}. For the sake of completeness, here we briefly recall the algorithm for the construction of these geometries (further details are provided in~\cite{larmier}).

\begin{figure}[t]
\begin{center}
\,\,\,\, ${\langle \Lambda \rangle}_{\infty} =1$ \,\,\,\,\\
\includegraphics[width=0.42\columnwidth]{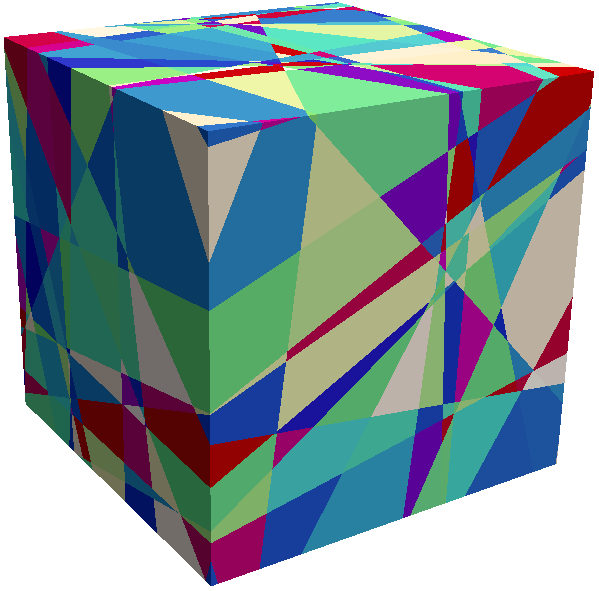}\,\,\,\,
\includegraphics[width=0.42\columnwidth]{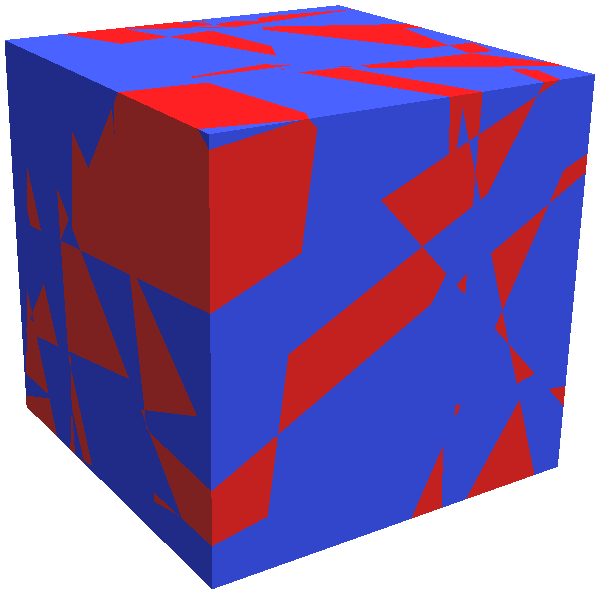}\\
\,\,\,\, ${\langle \Lambda \rangle}_{\infty} =0.5$ \,\,\,\,\\
\includegraphics[width=0.42\columnwidth]{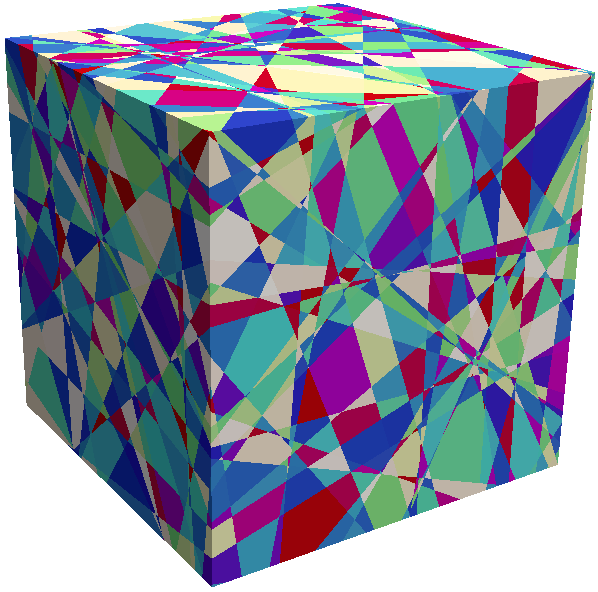}\,\,\,\,
\includegraphics[width=0.42\columnwidth]{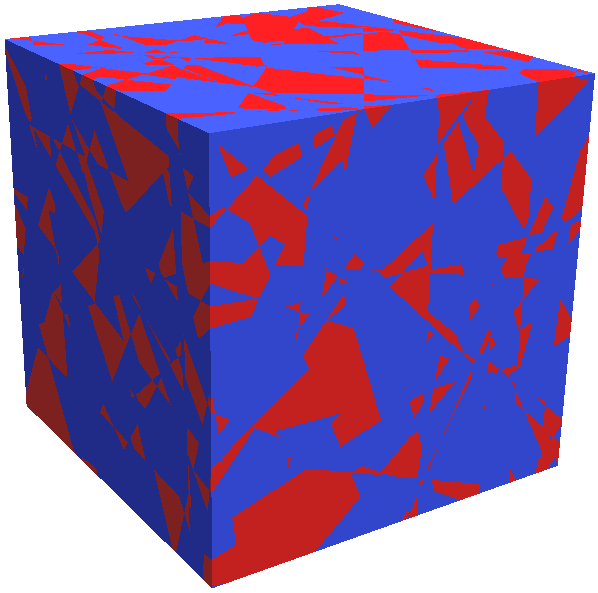}\\
\,\,\,\, ${\langle \Lambda \rangle}_{\infty} =0.1$ \,\,\,\,\\
\includegraphics[width=0.42\columnwidth]{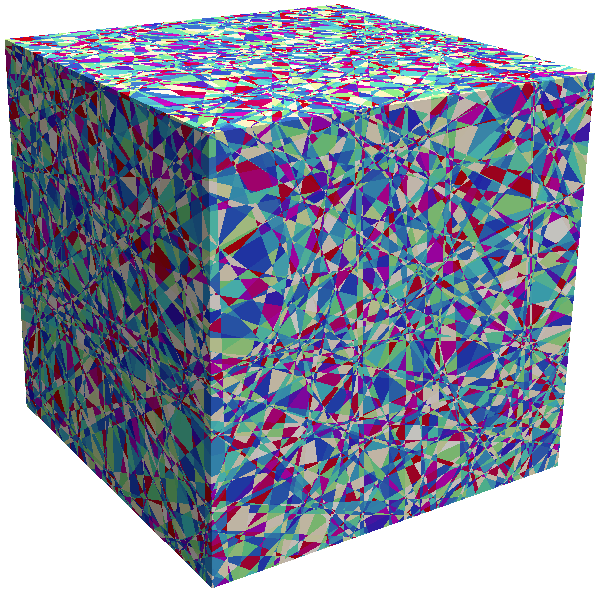}\,\,\,\,
\includegraphics[width=0.42\columnwidth]{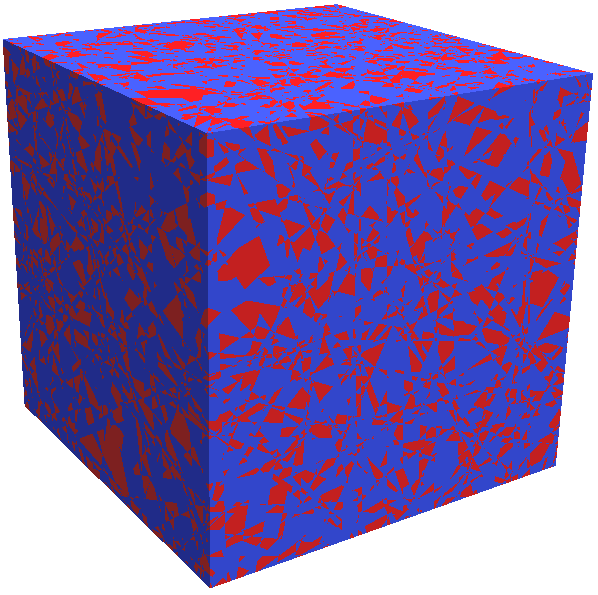}\\
\end{center}
\caption{Examples of realizations of homogeneous isotropic Poisson tessellations corresponding to the benchmark specifications, before (left) and after (right) attributing the material label, with probability $p=0.3$ of assigning the label $\alpha$. Red corresponds to the label $\alpha$ and blue to the label $\beta$. The size of the cube is $L=10$.}
\label{geo_poisson}
\end{figure}

We start by sampling a random number of hyper-planes $N_H$ from a Poisson distribution of intensity $4 \rho_{\cal P} R$, where $R$ is the radius of the sphere circumscribed to the cube and $\rho_{\cal P}$ is the (arbitrary) density of the tessellation, carrying the units of an inverse length. This normalization of the density $\rho_{\cal P}$ corresponds to the convention used in~\cite{santalo}, and is such that $\rho_{\cal P}$ yields the mean number of (d-1)-hyperplanes intersected by an arbitrary segment of unit length. Then, we generate the planes that will cut the cube. We choose a radius $r$ uniformly in the interval $[0,R]$ and then sample two additional parameters, namely, $\xi_1$ and $\xi_2$, from two independent uniform distributions in the interval $[0,1]$. A unit vector ${\mathbf n}=(n_1,n_2,n_3)^T$ with components
\begin{align}
n_1&=1-2\xi_1 \nonumber \\
n_2&=\sqrt{1-n_1^2}\cos{(2 \pi \xi_2)}\nonumber \\
n_3&=\sqrt{1-n_1^2}\sin{(2 \pi \xi_2)}\nonumber
\end{align}
is generated. Denoting by $\mathbf M$ the point such that ${\mathbf{ OM}}=r {\mathbf n}$, the random plane will finally obey $n_1 x + n_2 y +n_3 z =r$, passing trough $\mathbf M$ and having normal vector ${\mathbf n}$. By construction, this plane does intersect the circumscribed sphere of radius $R$ but not necessarily the cube. The procedure is iterated until $N_H$ random planes have been generated. The polyhedra defined by the intersection of such random planes are convex. Some examples of homogeneous isotropic Poisson tessellations are provided in Fig.~\ref{geo_poisson}. 

\begin{figure}[t]
\begin{center}
\,\,\,\, ${\langle \Lambda \rangle}_{\infty} =1$ \,\,\,\,\\
\includegraphics[width=0.42\columnwidth]{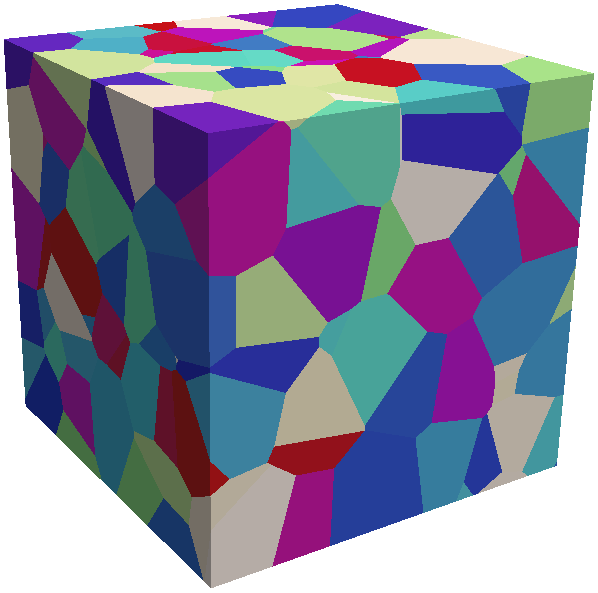}\,\,\,\,
\includegraphics[width=0.42\columnwidth]{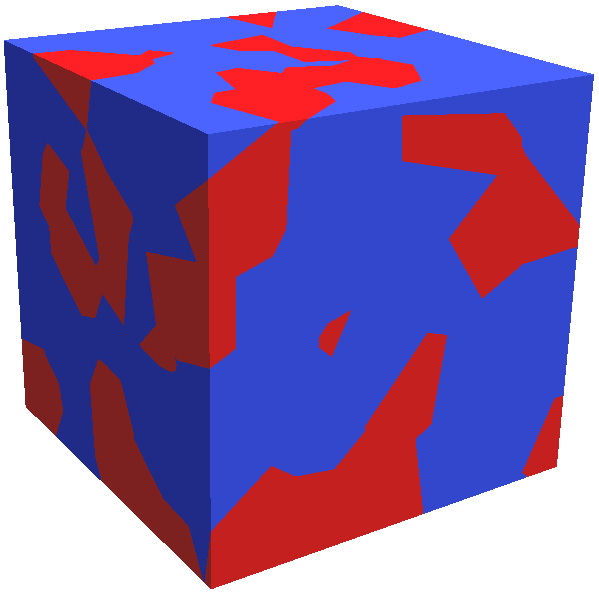}\\
\,\,\,\, ${\langle \Lambda \rangle}_{\infty} =0.5$ \,\,\,\,\\
\includegraphics[width=0.42\columnwidth]{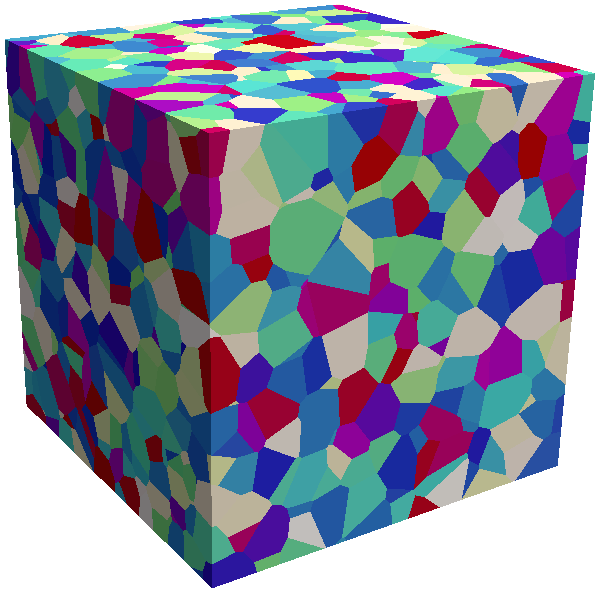}\,\,\,\,
\includegraphics[width=0.42\columnwidth]{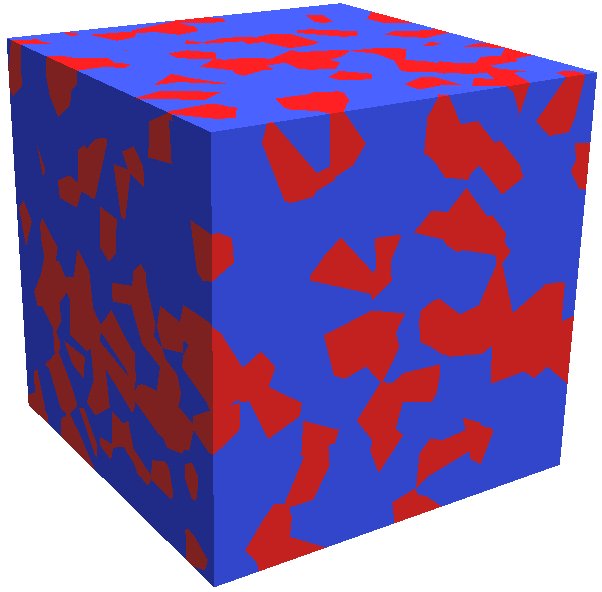}\\
\,\,\,\, ${\langle \Lambda \rangle}_{\infty} =0.1$ \,\,\,\,\\
\includegraphics[width=0.42\columnwidth]{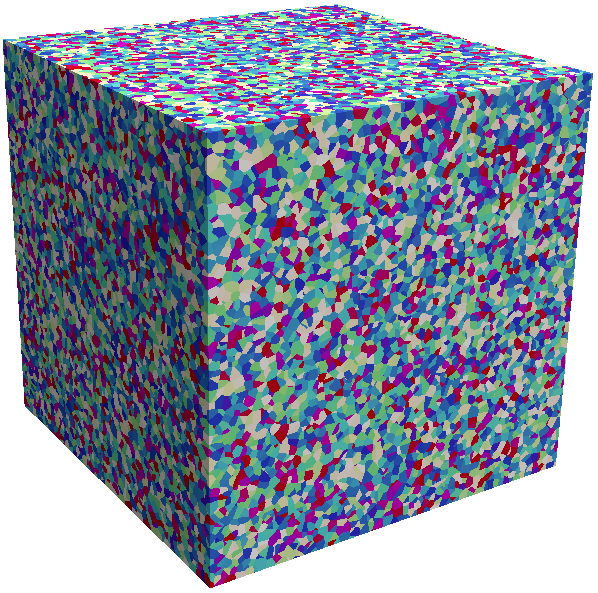}\,\,\,\,
\includegraphics[width=0.42\columnwidth]{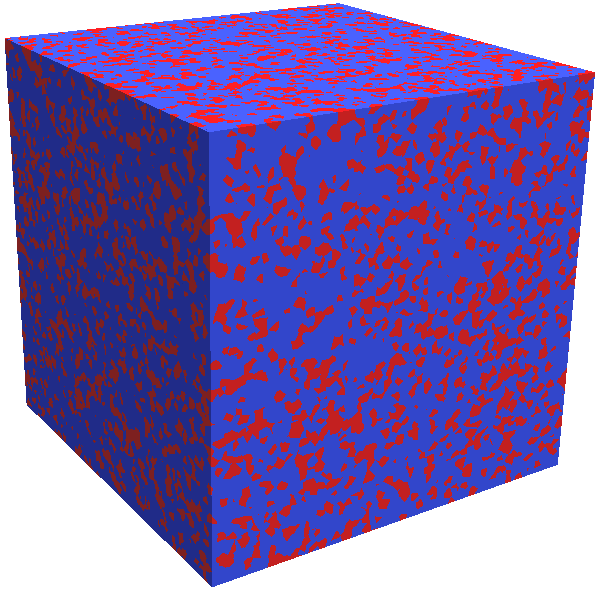}\\
\end{center}
\caption{Examples of realizations of Voronoi tessellations corresponding to the benchmark specifications, before (left) and after (right) attributing the material label, with probability $p=0.3$ of assigning the label $\alpha$. Red corresponds to the label $\alpha$ and blue to the label $\beta$.The size of the cube is $L=10$.}
\label{geo_voronoi}
\end{figure}

\begin{figure}[t]
\begin{center}
\,\,\,\, ${\langle \Lambda \rangle}_{\infty} =1$ \,\,\,\,\\
\includegraphics[width=0.42\columnwidth]{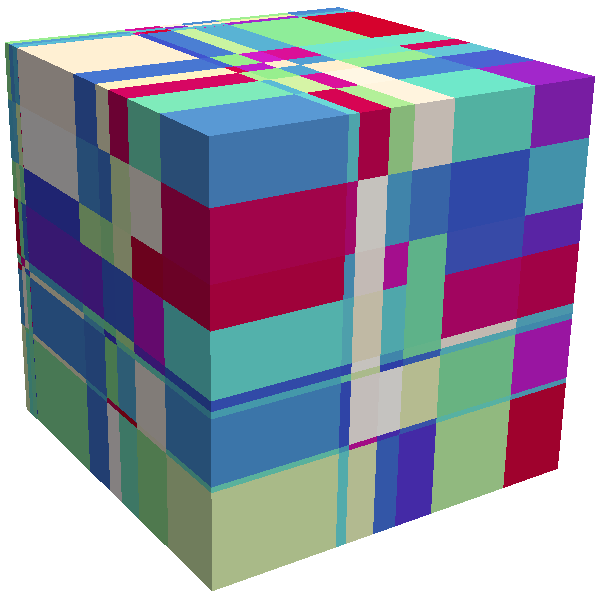}\,\,\,\,
\includegraphics[width=0.42\columnwidth]{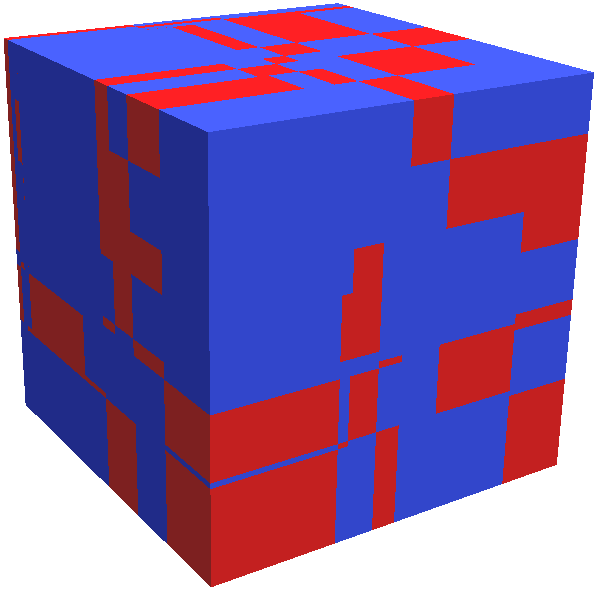}\\
\,\,\,\, ${\langle \Lambda \rangle}_{\infty} =0.5$ \,\,\,\,\\
\includegraphics[width=0.42\columnwidth]{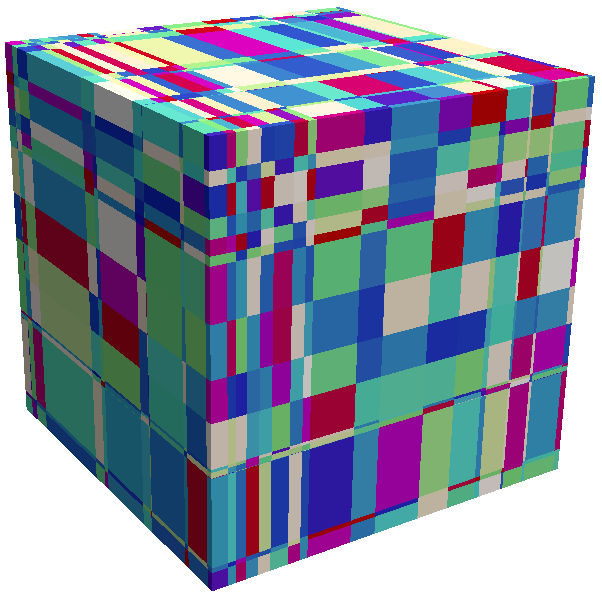}\,\,\,\,
\includegraphics[width=0.42\columnwidth]{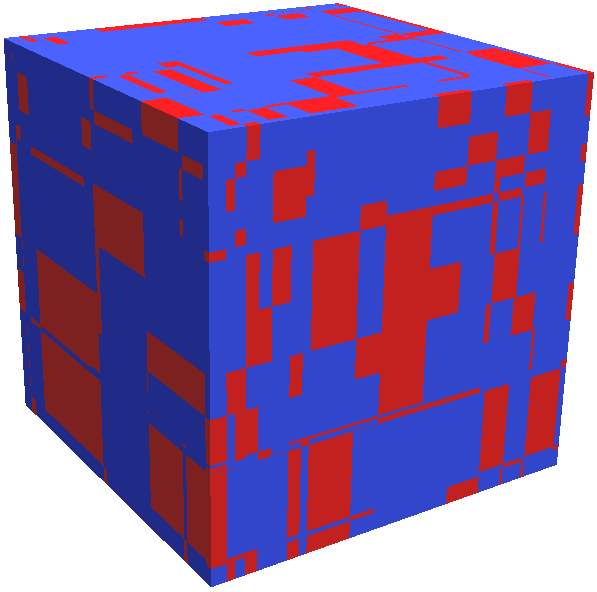}\\
\,\,\,\, ${\langle \Lambda \rangle}_{\infty} =0.1$ \,\,\,\,\\
\includegraphics[width=0.42\columnwidth]{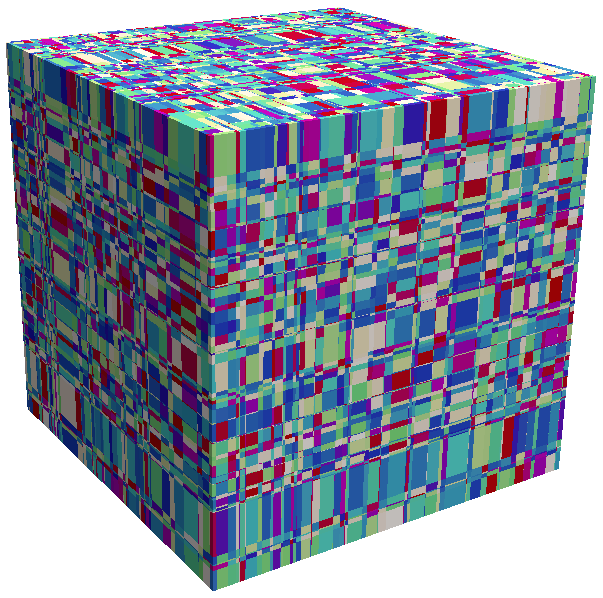}\,\,\,\,
\includegraphics[width=0.42\columnwidth]{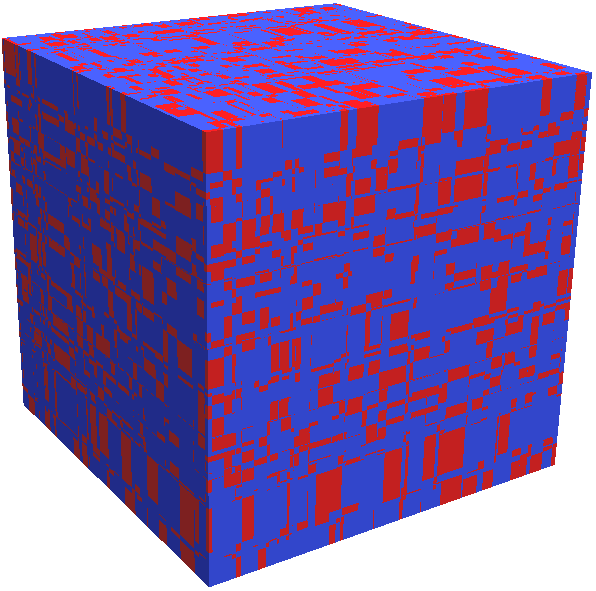}\\
\end{center}
\caption{Examples of realizations of Box tessellations corresponding to the benchmark specifications, before (left) and after (right) attributing the material label, with probability $p=0.3$ of assigning the label $\alpha$. Red corresponds to the label $\alpha$ and blue to the label $\beta$.The size of the cube is $L=10$.}
\label{geo_ortho}
\end{figure}

\subsection{Poisson-Voronoi tessellations}
\label{voronoi}

Voronoi tessellations refer to another prototype process for isotropic random division of space~\cite{santalo}. A portion of a space is decomposed into polyhedral cells by a partitioning process based on a set of random points, called `seeds'. From this set of seeds, the Voronoi decomposition is obtained by applying the following deterministic procedure: each seed is associated with a Voronoi cell, defined as the set of points which are nearer to this seed than to any other seed. Such a cell is convex, because obtained from the intersection of half-spaces.

In this paper, we will exclusively focus on Poisson-Voronoi tessellations, which form a subclass of Voronoi geometries~\cite{meijering, gilbert, miles1972}. The specificity of Poisson-Voronoi tessellations concerns the sampling of the seeds. In order to construct Poisson-Voronoi tessellations restricted to a cubic box of side $L$, we resort to the algorithm proposed in~\cite{miles1972}. First, we choose the random number of seeds $N_S$ from a Poisson distribution of parameter $(\rho_{\cal V} L)^3$, where $\rho_{\cal V}$ characterizes the density of the tessellation. Then, $N_S$ seeds are uniformly sampled in the box $[-L/2,L/2]^3$. For each seed, we compute the corresponding Voronoi cell as the intersection of half-spaces bounded by the mid-planes between the selected seed and any other seed. In order to avoid confusion with the Poisson tessellations described above, we will mostly refer to Poisson-Voronoi geometries simply as Voronoi tessellations in the following. Some examples of Voronoi tessellations are provided in Fig.~\ref{geo_voronoi}. 

\subsection{Poisson Box tessellations}
\label{box}

Box tessellations form a class of anisotropic stochastic geometries, composed of rectangular boxes with random sides. For the special case of Poisson Box tessellations (as proposed by~\cite{miles1972}), a domain is partitioned by $i)$ randomly generated planes orthogonal to the $x$-axis, through a Poisson process of intensity $\rho_x$; $ii)$ randomly generated planes orthogonal to the $y$-axis, through a Poisson process of intensity $\rho_y$; $iii)$ randomly generated planes orthogonal to the $z$-axis, through a Poisson process of intensity $\rho_z$. In the following, we will assume that the three parameters are equal, namely, $\rho_x=\rho_y=\rho_z= \rho_{\cal B}$.

In order to tessellate a cube of side $L$, the construction algorithm is the following: we begin by sampling a random number $N_x$ from a Poisson distribution of intensity $\rho_{\cal B} L$. Then, we sample $N_x$ points uniformly on the segment $[-L/2,L/2]$. For each point of this set, we cut the geometry with the plane orthogonal to the $x$-axis and passing through this point. We repeat this process for the $y$-axis and the $z$-axis. For the sake of conciseness, we will denote by Box tessellations these Poisson Box tessellations. Some examples of Box tessellations are provided in Fig.~\ref{geo_ortho}. 

\subsection{Statistical properties of infinite tessellations}

The observables of interest associated to the stochastic geometries, such as for instance the volume of a polyhedron, its surface, the number of faces, and so on, are random variables. With a few remarkable exceptions, their exact distributions are unfortunately unknown~\cite{santalo}. Nevertheless, exact results have been established for some low-order moments of the observables, in the limit case of domains having an infinite extension~\cite{santalo, kendall, miles1970}.

In this respect, Poisson tessellations have been shown to possess a remarkable property: in the limit of infinite domains, an arbitrary line will be cut by the hyperplanes of the tessellation into chords whose lengths are exponentially distributed with parameter $\rho_{\cal P}$ (whence the identification with Markovian mixing). Thus, in this case, the average chord length ${\langle \Lambda \rangle}_{\infty}$ satisfies ${\langle \Lambda \rangle}_{\infty}=\rho_{\cal P}^{-1}$, and its probability density $\Pi^{\cal P}(\Lambda)$ is given by
\begin{equation}
\Pi^{\cal P}(\Lambda)= \frac{1}{{\langle \Lambda \rangle}_{\infty}} e^{-\Lambda / \langle \Lambda \rangle_{\infty}}.
\end{equation}\label{markovian_density}
To the best of our knowledge, the exact distribution of the chord length for Voronoi and Box tessellations is not known. However, when lines are drawn uniformly and isotropically (formally, with a $\mu$-randomness~\cite{santalo, coleman}), it is possible to apply the Cauchy's formula
\begin{equation}
{\langle \Lambda \rangle}_{\infty} = 4 \dfrac{{\langle V \rangle}_{\infty}}{{\langle S \rangle}_{\infty}},
\end{equation}
which relates the average chord length to purely geometrical quantities, namely, the average volume ${\langle V \rangle}_{\infty}$ of a random polyhedron and its average surface ${\langle S \rangle}_{\infty}$~\cite{santalo}. This leads to exact expressions for the first moment of the chord length for Voronoi and Box tessellations, which are recalled in Tab.~\ref{tab_theory}. As discussed in the following sections, Monte Carlo simulations show that the chord length distribution for the Box tessellations is actually very close to that of Poisson tessellations. Nevertheless, it is easy to rule out the possibility of a complete equivalence. If Box tessellations were exactly Markovian, with an exponential distribution for $\Lambda$, then the fourth moment of the chord length should satisfy
\begin{equation}
{\langle \Lambda^4 \rangle}_{\infty} = 4! {\langle \Lambda \rangle}_{\infty}^4,
\label{moment4_0}
\end{equation}
which for $m={\cal B}$ would yield
\begin{equation}
{\langle \Lambda^4 \rangle}_{\infty} = \dfrac{128}{27 \rho_{\cal B}^{4}}.
\label{moment4}
\end{equation}
However, under the hypothesis of a $\mu$-randomness for the sampled lines, a remarkable formula from stochastic geometry holds for convex volumes, namely,
\begin{equation}
{\langle \Lambda^4 \rangle}_{\infty} = \dfrac{12}{\pi} \dfrac{{\langle V^2 \rangle}_{\infty}}{{\langle S \rangle}_{\infty}}
\label{inter}
\end{equation}
which again depends only on purely geometrical quantities~\cite{miles1972}. Now, since ${\langle V^k \rangle}_{\infty}=k!(\rho_{\cal B})^{-3k}$ for $m={\cal B}$~\cite{miles1972}, Eq.~\ref{inter} can be simplified to
\begin{equation}
{\langle \Lambda^4 \rangle}_{\infty} = \dfrac{16}{\pi \rho_{\cal B}^{4}},
\end{equation}
by resorting to the expressions given in Tab.~\ref{tab_theory}. This clearly contradicts Eq.~\ref{moment4}: the chord length for Box tessellations is not exponential. The same argument applies also for Voronoi tessellations, since we have ${\langle \Lambda^4 \rangle}_{\infty}=0.774 \rho_{\cal V}^{-4}$ and $4!{\langle \Lambda \rangle}^4_\infty \approx 5.352 \rho_{\cal V}^{-4}$.

\begin{table}[t]
\begin{center}
\begin{tabular}{ccccc}
\toprule
$m$ & ${\langle V \rangle}_{\infty}$ & ${\langle S \rangle}_{\infty}$ & ${\langle \Lambda \rangle}_{\infty} = 4 \frac{{\langle V \rangle}_{\infty}}{{\langle S \rangle}_{\infty}} $\\
 \midrule
${\cal P}$  & $(6/\pi)\rho_{\cal P}^{-3}$ & $(24/\pi){\rho_{\cal P}}^{-2}$ & ${\rho_{\cal P}}^{-1}$ \\
${\cal V}$ & ${\rho_{\cal V}}^{-3}$ & ${(256\pi/3)}^{\frac{1}{3}} \Gamma \left(\frac{5}{3}\right){\rho_{\cal V}}^{-2}$ & $0.6872{\rho_{\cal V}}^{-1}$ \\
${\cal B}$ & ${\rho_{\cal B}}^{-3}$ & $6{\rho_{\cal B}}^{-2}$ & $(2/3){\rho_{\cal B}}^{-1}$ \\
\bottomrule
\end{tabular}
\end{center}
\caption{Exact formulas for the average volume ${\langle V \rangle}_{\infty}$, the average total surface ${\langle S \rangle}_{\infty}$ and the average chord length ${\langle \Lambda \rangle}_{\infty}$ in infinite tessellations, for different mixing statistics $m$. Expressions are taken from~\cite{santalo, miles1972}.}
\label{tab_theory}
\end{table}

\begin{figure}[t]
\begin{center}
\includegraphics[scale=0.6]{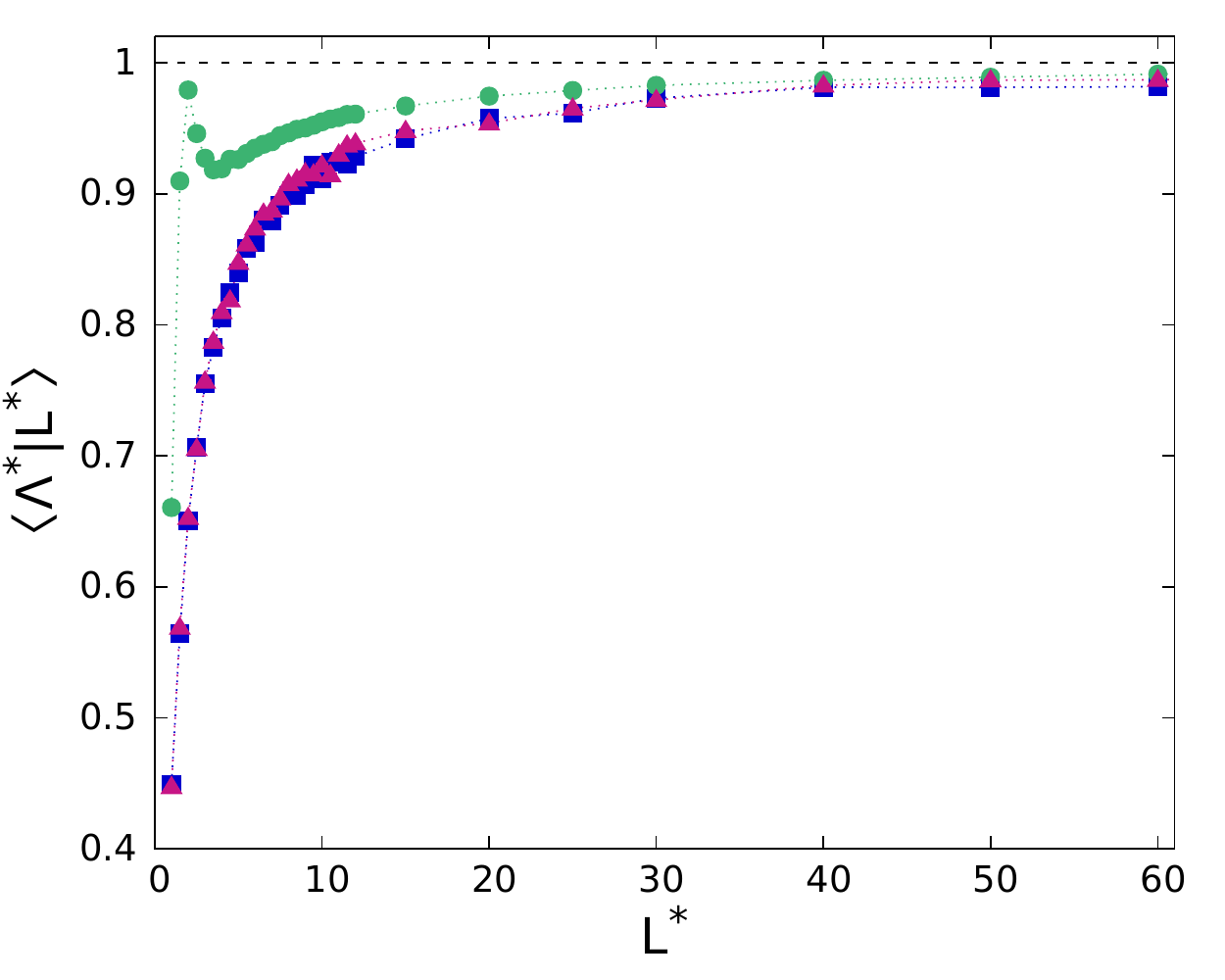}
\end{center}
\caption{The normalized average chord length $\langle \Lambda^* | L^*\rangle$, as a function of the normalized size $L^*$ of the domain and of the mixing statistics $m$. Symbols correspond to the Monte Carlo simulation results, with dotted lines added to guide the eye: blue squares denote $m={\cal P}$, green circles $m={\cal V}$, and red triangles $m={\cal B}$. The asymptotical value $\Lambda^*=1$ is displayed for reference with a black dashed line.}
\label{cl_size}
\end{figure}

\begin{figure}[t]
\begin{center}
\includegraphics[scale=0.6]{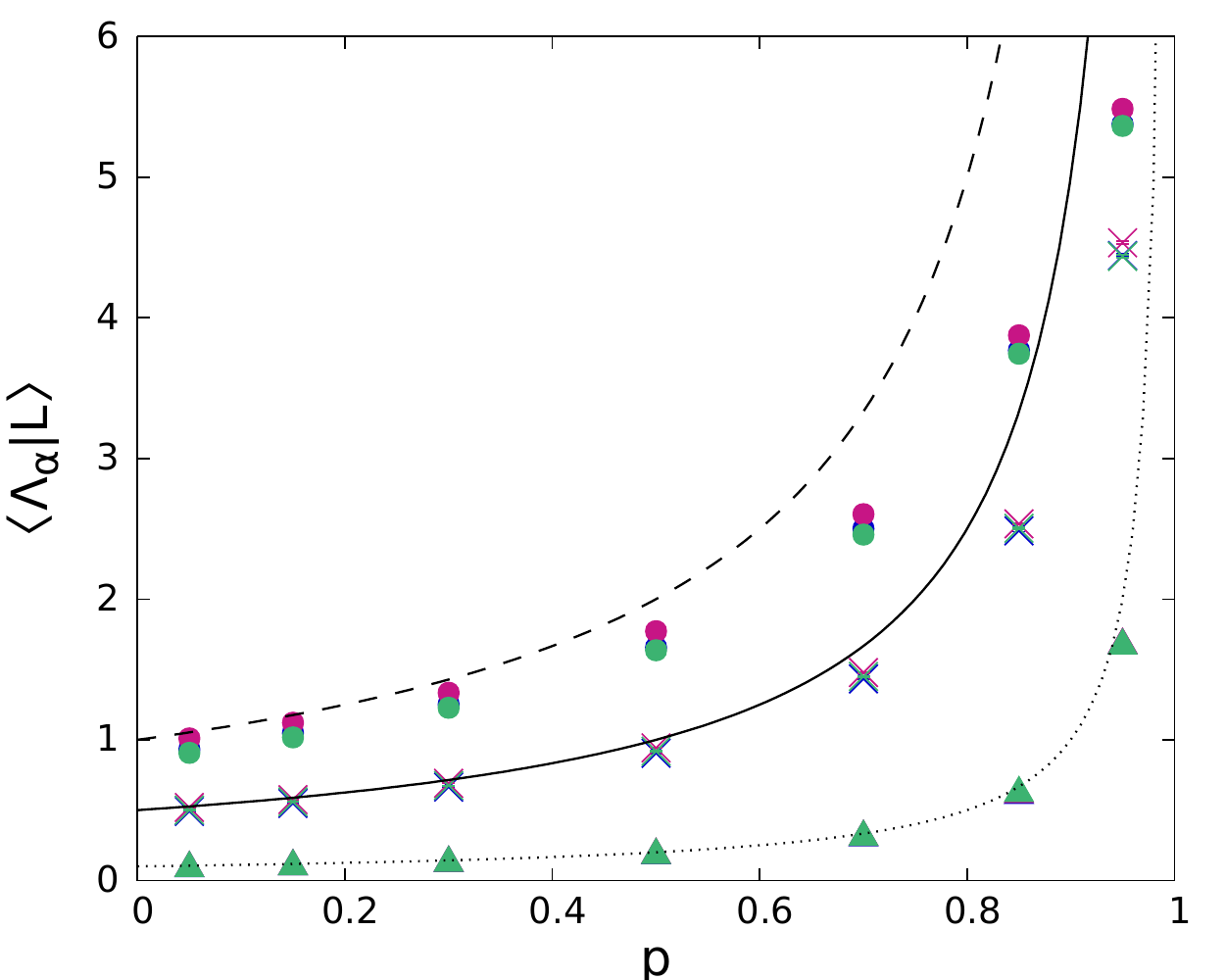}
\end{center}
\caption{The average chord length $\langle \Lambda_\alpha | L\rangle$ through clusters of composition $\alpha$, as a function of the probability $p$, of the mixing statistics $m$, and of the average chord length ${\langle \Lambda \rangle}_{\infty}$, for a domain of linear size $L=10$. Symbols correspond to the Monte Carlo simulation results: blue symbols are chosen for $m={\cal P}$, green for $m={\cal V}$, and red for $m={\cal B}$. Circles denote ${\langle \Lambda \rangle}_{\infty}=1$, crosses ${\langle \Lambda \rangle}_{\infty}=0.5$, and triangles ${\langle \Lambda \rangle}_{\infty}=0.1$. In order to provide a reference, the asymptotic function ${\langle \Lambda_{\alpha} \rangle}_{\infty}(p)={\langle \Lambda \rangle}_{\infty}/(1-p)$ is displayed for ${\langle \Lambda \rangle}_{\infty}=1$ (dashed line), ${\langle \Lambda \rangle}_{\infty}=0.5$ (continuous line) and ${\langle \Lambda \rangle}_{\infty}=0.1$ (dotted line).}
\label{cl_p}
\end{figure}

\subsection{Assigning material properties: colored geometries}

For the benchmark problems of transport in random media that will be discussed in this work, we will only consider binary stochastic mixtures, which are realized by resorting to the following procedure. First, Poisson, Voronoi or Box tessellations are constructed as described above. Then, each polyhedron of the geometry is assigned a material composition by formally attributing a distinct `label' (also called `color'), say `$\alpha$' or `$\beta$', with associated complementary probabilities. In the following, we will denote by $p$ the probability of assigning the label $\alpha$. We will define a `cluster' the collection of adjacent polyhedra sharing the same label.

When assigning colours to stochastic geometries, additional properties must be introduced, such as the average chord length through clusters with label $\alpha$, denoted by ${\langle \Lambda_\alpha \rangle}_{\infty}$. For infinite tessellations, it can be shown that ${\langle \Lambda_\alpha \rangle}_{\infty}$ is related to the average chord length ${\langle \Lambda \rangle}_{\infty}$ of the geometry via
\begin{equation}
{\langle \Lambda \rangle}_{\infty} = (1-p) {\langle \Lambda_\alpha \rangle}_{\infty},
\label{alpha}
\end{equation}
and for ${\langle \Lambda_\beta \rangle}_{\infty}$ we similarly have
\begin{equation}
{\langle \Lambda \rangle}_{\infty} =  p {\langle \Lambda_\beta \rangle}_{\infty}.
\label{beta}
\end{equation}
These properties of the average chord length through clusters with composition $\alpha$ or $\beta$ stems from the binomial distribution of the colouring procedure~\cite{haran, larmier}, and hold true for each tessellation $m$. Additionally, the corresponding probability density $\Pi^{\cal P}(\Lambda_{\alpha})$ is still exponential for infinite Poisson geometries, i.e.,
\begin{equation}
\Pi^{\cal P}(\Lambda_{\alpha})=\frac{1}{{\langle \Lambda_\alpha \rangle}_{\infty}}e^{- \Lambda_{\alpha} / {\langle \Lambda_\alpha \rangle}_{\infty}}
\label{markovian_density_alpha}
\end{equation}
For Voronoi and Box tessellations, the full probability densities $\Pi^{\cal V}(\Lambda_{\alpha})$ and $\Pi^{\cal B}(\Lambda_{\alpha})$ are not known.

\begin{figure}[t]
\begin{center}
\includegraphics[scale=0.6]{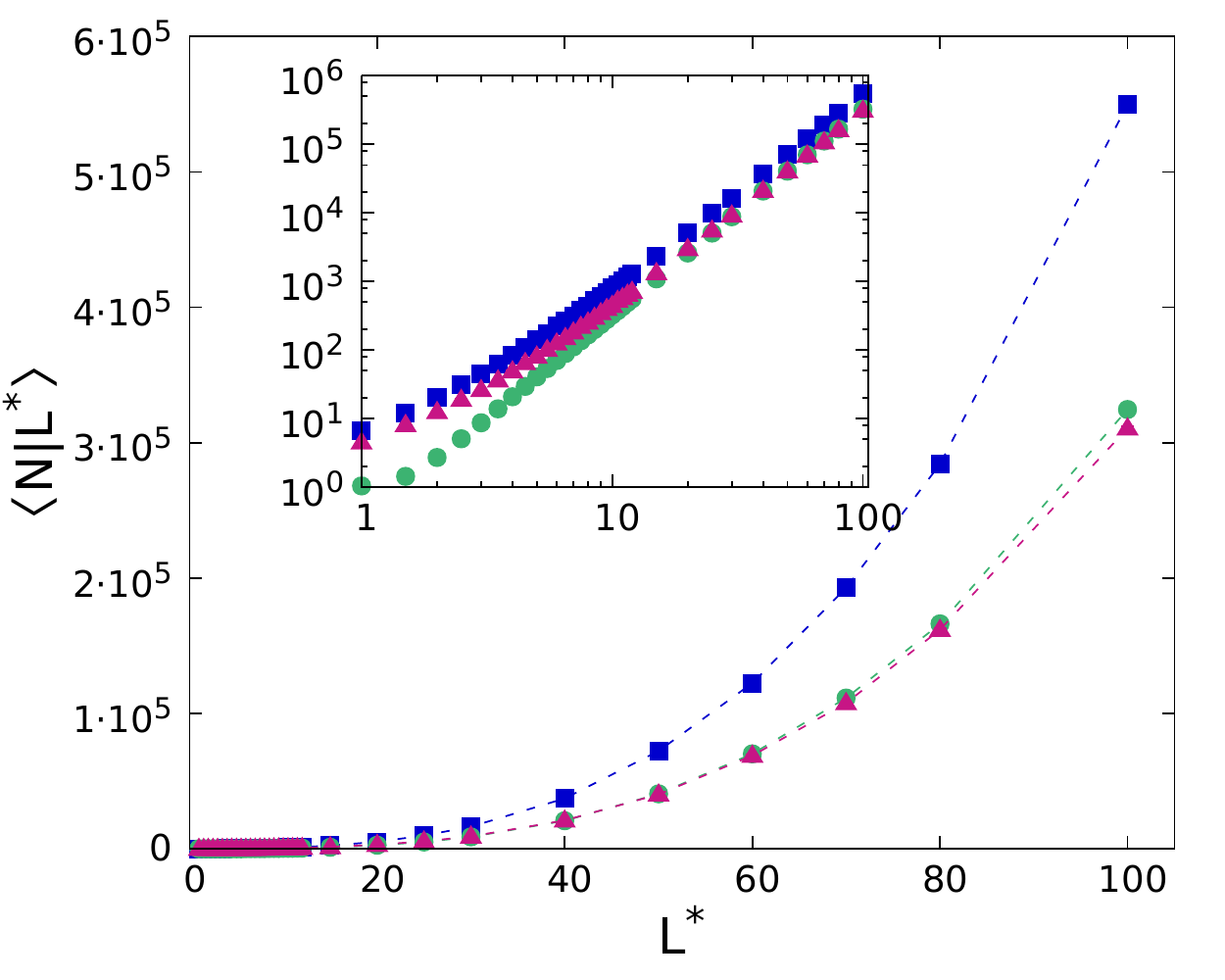}
\end{center}
\caption{The average number $\langle N | L^*\rangle$ of polyhedra composing the tessellation, as a function of the normalized size $L^*$ of the domain and of the mixing statistics $m$. Symbols correspond to the Monte Carlo simulation results, with dashed lines added to guide the eye: blue squares denote $m={\cal P}$, green circles $m={\cal V}$, and red triangles $m={\cal B}$. The inset displays the same data in log-log scale.}
\label{nbp_size}
\end{figure}

\begin{figure}[t]
\begin{center}
\includegraphics[scale=0.6]{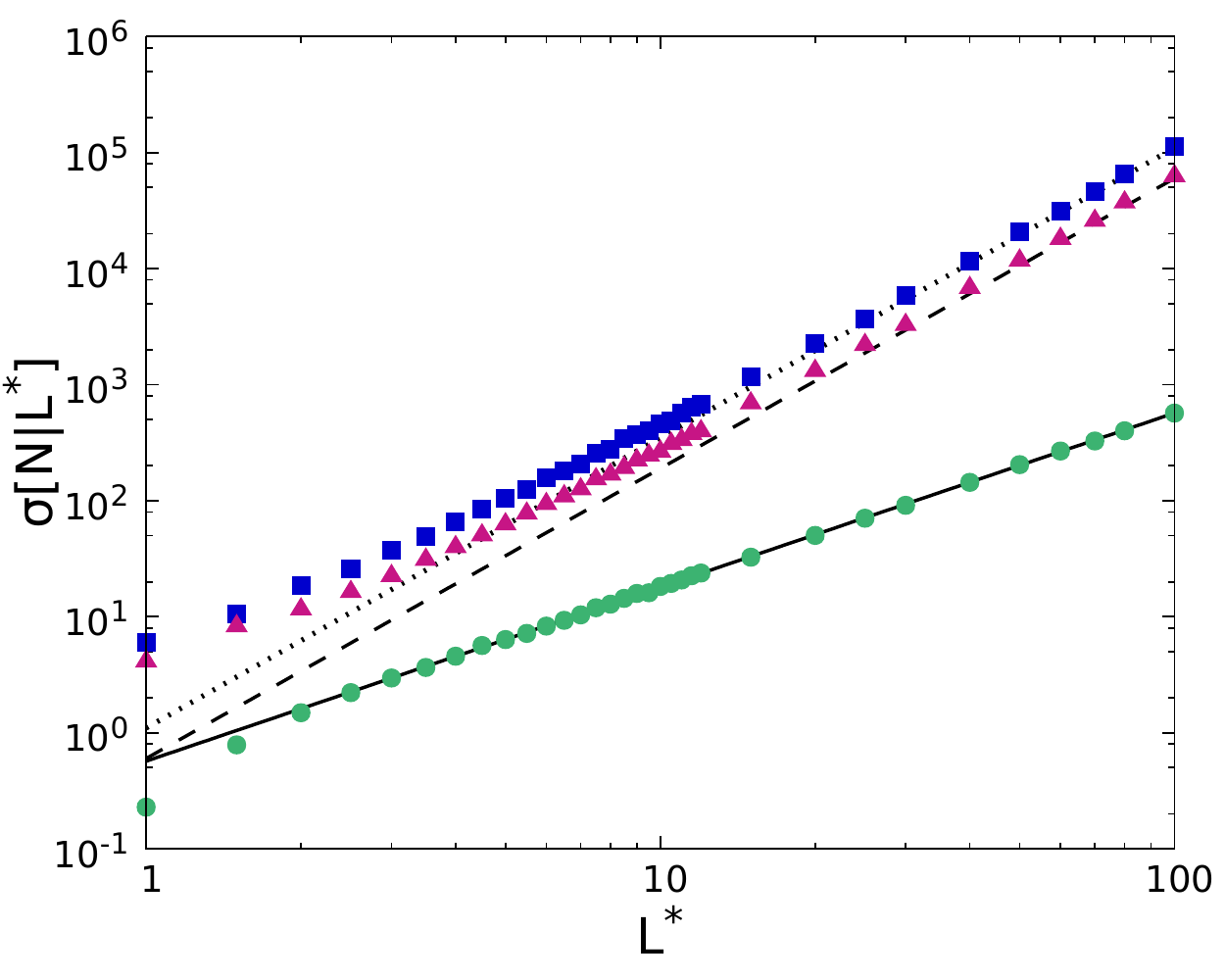}
\end{center}
\caption{The standard deviation $\sigma[N|L^*]$ of the number of polyhedra  composing the tessellation, as a function of the normalized size $L^*$ of the domain and of the mixing statistics $m$. Symbols correspond to the Monte Carlo simulation results: blue squares denote $m={\cal P}$, green circles $m={\cal V}$, and red triangles $m={\cal B}$. The scaling law $(L^*)^{5/2}$ is displayed for reference with dashed or dotted lines; the scaling law $(L^*){3/2}$ is displayed with a continuous line.}
\label{std_nbp_size}
\end{figure}

\subsection{Setting the model parameters}

In order to compare the effects of the underlying mixing statistics $m={\cal P}$, ${\cal V}$ or ${\cal B}$ on particle transport in random media, a mandatory requirement is to determine a criterion on whose basis the tessellations can be considered statistically `equivalent' with respect to some physical property. An important point is that the three models examined here depend on a single free parameter, namely the average chord length ${\langle \Lambda \rangle}_{\infty}$ of the tessellation, plus the colouring probability $p$. In the Markovian binary mixtures, the average chord length of the Poisson tessellation and the colouring probability are chosen so that the resulting average chord lengths in the coloured clusters, ${\langle \Lambda_{\alpha} \rangle}_{\infty}$ and ${\langle \Lambda_{\beta} \rangle}_{\infty}$, intuitively represent the typical scale of the disorder in the random media, to be compared with the average mean free paths of the particles traversing the geometry~\cite{pomraning}.

By analogy with the Markovian binary mixing, we choose therefore to set ${\langle \Lambda_{\alpha} \rangle}_{\infty}$ and ${\langle \Lambda_{\beta} \rangle}_{\infty}$ to be equal for the two other mixing statistics $m={\cal V}$ or ${\cal B}$. According to Eqs.~\ref{alpha} and~\ref{beta}, this can be achieved by choosing the same parameters ${\langle \Lambda \rangle}_{\infty}$ and $p$ for any tessellation. Correspondingly, we have a constraint on the densities $\rho_{\cal P}$, $\rho_{\cal V}$ and $\rho_{\cal B}$ of the tessellations, which must now satisfy
\begin{equation}
\frac{1}{{\langle \Lambda \rangle}_{\infty}} =\rho_{\cal P}=0.6872 \rho_{\cal V} =\frac{2}{3} \rho_{\cal B}.
\label{calib}
\end{equation}
In practice, one is often lead to simulate tessellations restricted to some bounded regions of linear size $L$: finite-size effects typically emerge, and the relation~\eqref{calib} would not be strictly valid. Indeed, in finite geometries the average chord length ${\langle \Lambda | L \rangle}$ differs from the corresponding ${\langle \Lambda \rangle}_{\infty}$ (see also Sec.~\ref{size_effects}). However, although a large $L$ is typically required in order for ${\langle \Lambda  | L \rangle}$ to converge to the asymptotic value ${\langle \Lambda \rangle}_{\infty}$, the variability of ${\langle \Lambda | L \rangle}$ between tessellations vanishes for comparatively smaller $L$, as illustrated in Fig.~\ref{cl_size}. The same remark applies to ${\langle \Lambda_{\alpha} | L\rangle}$ and ${\langle \Lambda_{\beta} | L\rangle}$, as shown in Fig.~\ref{cl_p}. For the sake of simplicity, we will thus neglect such finite-size effects and use Eq.~\eqref{calib} to calibrate the model parameters.

\begin{figure}[t]
\begin{center}
\includegraphics[scale=0.6]{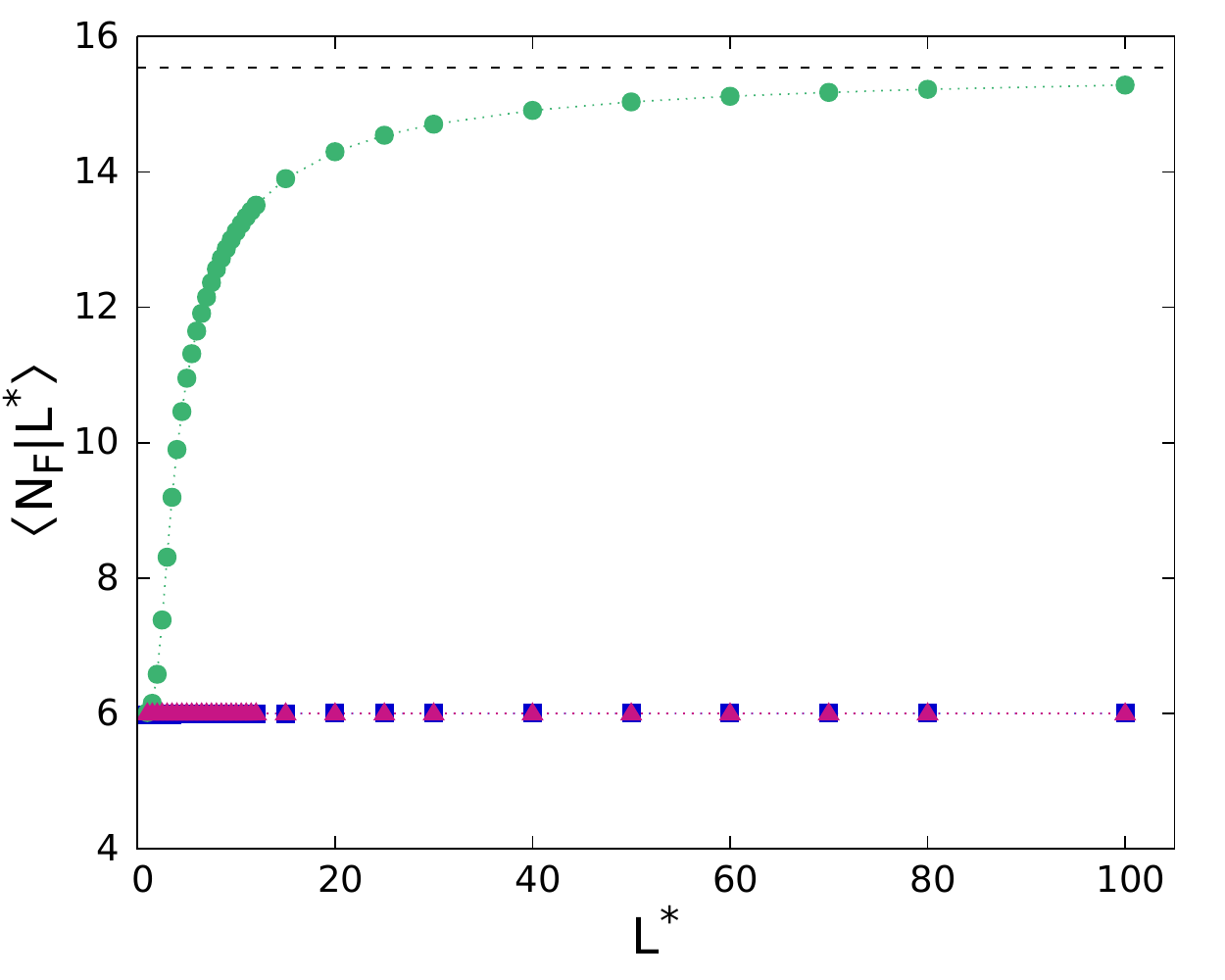}
\end{center}
\caption{The average number of faces by polyhedron $\langle N_F | L^*\rangle$, as a function of the normalized size $L^*$ of the domain and of the mixing statistics $m$. Symbols correspond to the Monte Carlo simulation results, with dotted lines added to guide the eye: blue squares denote $m={\cal P}$, green circles $m={\cal V}$, and red triangles $m={\cal B}$. The asymptotic value for $m={\cal V}$, ${\langle N_F \rangle}_{\infty}=15.54$, is displayed for reference with a black dashed line.}
\label{connect_size}
\end{figure}

\section{Statistical properties and finite-size effects}
\label{size_effects}

In the following, we will focus on tessellations restricted to a cubic box of linear size $L$, and investigate the impact of finite-size effects on the statistical properties of such random media. By generating a large number of random tessellations by Monte Carlo simulation via the algorithms described above, we can assess the convergence of the moments and distributions of arbitrary physical observables to their limit behaviour for infinite domains. A Monte Carlo code capable of generating Poisson, Voronoi and Box tessellations with arbitrary average chord length ${\langle \Lambda \rangle}_{\infty}$ has been developed to this aim. 

\subsection{Number of polyhedra}

The number $N$ of polyhedra composing a tessellation provides a measure of the complexity of the resulting geometries. We will analyse the growth of this quantity in Poisson, Voronoi or Box tessellations as a function of the normalized size $L^*=L/{\langle \Lambda \rangle}_{\infty}$ of the domain (with both $L$ and ${\langle \Lambda \rangle}_{\infty}$ expressed in arbitrary units). The simulation findings for the average number $\langle N | L^*\rangle$  of polyhedra are illustrated in Fig.~\ref{nbp_size}, whereas Fig.~\ref{std_nbp_size} displays results concerning the standard deviation $\sigma[N|L^*]$ of $N$. To begin with, we observe that $\langle N | L^* \rangle$ is smaller in Voronoi and Box tessellations than in Poisson tessellations. However, for large $L$, we find a common asymptotic scaling law $\langle N | L^* \rangle \sim {(L^*)}^3$. The growth of the dispersion $\sigma[N|L^*]$ differs considerably between Voronoi tessellations and the two other models. In the case of Voronoi mixing statistics, we asymptotically find $\sigma[N|L^*] \sim (L^*)^\frac{3}{2}$: this is not surprising, since by construction $N$ coincides with the number of seeds $N_S$ (unless $N_S=0$) and is assumed to follow a Poisson distribution of intensity proportional to $(L^*)^3$, according to Eq.~\ref{calib}. For Poisson and Box tessellations, the dispersion is significantly larger and the asymptotic scaling law becomes $\sigma[N|L^*] \sim (L^*)^\frac{5}{2}$. Therefore, the distribution of $N$ is peaked around its average value in Voronoi tessellations, whereas it is more dispersed in Poisson and Box geometries.

\subsection{Connectivity}

The number of faces $N_F$ of each polyhedron intuitively represents the connectivity degree of the tessellation. For this observable, Voronoi geometries are expected to have a peculiar behaviour, since the average number of faces amounts to $\langle N_F \rangle_\infty \simeq 15.54$ (for infinite domains), which is much larger than the value $\langle N_F \rangle_\infty = 6$ for Poisson and Box geometries (see, e.g.,~\cite{santalo, miles1972} and references therein). In order to verify this behaviour in finite geometries, we have numerically computed by Monte Carlo simulation the average number of faces in Poisson, Voronoi and Box tessellations, as a function of the normalized size $L^*$ of the domain. The numerical results are illustrated in Fig.~\ref{connect_size}. The average number of faces converges towards the expected value for each tessellation; nevertheless, Voronoi tessellations need much larger $L^*$ to attain the asymptotic behaviour.

\subsection{Chord lengths}

In order to investigate finite-size effects on chord lengths, we have numerically computed by Monte Carlo simulation the distribution and the average of the chord length for each mixing statistics $m$. To this aim, we resort to the following method. A random tessellation is first generated, and a line obeying the $\mu$-randomness is then drawn according to the prescriptions in~\cite{coleman}. The intersections of the line with the polyhedra of the geometry are computed, and the resulting segment lengths are recorded. This step is repeated for a large number of random lines. Then, a new geometry is generated and the whole procedure is iterated for several geometries, in order to get satisfactory statistics.

\begin{figure*}[t]
\begin{center}
\,\,\,\, $L=1$ \,\,\,\,\\
\includegraphics[width=0.7\columnwidth]{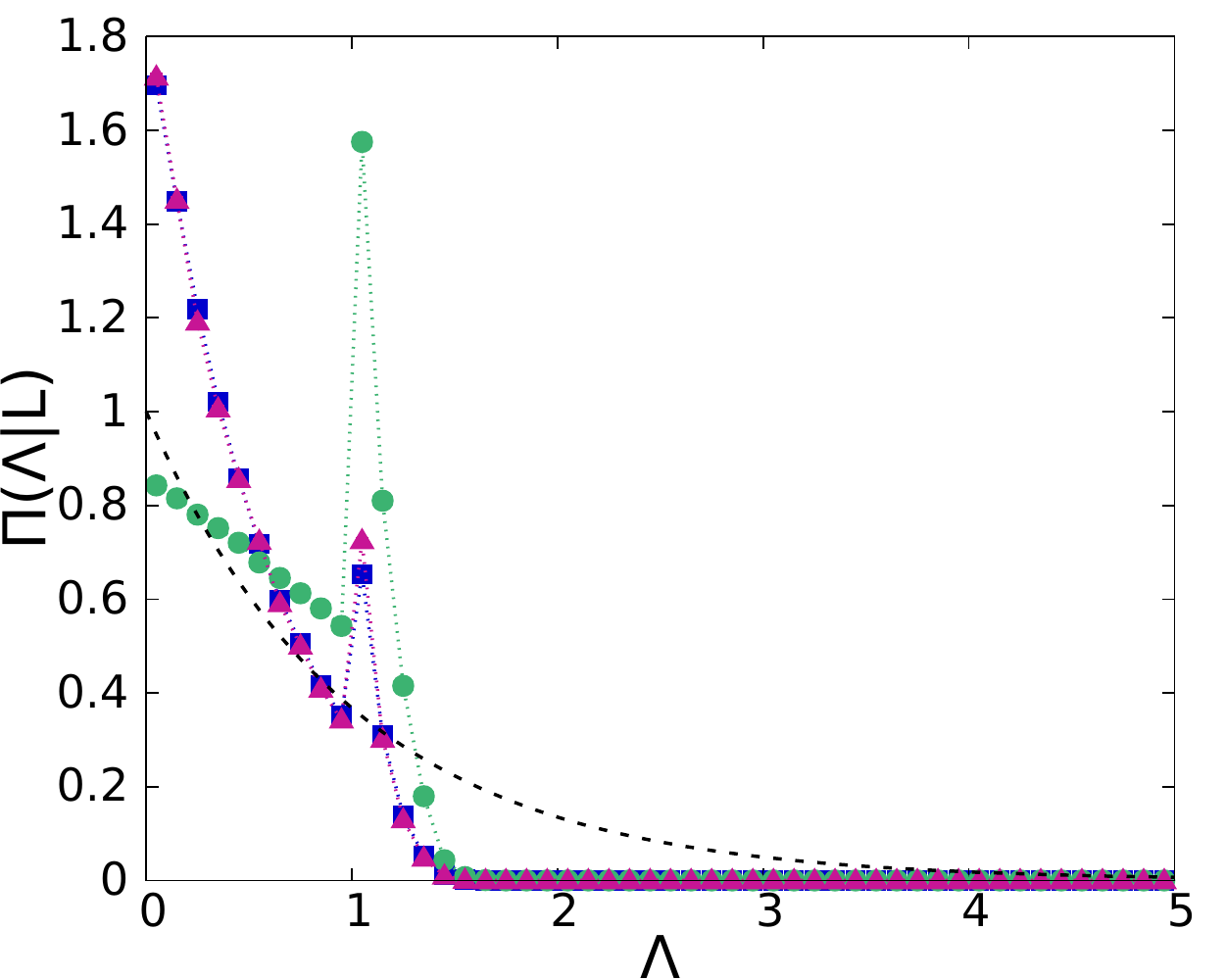}\\
\,\,\,\, $L=2$ \,\,\,\,\\
\includegraphics[width=0.7\columnwidth]{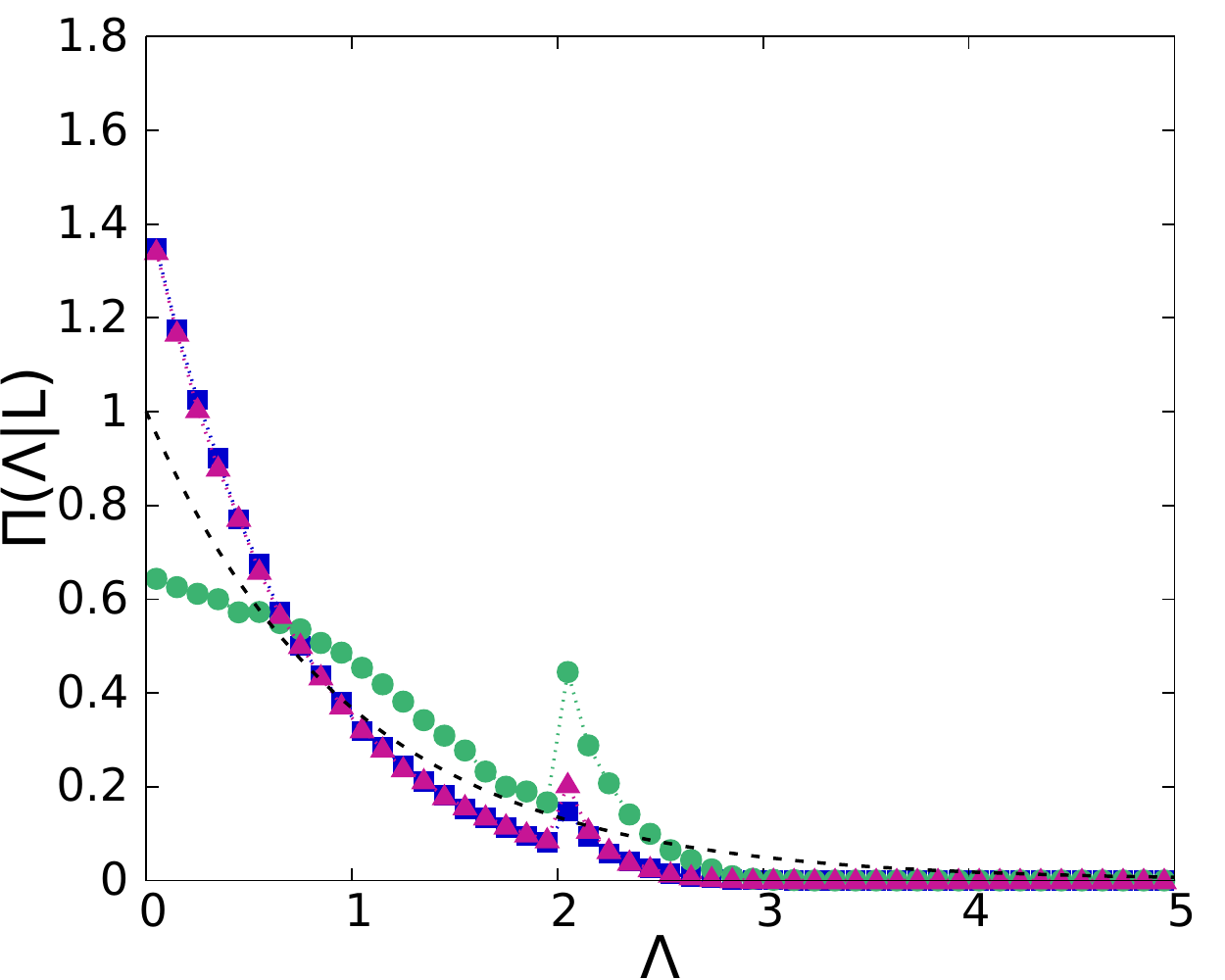}\\
\,\,\,\, $L=10$ \,\,\,\,\\
\includegraphics[width=0.7\columnwidth]{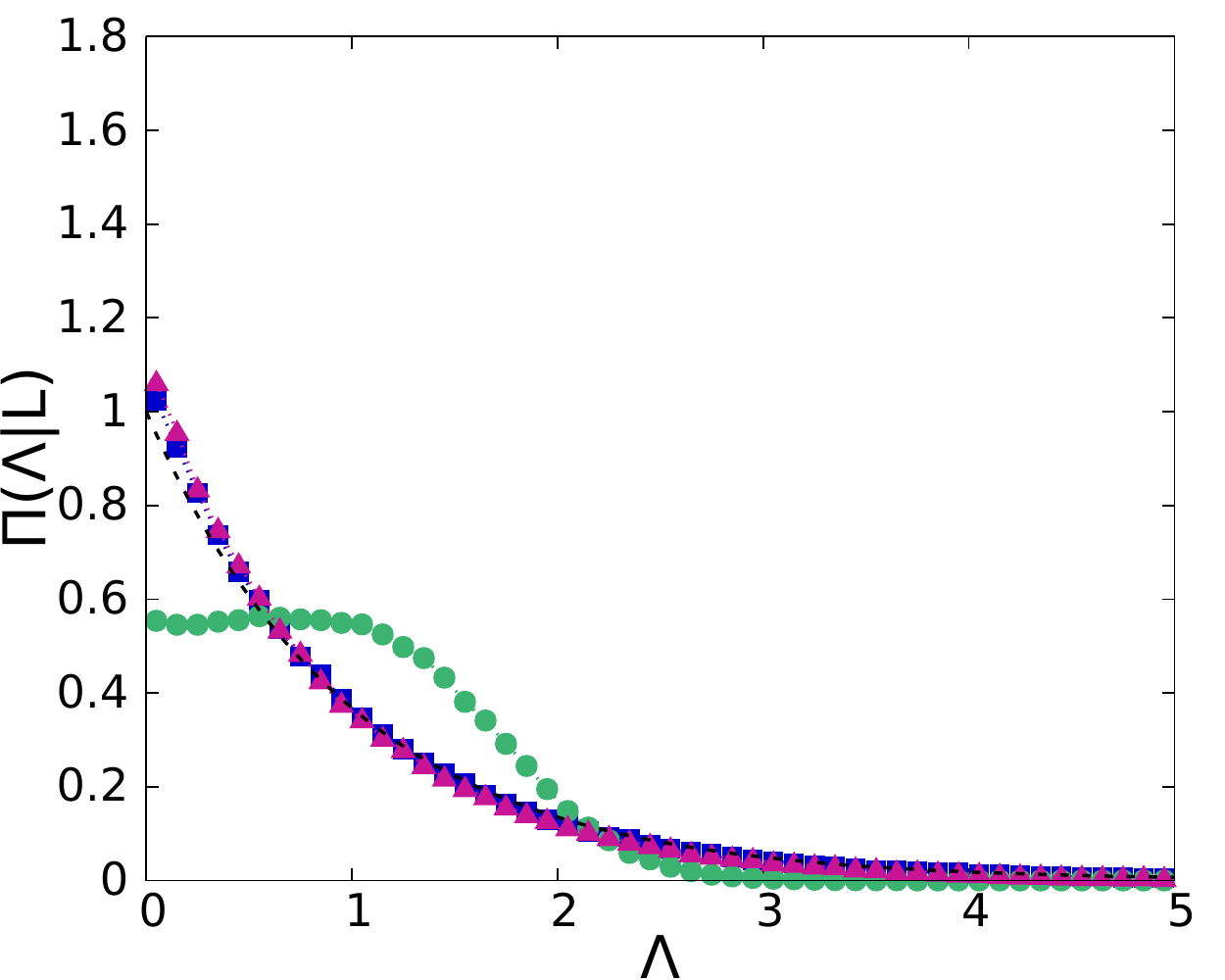}\\
\,\,\,\, $L=100$ \,\,\,\,\\
\includegraphics[width=0.7\columnwidth]{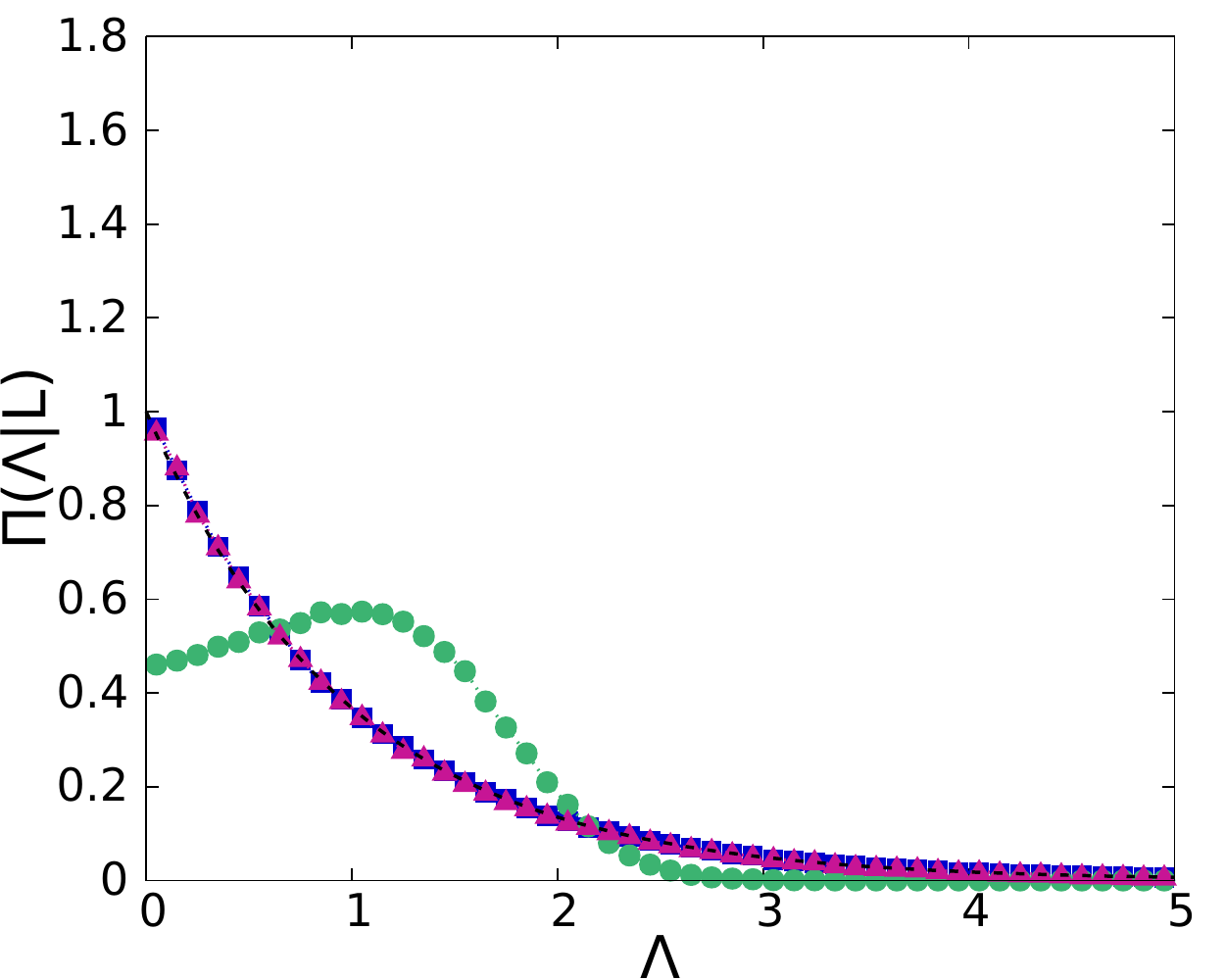}\\
\end{center}
\caption{The probability density $\Pi(\Lambda | L)$ of the chord length, as a function of the mixing statistics $m$, for ${\langle \Lambda \rangle}_{\infty}=1$, for several values of the linear size $L$ of the domain: $L=1$, $L=2$, $L=10$, $L=100$ (from top to bottom). Symbols correspond to the Monte Carlo simulation results, with lines added to guide the eye; blue squares denote $m={\cal P}$, green circles $m={\cal V}$, and red triangles $m={\cal B}$. For $m={\cal P}$, the asymptotic (i.e., $L \rightarrow \infty $) exponential distribution given in Eq.\ref{markovian_density} is displayed as a black dashed line for reference.}
\label{cl_distrib}
\end{figure*}

The numerical results for the normalized average chord length ${\langle \Lambda^* | L^* \rangle}={\langle \Lambda | L^* \rangle} / {\langle \Lambda \rangle}_{\infty}$ as a function of the normalized size $L^*$ of the domain are illustrated in Fig.~\ref{cl_size} for different mixing statistics $m$. Monte Carlo simulation results for the chord length distribution are shown in Fig.~\ref{cl_distrib}, for ${\langle \Lambda \rangle}_{\infty}=1$ and for several values of $L$. For small $L$, finite-size effects are visible in the chord length distribution: indeed, the longest length that can be drawn across a box of linear size $L$ is $\sqrt{3}L$, which thus induces a cut-off on the distribution. For large $L$, the finite-size effects due to the cut-off fade away. In particular, the probability density for Poisson tessellations eventually converges to the expected exponential behaviour. Simulations show that the chord length distributions in Box tessellations and in Poisson tessellations are very close, which is consistent with the observations in~\cite{mikhailov}. On the contrary, in Voronoi tessellations, the probability density has a distinct non-exponential functional form.

\begin{figure*}
\begin{center}
${\langle \Lambda \rangle}_{\infty}=1$\,\,\,-\,\,\,$p=0.15$\,\,\,\,\,\,\,\,\,\,\,\,\,\,\,\,\,\,\,\,\,\,\,\,\,\,\,\,\,\,\,\,\,\,\,\,\,\,\,\,\,\,\,${\langle \Lambda \rangle}_{\infty}=0.5$\,\,\,-\,\,\,$p=0.15$\,\,\,\,\,\,\,\,\,\,\,\,\,\,\,\,\,\,\,\,\,\,\,\,\,\,\,\,\,\,\,\,\,\,\,\,\,\,\,\,\,\,\,\,\,\,\,\,${\langle \Lambda \rangle}_{\infty}=0.1$\,\,\,-\,\,\,$p=0.15$\\
\includegraphics[width=0.65\columnwidth]{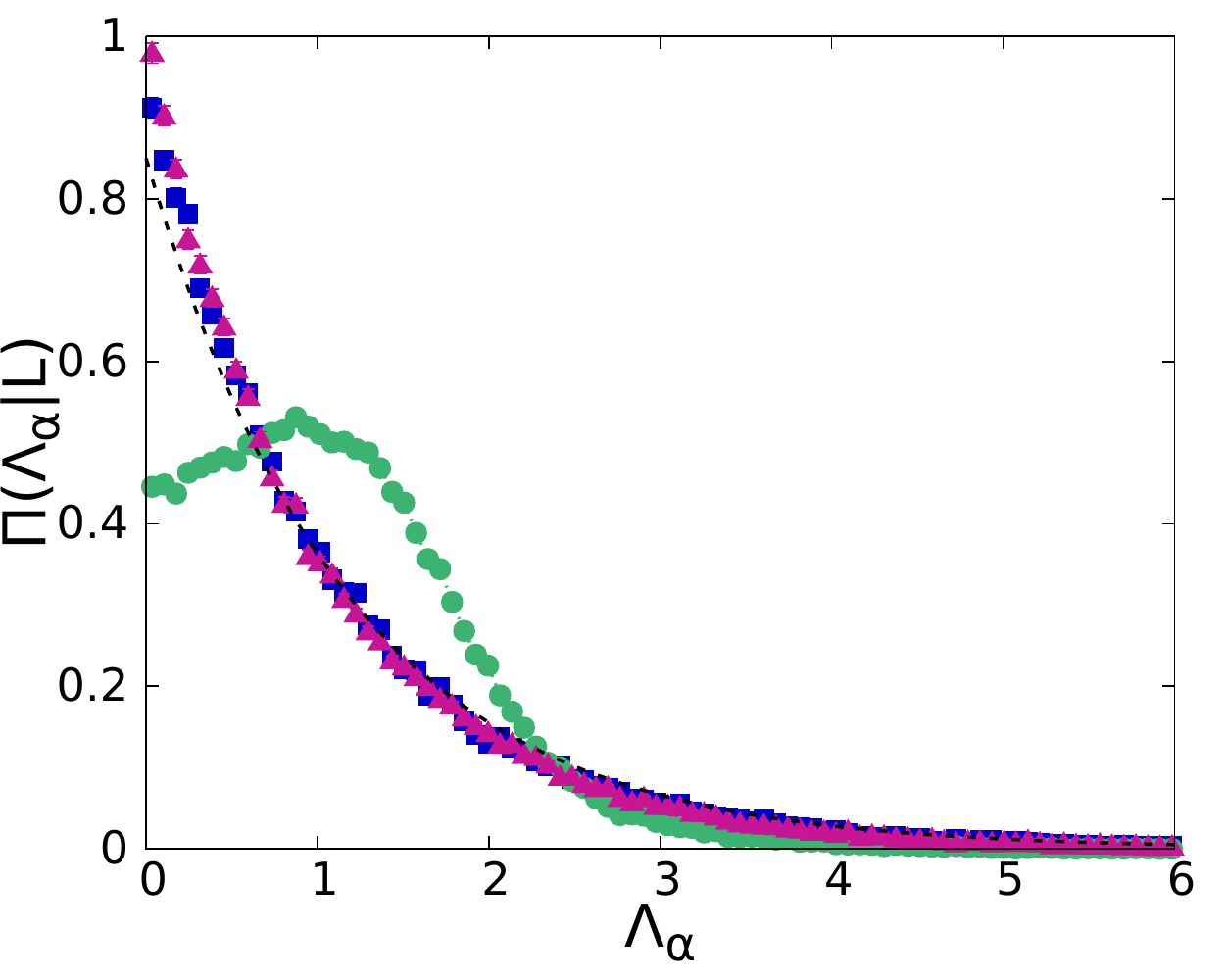}\,\,\,\,
\includegraphics[width=0.65\columnwidth]{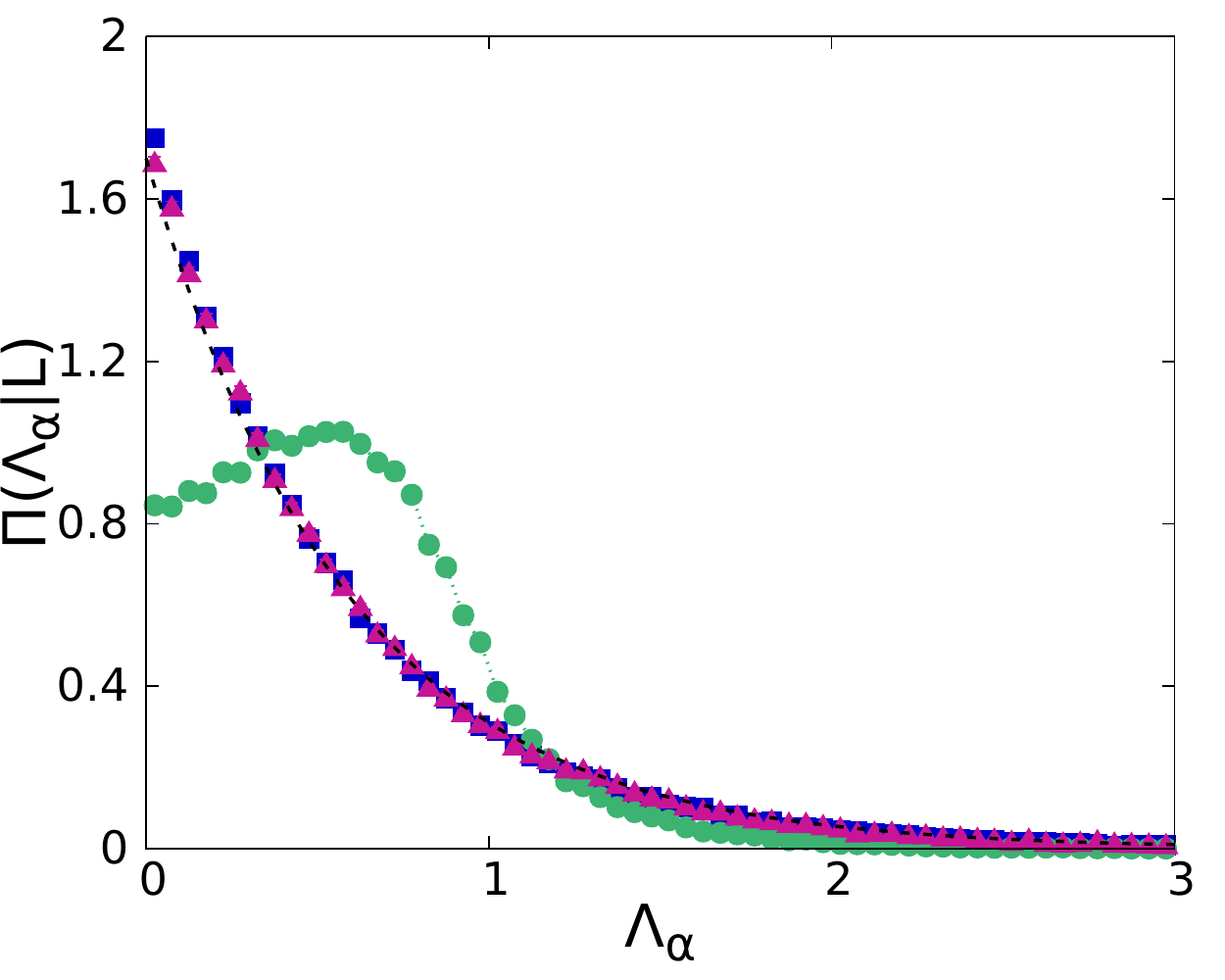}\,\,\,\,
\includegraphics[width=0.65\columnwidth]{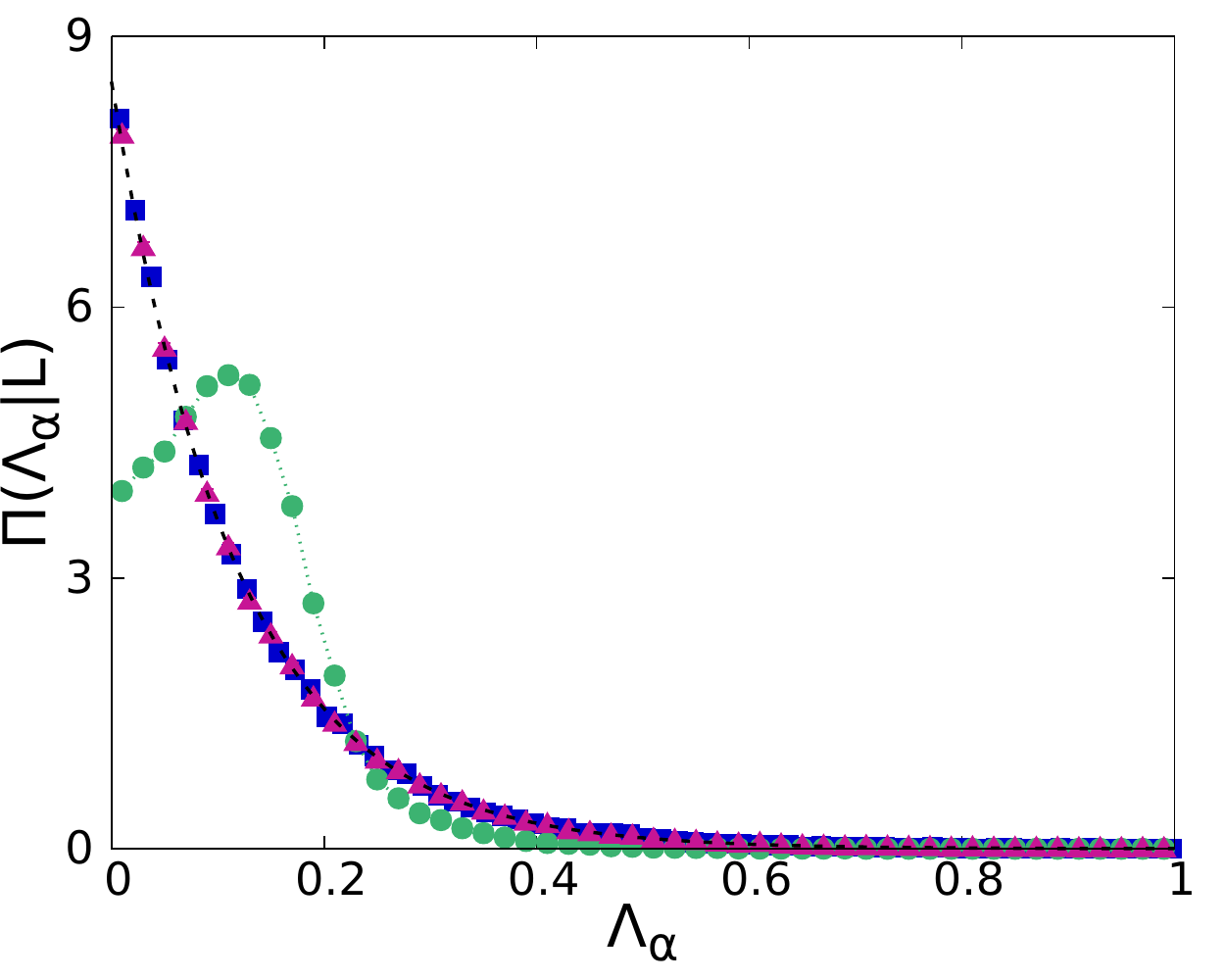}\\
${\langle \Lambda \rangle}_{\infty}=1$\,\,\,-\,\,\,$p=0.5$\,\,\,\,\,\,\,\,\,\,\,\,\,\,\,\,\,\,\,\,\,\,\,\,\,\,\,\,\,\,\,\,\,\,\,\,\,\,\,\,\,\,\,${\langle \Lambda \rangle}_{\infty}=0.5$\,\,\,-\,\,\,$p=0.5$\,\,\,\,\,\,\,\,\,\,\,\,\,\,\,\,\,\,\,\,\,\,\,\,\,\,\,\,\,\,\,\,\,\,\,\,\,\,\,\,\,\,\,\,\,\,\,\,${\langle \Lambda \rangle}_{\infty}=0.1$\,\,\,-\,\,\,$p=0.5$\\
\includegraphics[width=0.65\columnwidth]{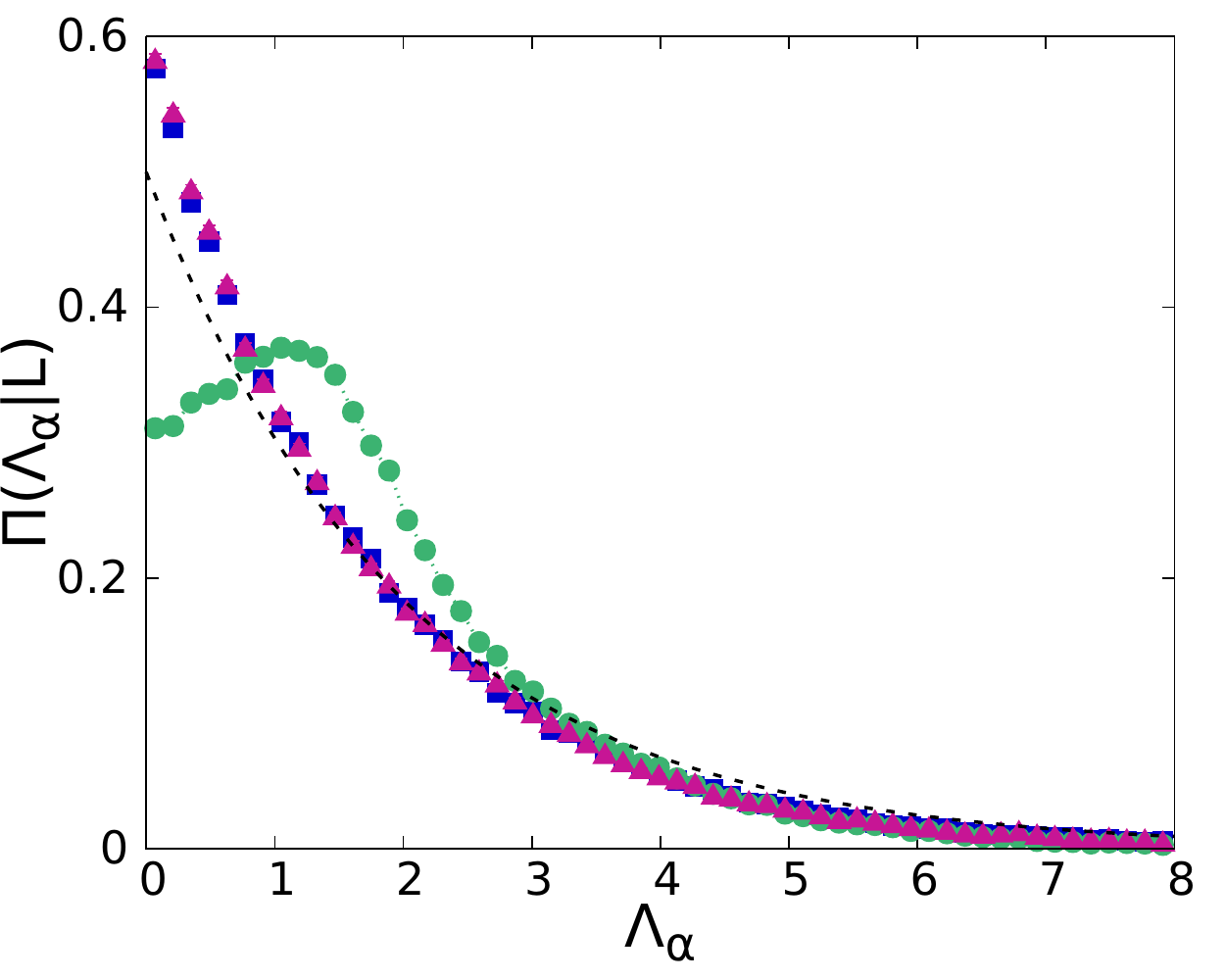}\,\,\,\,
\includegraphics[width=0.65\columnwidth]{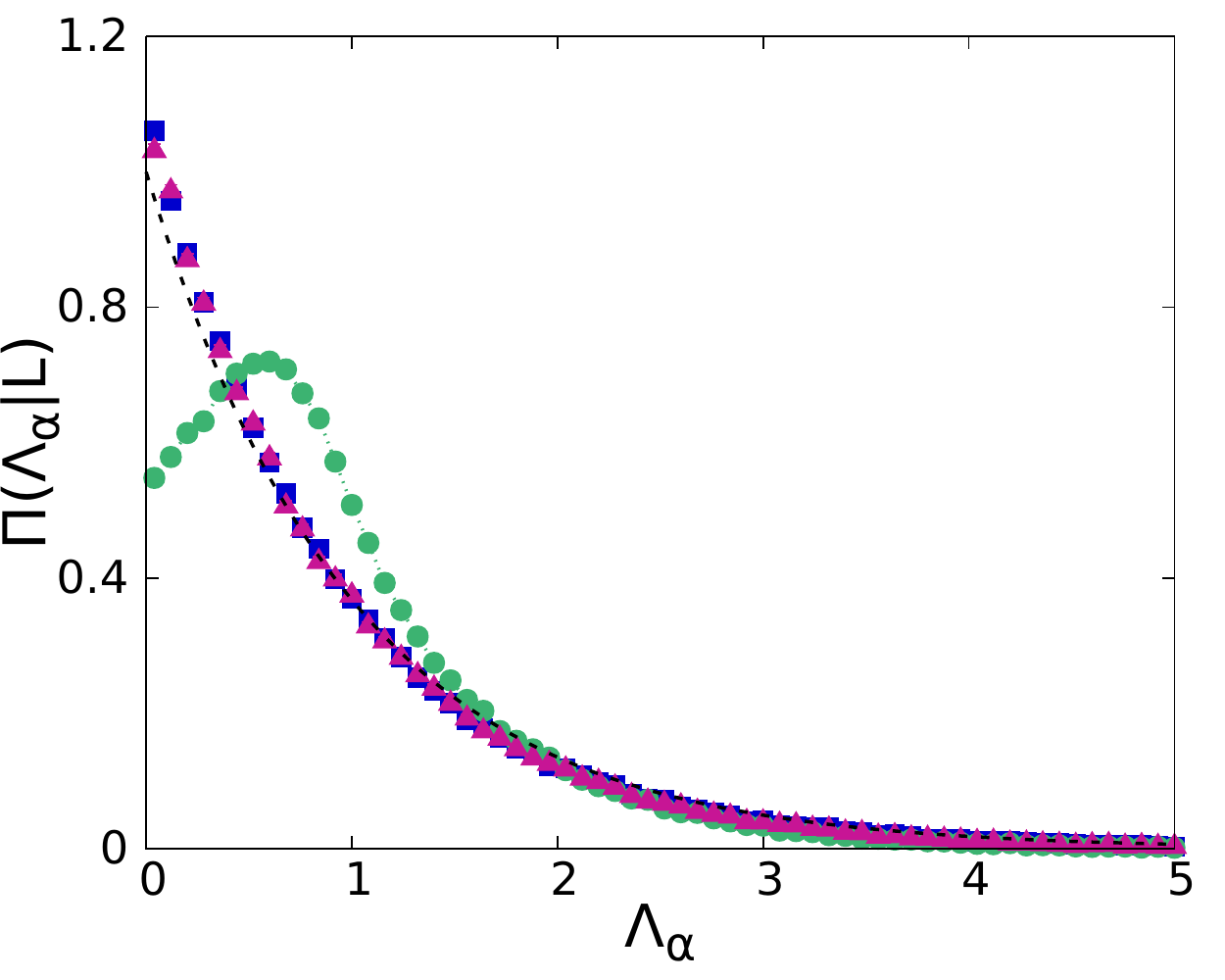}\,\,\,\,
\includegraphics[width=0.65\columnwidth]{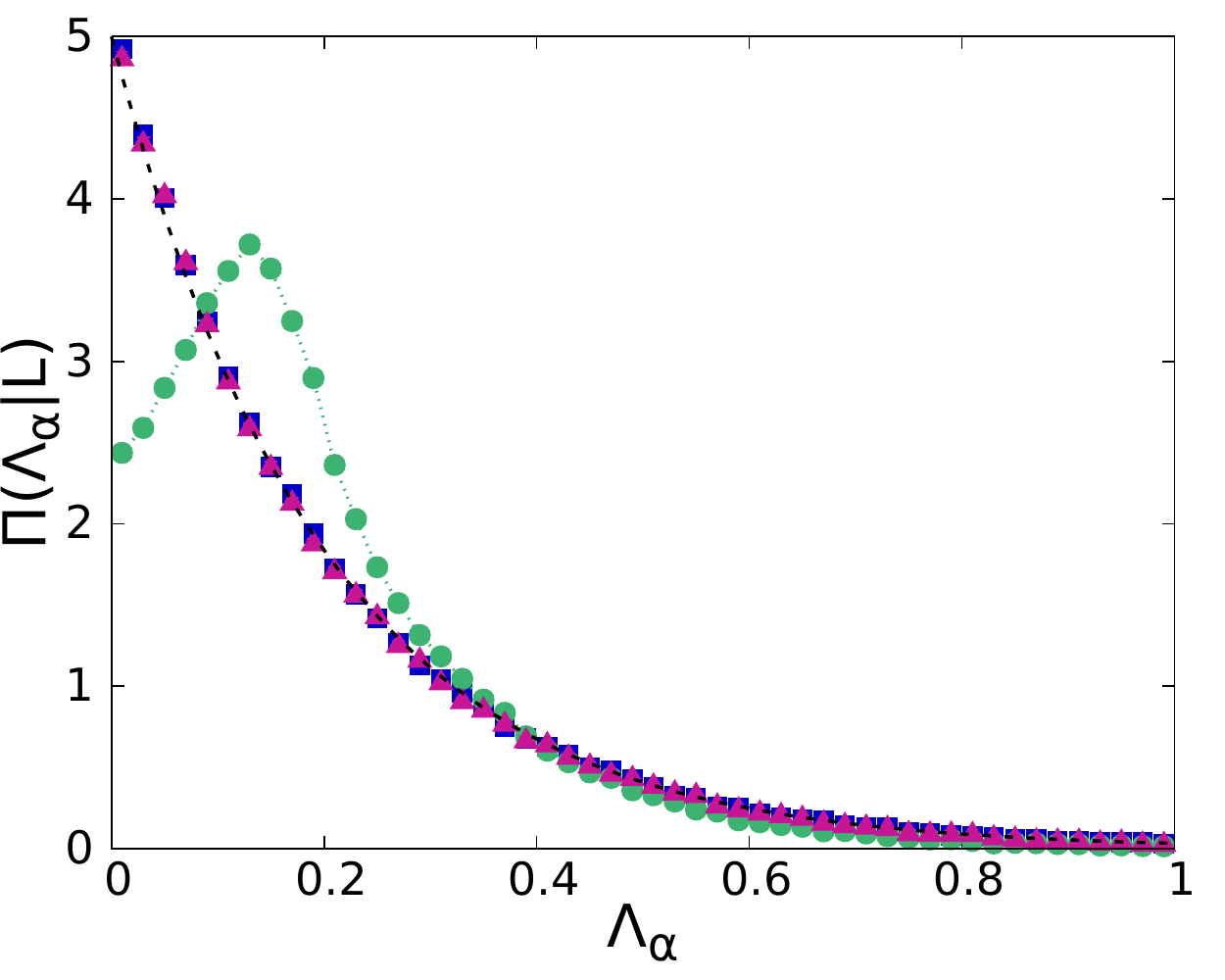}\\
${\langle \Lambda \rangle}_{\infty}=1$\,\,\,-\,\,\,$p=0.85$\,\,\,\,\,\,\,\,\,\,\,\,\,\,\,\,\,\,\,\,\,\,\,\,\,\,\,\,\,\,\,\,\,\,\,\,\,\,\,\,\,\,\,${\langle \Lambda \rangle}_{\infty}=0.5$\,\,\,-\,\,\,$p=0.85$\,\,\,\,\,\,\,\,\,\,\,\,\,\,\,\,\,\,\,\,\,\,\,\,\,\,\,\,\,\,\,\,\,\,\,\,\,\,\,\,\,\,\,\,\,\,\,\,${\langle \Lambda \rangle}_{\infty}=0.1$\,\,\,-\,\,\,$p=0.85$\\
\includegraphics[width=0.65\columnwidth]{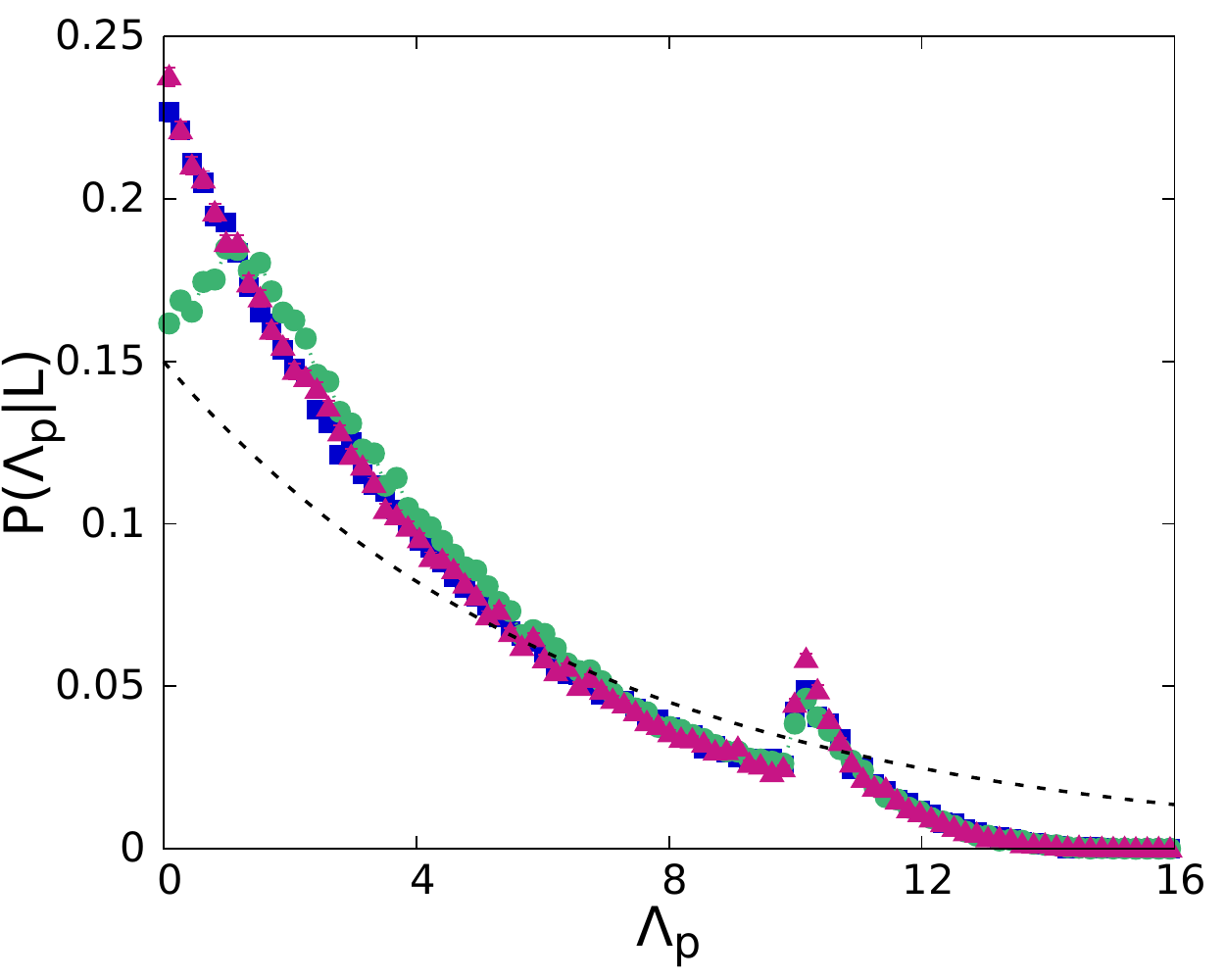}\,\,\,\,
\includegraphics[width=0.65\columnwidth]{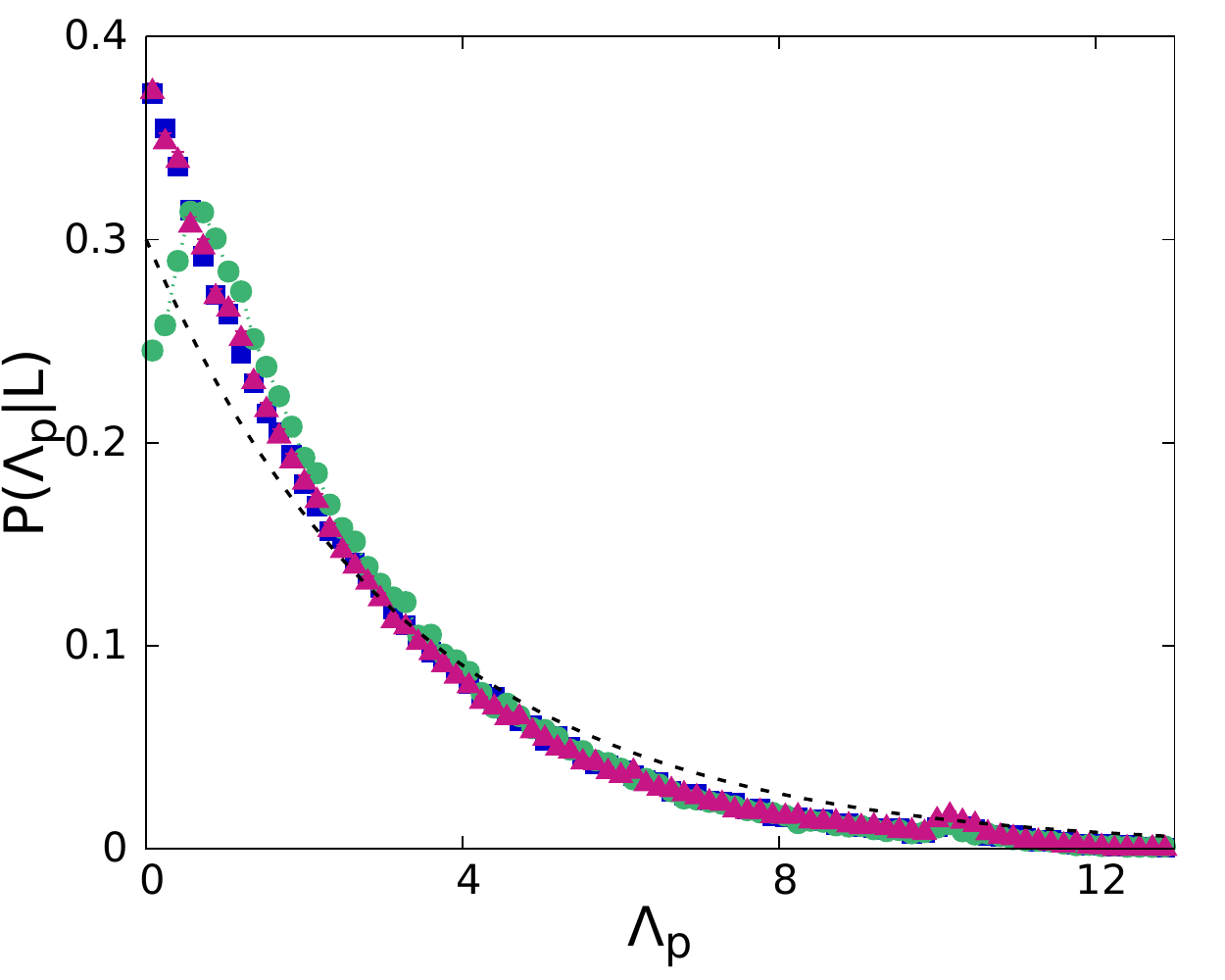}\,\,\,\,
\includegraphics[width=0.65\columnwidth]{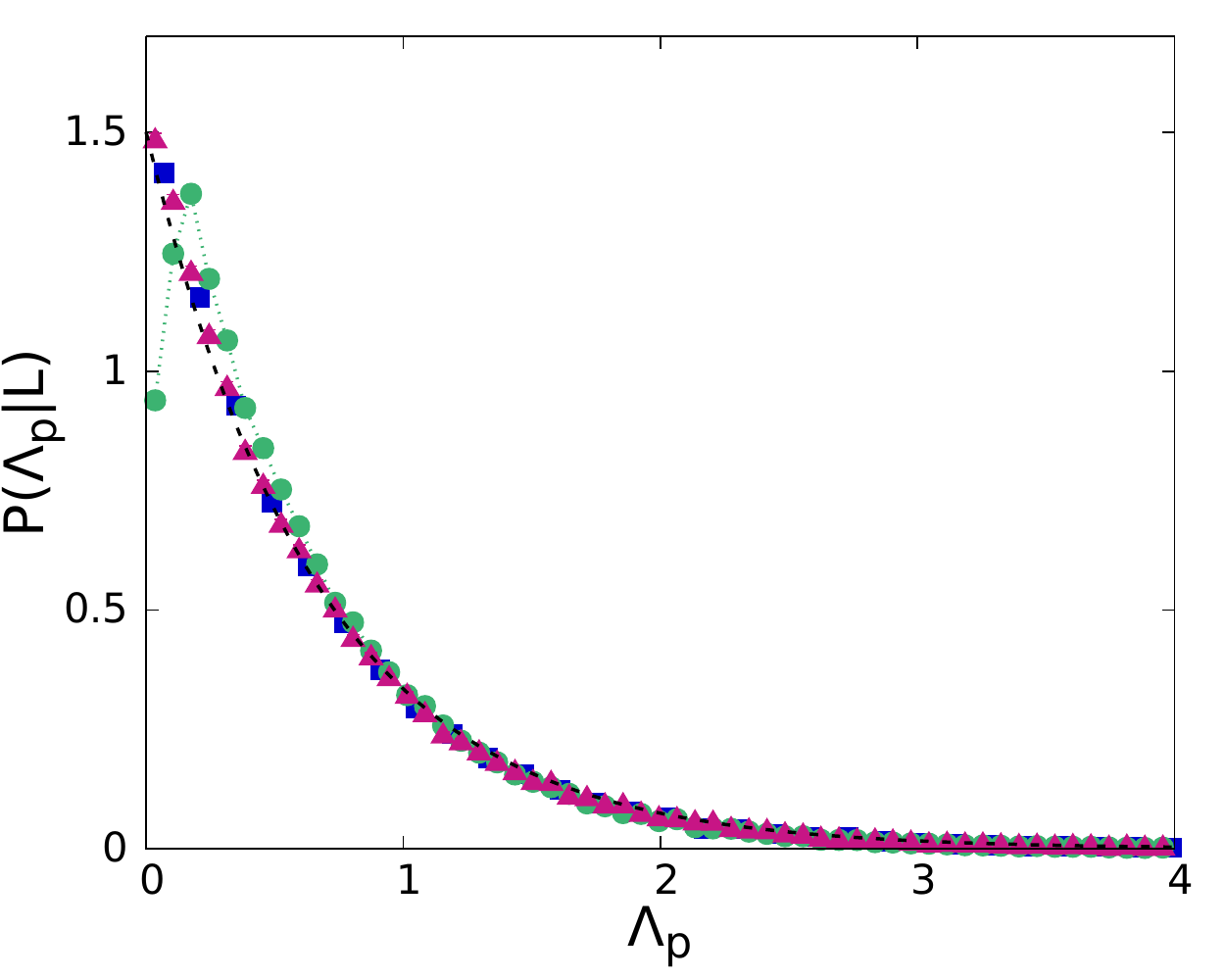}\\
${\langle \Lambda \rangle}_{\infty}=1$\,\,\,-\,\,\,$p=0.95$\,\,\,\,\,\,\,\,\,\,\,\,\,\,\,\,\,\,\,\,\,\,\,\,\,\,\,\,\,\,\,\,\,\,\,\,\,\,\,\,\,\,\,${\langle \Lambda \rangle}_{\infty}=0.5$\,\,\,-\,\,\,$p=0.95$\,\,\,\,\,\,\,\,\,\,\,\,\,\,\,\,\,\,\,\,\,\,\,\,\,\,\,\,\,\,\,\,\,\,\,\,\,\,\,\,\,\,\,\,\,\,\,\,${\langle \Lambda \rangle}_{\infty}=0.1$\,\,\,-\,\,\,$p=0.95$\\
\includegraphics[width=0.65\columnwidth]{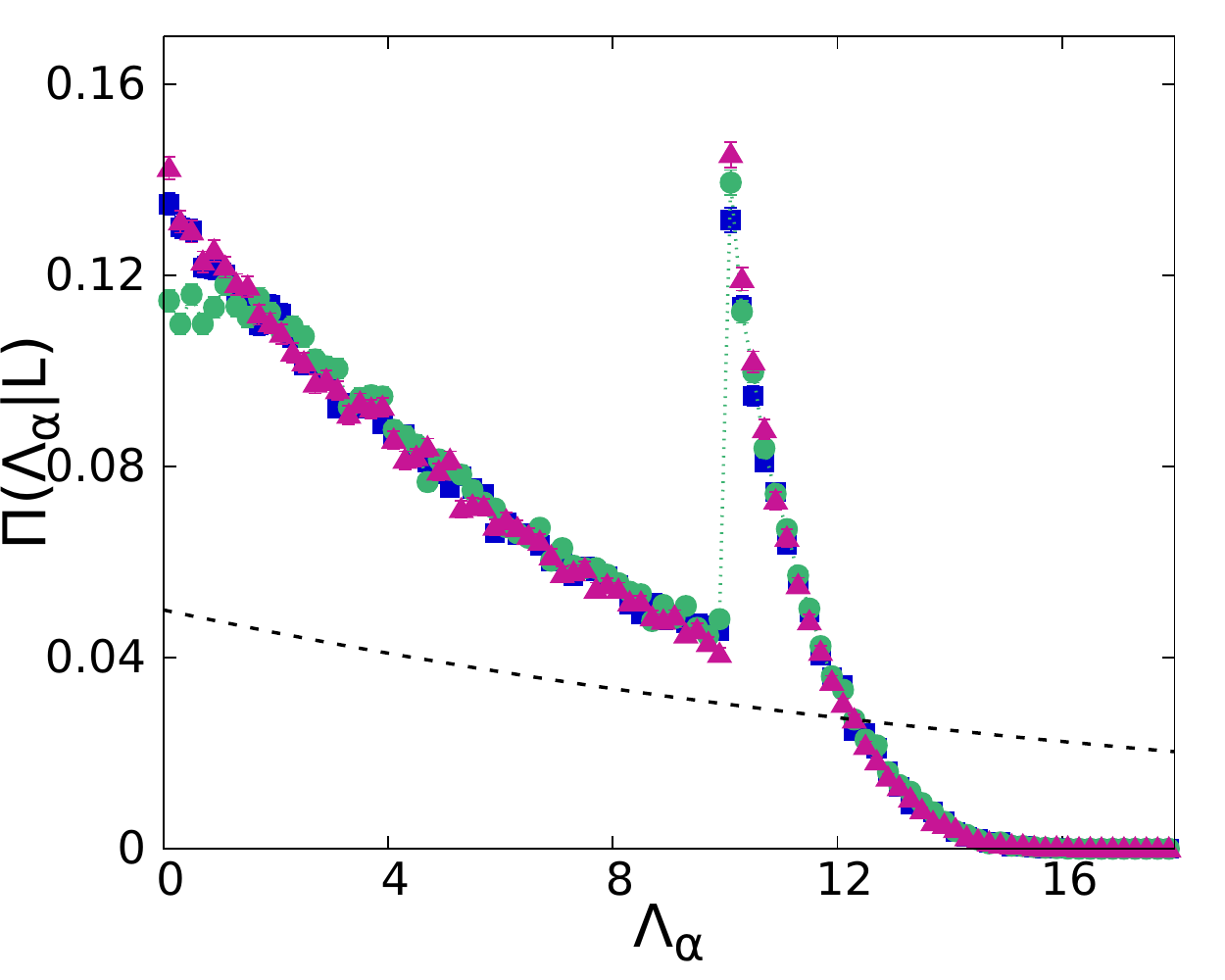}\,\,\,\,
\includegraphics[width=0.65\columnwidth]{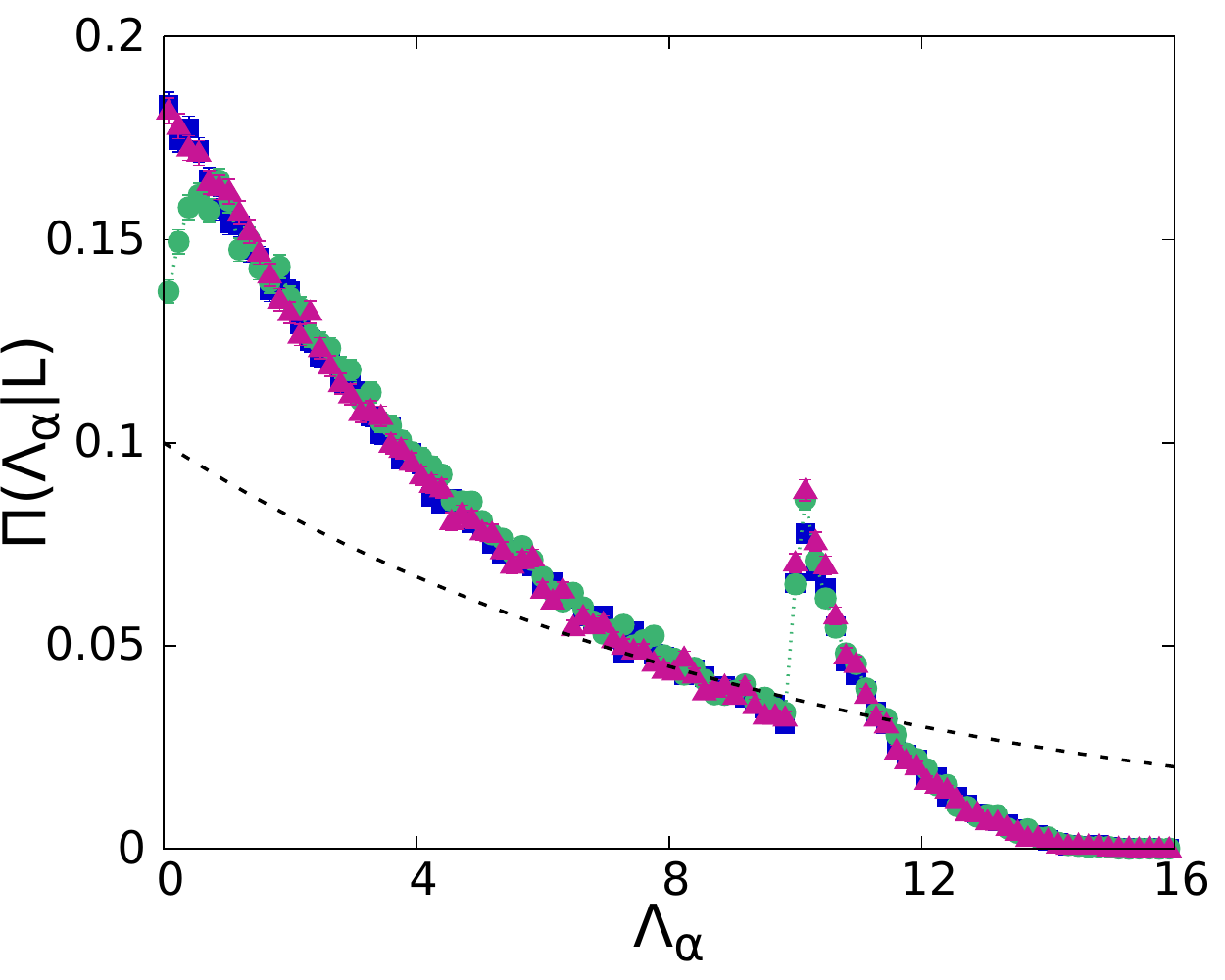}\,\,\,\,
\includegraphics[width=0.65\columnwidth]{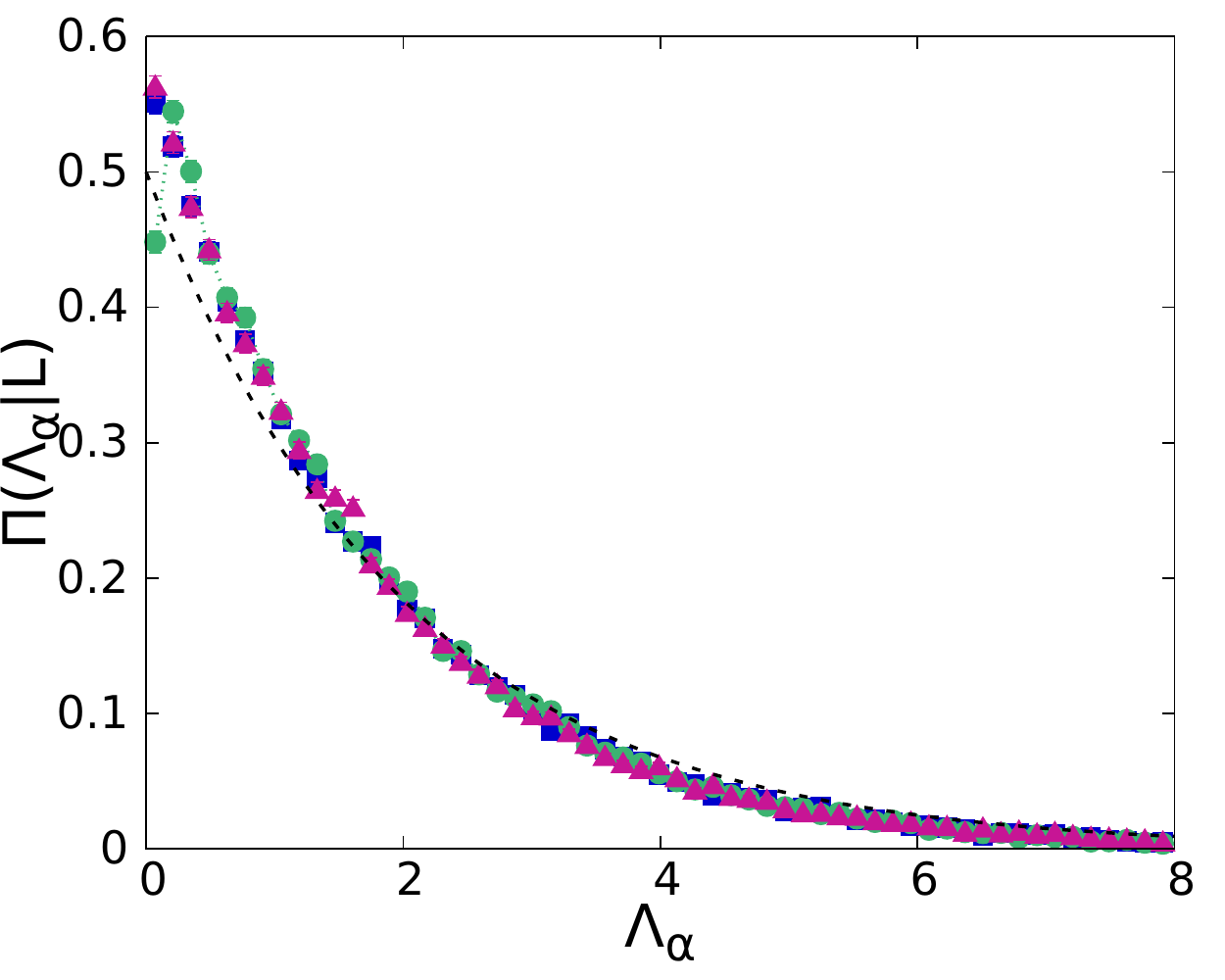}\\
\end{center}
\caption{The probability density $\Pi(\Lambda_{\alpha} | L)$ of the chord length through clusters of composition $\alpha$, as a function of the mixing statistics $m$, for several values of ${\langle \Lambda \rangle}_{\infty}$: ${\langle \Lambda \rangle}_{\infty}=1$ (left), ${\langle \Lambda \rangle}_{\infty}=0.5$ (center), ${\langle \Lambda \rangle}_{\infty}=0.1$ (right), and for several values of $p$: $p=0.15$, $p=0.5$, $p=0.85$, $p=0.95$ (from top to bottom). The domain has a linear size $L=10$. Symbols correspond to the Monte Carlo simulation results, with lines added to guide the eye: blue squares denote $m={\cal P}$, green circles $m={\cal V}$, and red triangles $m={\cal B}$. For $m={\cal P}$, the asymptotic (i.e., $L \rightarrow \infty $) exponential distribution given in Eq.~\ref{markovian_density_alpha} is displayed as a black dashed line for reference.}
\label{histo_cl_p}
\end{figure*}

A similar investigation can be conducted for $\langle \Lambda_{\alpha} | L \rangle$, the chord length through clusters with material composition $\alpha$. We have computed by Monte Carlo simulation the average value $\langle \Lambda_{\alpha} | L \rangle$ as a function of $p$, ${\langle \Lambda \rangle}_{\infty}$ and the mixing statistics $m$, for a given domain size $L=10$. Numerical findings are displayed in Fig.~\ref{cl_p}. Theoretical results in the limit of infinite domains are also provided. Finite-size effects are apparent, and their impact increases with increasing ${\langle \Lambda \rangle}_{\infty}$ and $p$. However, the discrepancy due to mixing statistics is rather weak. The distribution of the chord lengths through material of composition $\alpha$ is illustrated in Fig.~\ref{histo_cl_p}. Here again, the finite-size effects vanish for small $p$ and ${\langle \Lambda \rangle}_{\infty}$.

\subsection{Percolation properties}

For binary mixtures, percolation statistics plays an important role~\cite{larmier, lepage}. To fix the ideas, we will consider the percolation properties of the clusters of composition $\alpha$. In the limit of infinite geometries, the (site) percolation threshold $p_c$ is defined as the probability $p_\alpha$ above which there exists a `giant connected cluster', i.e., a collection of connected red polyhedra spanning the entire geometry~\cite{percolation_book}. The percolation probability $P_c(p)$, i.e., the probability that there exists such a connected percolating cluster, has thus a step behaviour as a function of the colouring probability $p$, i.e., $P_C (p) = 0$ for $ p < p_c$ , and $P_c(p) = 1$ for $p > p_c$. Actually, for any finite $L$, there exists a finite probability that a percolating cluster exists below $p = p_c$, due to finite-size effects.

\begin{figure}[t]
\begin{center}
\includegraphics[scale=0.6]{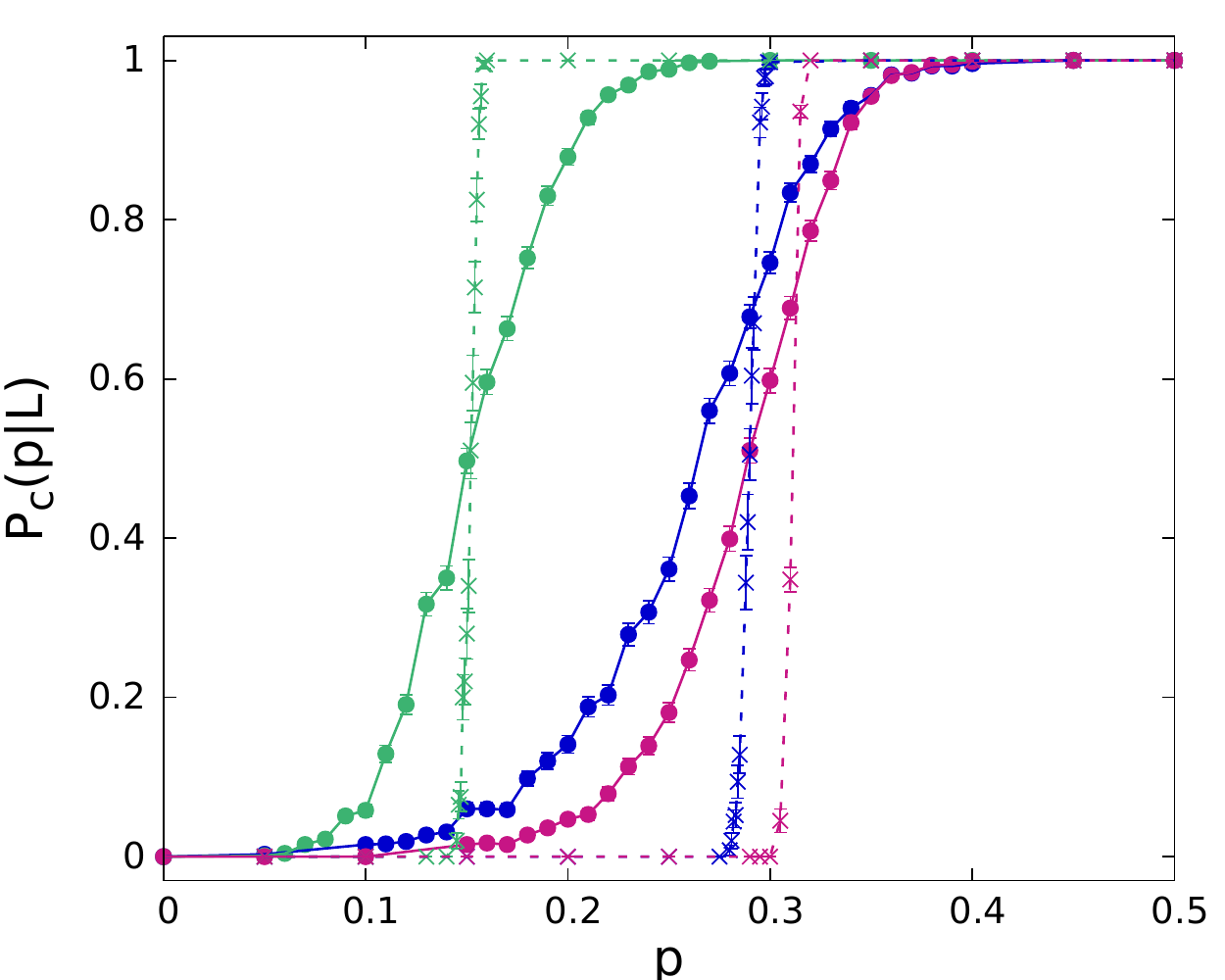}
\end{center}
\caption{The average percolation probability $P_c(p|L)$ as a function of $p$ and of ${\langle \Lambda \rangle}_{\infty}$, for a domain of linear size $L=10$. Symbols correspond to the Monte Carlo simulation results with added lines to guide the eye: blue is chosen for $m={\cal P}$, green for $m={\cal V}$, and red for $m={\cal B}$. Circles with a continuous line denote ${\langle \Lambda \rangle}_{\infty}=1$, and crosses with a dashed line ${\langle \Lambda \rangle}_{\infty}=0.1$.}
\label{perco}
\end{figure}

The site percolation properties of three-dimensional Poisson binary mixtures have been previously addressed in~\cite{larmier}. The percolation threshold of three-dimensional Voronoi tessellations has also been estimated~\cite{voronoi_perco}. Furthermore, the percolation threshold of a Box tessellation can be mapped to that of a cubic network, which has been widely studied~\cite{ortho_perco}. Tab.~\ref{tab_perco} resumes the values of the percolation threshold $p_c$ for each mixing statistics. In order to illustrate the finite-size effects related to percolation, Monte Carlo simulation results for percolation probabilities as a function of $p$, ${\langle \Lambda \rangle}_{\infty}$ and $m$ are shown in Fig.~\ref{perco}. For large domain size $L$, the percolation threshold estimated by Monte Carlo simulation converges to the asymptotic values reported in Tab.~\ref{tab_perco}.

\begin{table}[b]
\begin{center}
\begin{tabular}{cc}
\toprule
$m$ & $p_c$\\
\midrule
${\cal P}$ & $0.290 \pm 7.10^{-3}$ \\
${\cal V}$ &  $0.1453 \pm 2.10^{-3}$ \\
${\cal B}$ & $0.3116077 \pm 4.10^{-7} $ \\
\bottomrule
\end{tabular}
\end{center}
\caption{Estimation of the percolation threshold, as a function of the mixing statistics $m$. Estimates are taken from~\cite{larmier, voronoi_perco, ortho_perco}.}
\label{tab_perco}
\end{table}

\section{Particle transport in benchmark configurations}
\label{transport_benchmark}

Once the key statistical features of the random media have been characterized, we can now turn our attention to the properties of particle transport through such tessellations. We propose two benchmark configurations for single-speed linear particle transport through non-multiplying stochastic binary mixtures composed of materials $\alpha$ and $\beta$. For both configurations we will set the same geometry and source specifications. The geometry will consists of a cubic box of side $L = 10$, with reflective boundary conditions on all sides of the box except on two opposite faces (say those perpendicular to the x-axis), where leakage boundary conditions are imposed (particles that leave the domain through these faces can not re-enter). We will apply a normalized incident angular flux on the leakage surface at $x = 0$ (with zero interior sources). These specifications are inspired by our previous work~\cite{larmier_benchmark} on a $d$-dimensional generalization of the benchmark proposed by Adams, Larsen and Pomraning~\cite{benchmark_adams}.

In benchmark case $1$, material $\alpha$ is void, and material $\beta$ is purely scattering; in benchmark case $2$, material $\alpha$ is purely absorbing, and material $\beta$ is purely scattering. The former case could represent for instance light propagation through turbid media, or to neutron transport in water-steam mixtures, the probability $p$ determining the fraction of voids. The latter case could represent for instance neutron diffusion in the presence of randomly distributed traps, such as Boron or Gadolinium grains, the probability $p$ determining the fraction of absorbers. The compositions for the two benchmarks are provided in Tab.~\ref{tab_compo}. For each case, we consider two sub-cases $a$ and $b$ by varying the scattering macroscopic cross sections: we set a scattering cross section $\Sigma_{\beta}^s=1$ for cases $1a$ and $2a$, and $\Sigma_{\beta}^s=10$ for cases $1b$ and $2b$. The absorbing cross section for material $\beta$ is zero for all cases and sub-cases. For material $\alpha$, we set $\Sigma_{\alpha}^a=10$ for case $2a$ and $2b$ (absorber), and $\Sigma_{\alpha}^a=0$ elsewhere. Scattering is assumed to be isotropic.

Following~\cite{brantley_benchmark, larmier_benchmark}, the physical observables that we would like to determine are the ensemble-averaged reflection probability $\langle R \rangle $ on the face where the incident flux is imposed, the ensemble-averaged transmission probability $\langle T \rangle$ on the opposite face, and the ensemble-averaged scalar particle flux $\langle \varphi \rangle $ within the box. Observe that the flux $\langle \varphi \rangle $ integrated over the box has a clear physical interpretation: actually, the integral flux is equal to the average length $\langle \ell_V \rangle$ travelled by the particles within the geometry: see, e.g., the considerations in~\cite{blanco, zoia_cauchy_1, zoia_cauchy_2, zoia_cauchy_3}. Since we are considering single-speed transport, due to the geometrical configuration of our benchmark $\langle \varphi \rangle $ is thus also proportional to the residence time spent by the particles in the box. In the absence of absorption (case $1$), the residence time can be identified with the first passage time from the source to the leakage boundaries.

\begin{table}[t]
\begin{center}
\begin{tabular}{cccccc}
\toprule
Case & $\Sigma_{\alpha}^s$ & $\Sigma_{\alpha}^a$ & $\Sigma_{\beta}^s$ & $\Sigma_{\beta}^a$\\
 \midrule
$1a$ & $0$ & $0$ & $1$ & $0$ \\
$1b$ & $0$ & $0$ & $10$ & $0$ \\
 \midrule
$2a$ & $0$ & $10$ & $1$ & $0$\\
$2b$ & $0$ & $10$ & $10$ & $0$\\
\bottomrule
\end{tabular}
\end{center}
\caption{Material parameters for the two cases $1,2$ of benchmark configurations and the two sub-cases $a,b$.}
\label{tab_compo}
\end{table}

\begin{table}[t]
\begin{center}
\begin{tabular}{cccc}
\toprule
${\langle \Lambda \rangle}_{\infty}$ & $\rho_{\cal P}$ & $\rho_{\cal V}$ & $\rho_{\cal B}$\\
 \midrule
$1$ & $1$ & $0.6872$ & $0.66667$ \\
$0.5$ & $2$ & $1.3744$ & $1.33333$ \\
$0.1$ & $10$ & $6.872$ & $6.66667$\\
\bottomrule
\end{tabular}
\end{center}
\caption{Parameters $\rho_{\cal P}$, $\rho_{\cal V}$ and $\rho_{\cal B}$ chosen for the benchmark configurations, as a function of the average chord length ${\langle \Lambda \rangle}_{\infty}$.}
\label{tab_param_geo}
\end{table}

In order to study the impact of random media on particle transport, for each benchmark case we consider three types of tessellations: Poisson, Voronoi and Box; three average chord lengths: ${\langle \Lambda \rangle}_{\infty}=1$, ${\langle \Lambda \rangle}_{\infty}=0.5$ and ${\langle \Lambda \rangle}_{\infty}=0.1$; and seven probabilities of assigning a label $\alpha$: $p=0.05$, $p=0.15$, $p=0.30$, $p=0.50$, $p=0.70$, $p=0.85$ and $p=0.95$. The tessellation densities $\rho_{\cal P}$, $\rho_{\cal V}$ and $\rho_{\cal B}$ can be easily derived based on Eq.~\ref{calib} and are resumed in Tab.~\ref{tab_param_geo}. For the purpose of illustration, examples of realizations of Poisson tessellations for the benchmark configurations are displayed in Fig.~\ref{geo_poisson}, Voronoi tessellations in Fig.~\ref{geo_voronoi} and Box tessellations in Fig.~\ref{geo_ortho}.
 
For the sake of completeness, we have also considered the so-called atomic mix model~\cite{pomraning}: the statistical disorder is approximated by simply taking a full homogenization of the physical properties based on the ensemble-averaged cross sections. The atomic mix approximation is known to be valid when the chunks of each material are optically thin, i.e., $\Sigma_{i}^t {\langle \Lambda_{i} \rangle}_{\infty} \ll 1$ for $i=\alpha,\beta$. For each case, we compute the corresponding ensemble-averaged scattering cross section $\langle \Sigma^s\rangle $ and ensemble-averaged absorbing cross section $\langle \Sigma^a\rangle$ as follows
\begin{equation}
\langle \Sigma^s \rangle =p \Sigma_{\alpha}^s + (1-p) \Sigma_{\beta}^s
\end{equation}
and
\begin{equation}
\langle \Sigma^a \rangle =p \Sigma_{\alpha}^a + (1-p) \Sigma_{\beta}^a.
\end{equation}

\subsection{Monte Carlo simulation parameters}
\label{simulation_results}

The reference solutions for the probabilities $\langle R \rangle$ and $\langle T \rangle$ and the ensemble-averaged scalar particle flux $\langle \varphi(x) \rangle$ have been computed as follows. For each configuration, a large number $M$ of geometries has been generated, and the material properties have been attributed to each volume as described above. Then, for each realization $k$ of the ensemble, linear particle transport has been simulated by resorting to the production Monte Carlo code \tripoli{}, developed at CEA~\cite{T4}. \tripoli{} is a general-purpose stochastic transport code capable of simulating the propagation of neutral and charged particles with continuous-energy cross sections in arbitrary geometries. In order to comply with the benchmark specifications, constant cross sections adapted to mono-energetic transport and isotropic angular distributions have been prepared. The number of simulated particle histories per configuration is $10^6$. For a given physical observable ${\cal O}$, benchmark solutions are obtained by taking the ensemble average
\begin{equation}
\langle {\cal O} \rangle = \frac{1}{M} \sum_{k=1}^M {\cal O}_k,
\end{equation}
where ${\cal O}_k$ is the Monte Carlo estimate for the observable ${\cal O}$ obtained for the $k$-th realization. Specifically, the particle currents $R_k$ and $T_k$ at the respective surface are estimated by summing the statistical weights of the particles leaking from that surface. Scalar fluxes $\varphi_k(x)$ have been recorded by resorting to the standard track length estimator over a pre-defined spatial grid containing $10^2$ uniformly spaced meshes along the $x$ axis.

The error affecting the average observable $\langle {\cal O} \rangle$ results from two separate contributions, namely, the dispersion
\begin{equation}
\sigma^2_G = \frac{1}{M} \sum_{k=1}^M {{\cal O}_k}^2 - {\langle {\cal O} \rangle}^2
\end{equation}
of the observables exclusively due to the stochastic nature of the geometries and of the material compositions, and 
\begin{equation}
\sigma^2_{{\cal O}}=\frac{1}{M} \sum_{k=1}^M \sigma_{{\cal O}_k}^2,
\end{equation}
which is an estimate of the variance due to the stochastic nature of the Monte Carlo method for the particle transport, $\sigma_{{\cal O}_k}^2$ being the dispersion of a single calculation~\cite{donovan, sutton}. The statistical error on $\langle {\cal O} \rangle$ is then estimated as
\begin{equation}
\sigma[ \langle {\cal O}\rangle ] = \sqrt{\frac{\sigma^2_G}{M}+\sigma^2_{{\cal O}}}.
\end{equation}

The number $M$ of realizations used for the Monte Carlo simulations has been chosen as follows. Since Poisson, Voronoi and Box tessellations have ergodic properties~\cite{miles1970, miles1972}, tessellations with smaller average chord length ${\langle \Lambda \rangle}_{\infty}$ require a lower number of realizations to achieve statistically stable ensemble averages, as discussed in a previous work~\cite{larmier_benchmark}. Thus, for configurations where ${\langle \Lambda \rangle}_{\infty}=1$ we have performed $10^3$ realizations; for ${\langle \Lambda \rangle}_{\infty}=0.5$ we have taken $8\times 10^2$ realizations; and for ${\langle \Lambda \rangle}_{\infty}=0.1$ we have taken $2\times 10^2$ realizations. All the data sets considered in the following sections are available from the authors upon request.

\begin{table}
\begin{center}
\begin{tabular}{cccccc}
\toprule
$m$ & ${\langle \Lambda \rangle}_{\infty}$ & $\langle N | L \rangle$ & $\sigma [N | L]$ & $\langle t \rangle$ & $\sigma [t]$\\
\midrule
AM & & & & $145$ & $0$\\ 
\midrule
& $1$ & $784$ & $448$ & $176$ & $8$  \\
${\cal P}$ & $0.5$ & $5278$ & $2316$ & $217$ & $14$ \\
& $0.1$ & $561163$ & $117382$ & $7705$ & $2062$ \\
\midrule
& $1$ & $325$ & $17$ & $185$ & $2$ \\
${\cal V}$ & $0.5$ & $2596$ & $51$ & $258$ & $4$ \\
& $10$ & $324557$ & $563$ & $2511$ & $208$ \\
\midrule
& $1$ & $473$ & $285$ & $169$  & $7$ \\
${\cal B}$ & $0.5$ & $2975$ & $1344$ & $200$ & $11$ \\
& $0.1$ & $309264$ & $66029$ & $2212$ & $552$ \\
\bottomrule
\end{tabular}
\end{center}
\caption{Complexity of the tessellations used for the benchmark configurations, as a function of the mixing statistics $m$ (AM stands for atomic mix) and of the average chord length ${\langle \Lambda \rangle}_{\infty}$ of the tessellation, for a domain of linear size $L=10$. $\langle N | L \rangle$ denotes the average number of polyhedra composing the tessellation, whereas $\langle t \rangle$ denotes the average computer time (expressed in seconds) for a transport simulation of the benchmark configuration $1a$, with $p=0.05$.}
\label{tab_complexity}
\end{table}

\subsection{Computer time}

The average computer time of a transport simulation increases significantly for decreasing average chord length ${\langle \Lambda \rangle}_{\infty}$ of the tessellation, as shown Tab.~\ref{tab_complexity}. For the calculations discussed here we have largely benefited from a feature that has been recently revised and enhanced in the code \tripoli{}, namely the possibility of reading pre-computed connectivity maps for the volumes composing the geometry. During the generation of the tessellations, care has been taken so as to store the indexes of the neighbouring volumes for each realization, which means that during the geometrical tracking a particle will have to find the next crossed volume in a list that might be considerably smaller than the total number of random volumes composing the box. When provided to the transport code, such connectivity maps allow thus for considerable speed-ups for the most fragmented geometries, up to a factor of one hundred.

Transport calculations have been run on a cluster based at CEA, with Xeon E5-2680 V2 2.8 GHz processors. The average computer time $t$ for case $1a$ with $p=0.05$ is displayed in Tab.~\ref{tab_complexity} as a function of the mixing statistics $m$ and of the average chord length ${\langle \Lambda \rangle}_{\infty}$ of the tessellation. It is apparent that $t$ increases with the complexity of the system, i.e., the number of polyhedra composing the tessellation. Nevertheless, this is not true for geometries with higher ${\langle \Lambda \rangle}_{\infty}$: simulations in Voronoi tessellations are longer than those in Poisson and Box tessellations, in spite of a lower number of polyhedra. This is likely due to the larger average number of faces in Voronoi geometries, which slows down particle tracking. For geometries composed of a large number of polyhedra, the complexity of the system outweighs this effect. Moreover, the dispersion on the simulation time seems correlated to the dispersion on the number of polyhedra: thus, this dispersion may become very large, and even be comparable to the average $t$, for Poisson tessellations. Atomic mix simulations are based on a single homogenized realization, thus there is no associated dispersion.

\subsection{Reflection, transmission and integral flux}

The statistical properties of the random media adopted for the benchmark configurations, including the average chord length, the average volume and surface, and the number of faces, have been determined by Monte Carlo simulation and are resumed in Tab.~\ref{tab_geo} for reference, which allows probing finite-size effects.

\begin{figure*}
\begin{center}
$1a$\,\,\,\,\,\,\,\,\,\,\,\,\,\,\,\,\,\,\,\,\,\,\,\,\,\,\,\,\,\,\,\,\,\,\,\,\,\,\,\,\,\,\,\,\,\,\,\,\,\,\,\,\,\,\,\,\,\,\,\,\,\,\,\,\,\,\,\,\,\,\,\,\,\,\,\,\,\,\,\,\,\,\,\,\,\,\,\,\,\,\,\,\,\,\,\,\,\,\,\,\,\,\,\,\,\,\,\,\,\,\,\,\,\,\,\,\,\,\,\,\,\,\,\,\,\,\,\,\,\,\,\,\,\,\,\,\,\,\,\,\,\,\,\,\,\,\,\,\,\,\,\,\,\,\,\,$1b$\\
\includegraphics[width=1\columnwidth]{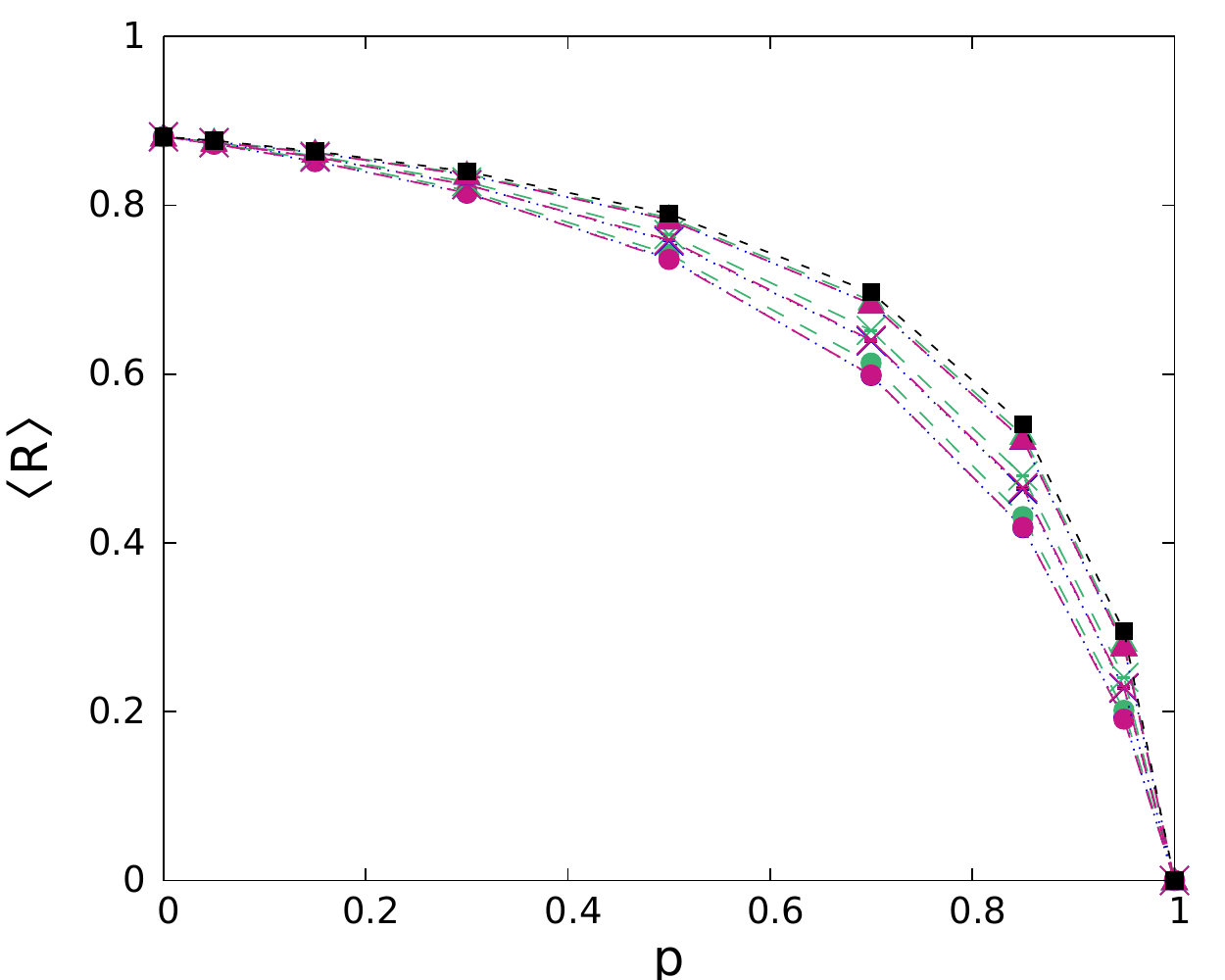}\,\,\,\,
\includegraphics[width=1\columnwidth]{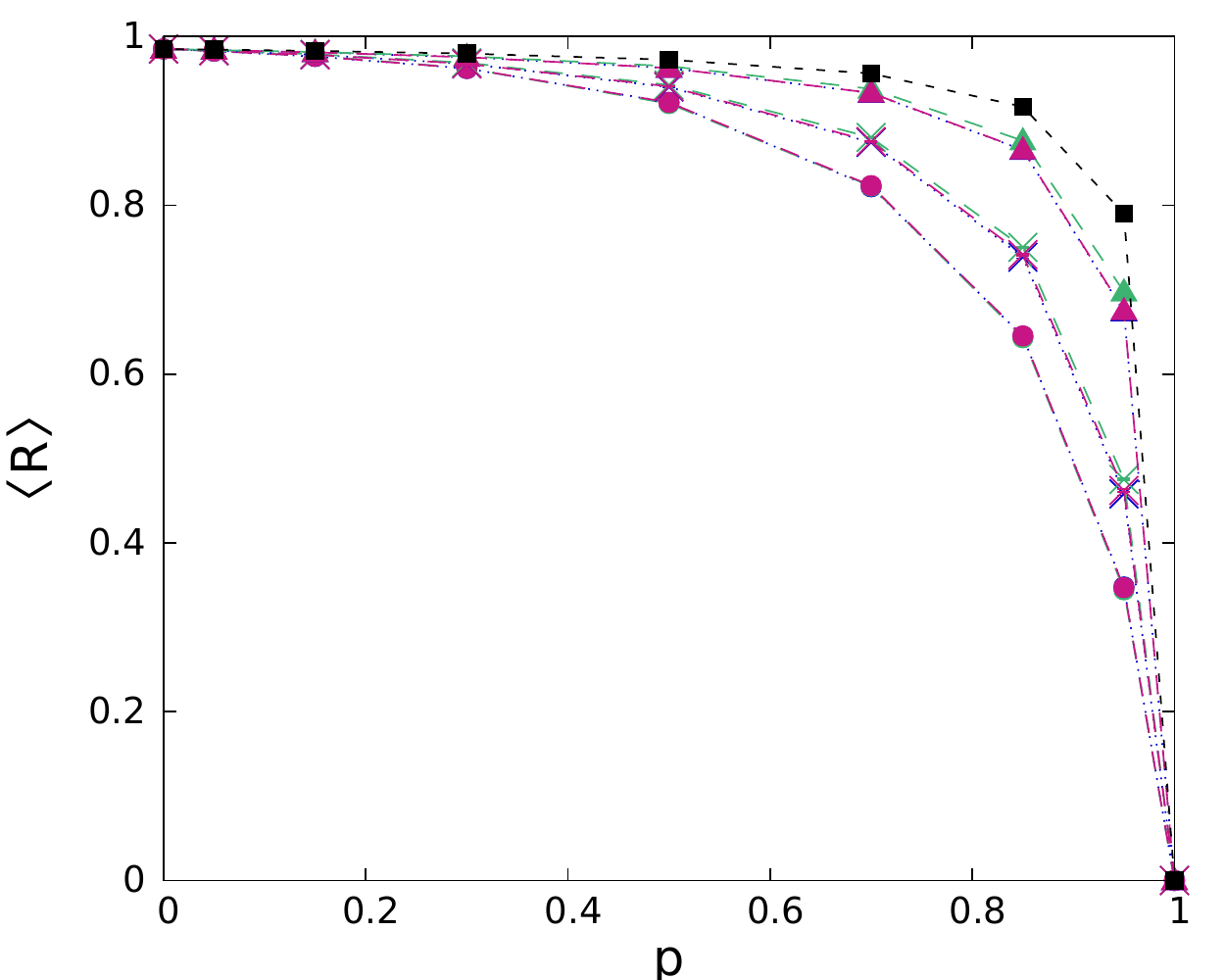}\\
\includegraphics[width=1\columnwidth]{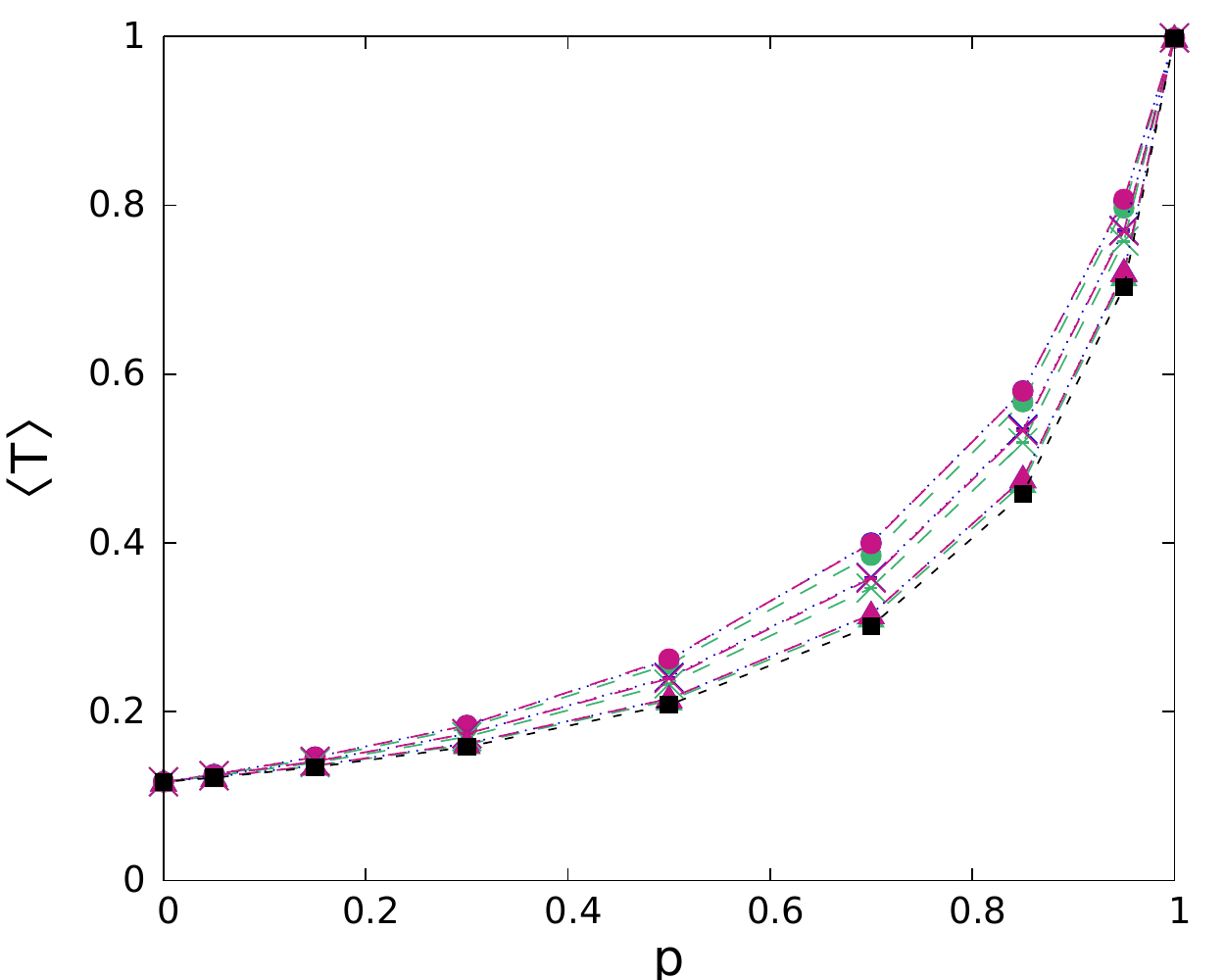}\,\,\,\,
\includegraphics[width=1\columnwidth]{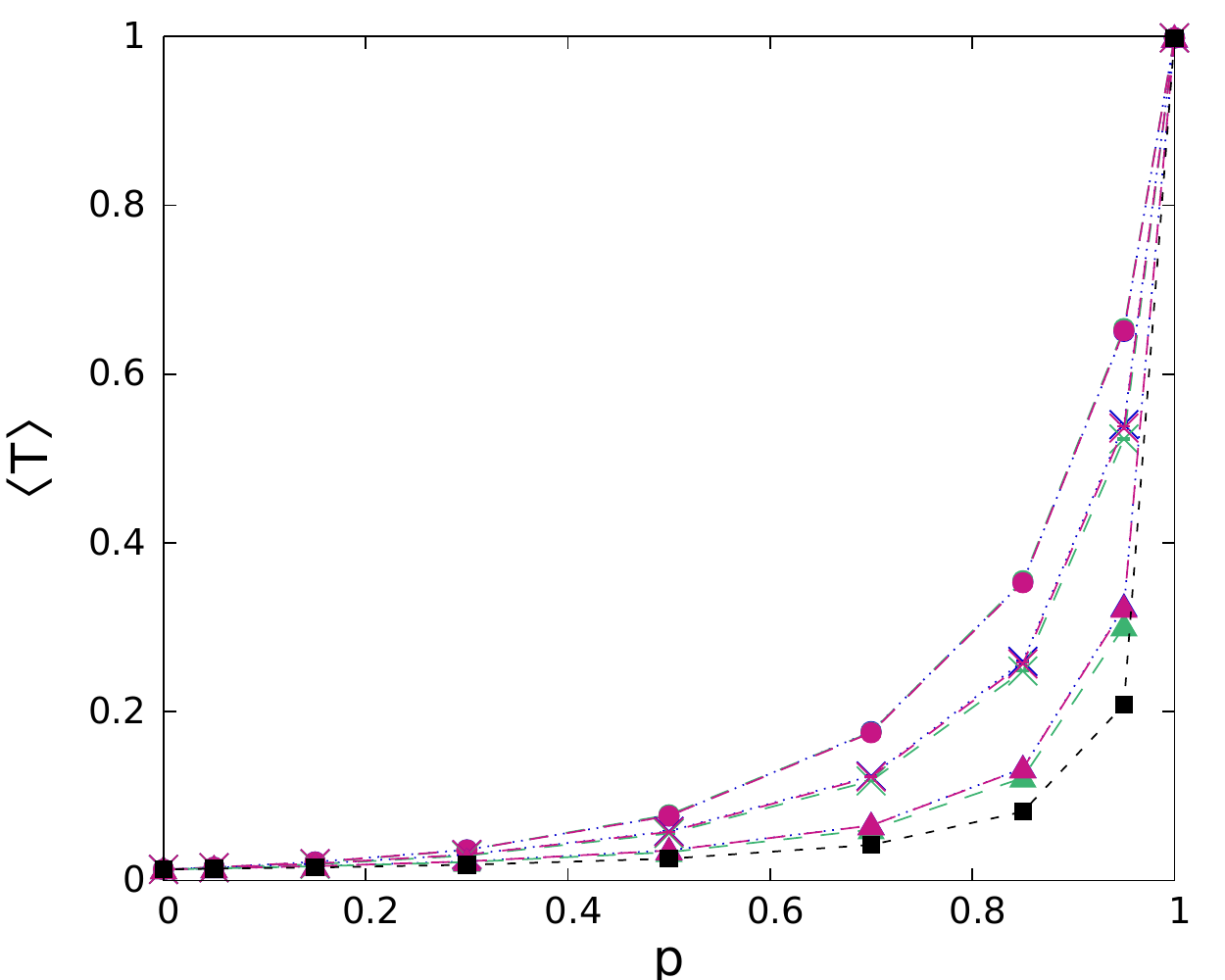}\\
\includegraphics[width=1\columnwidth]{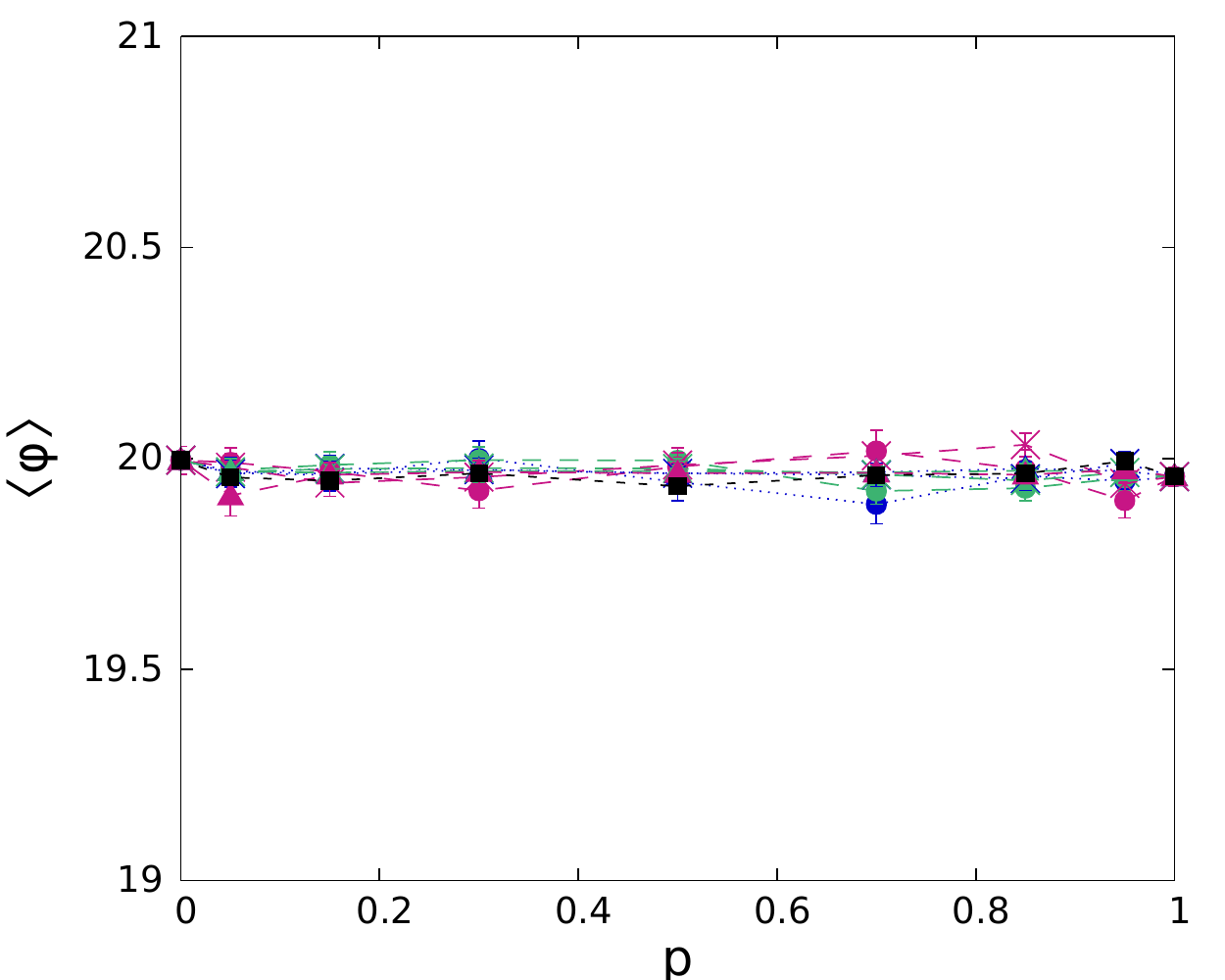}\,\,\,\,
\includegraphics[width=1\columnwidth]{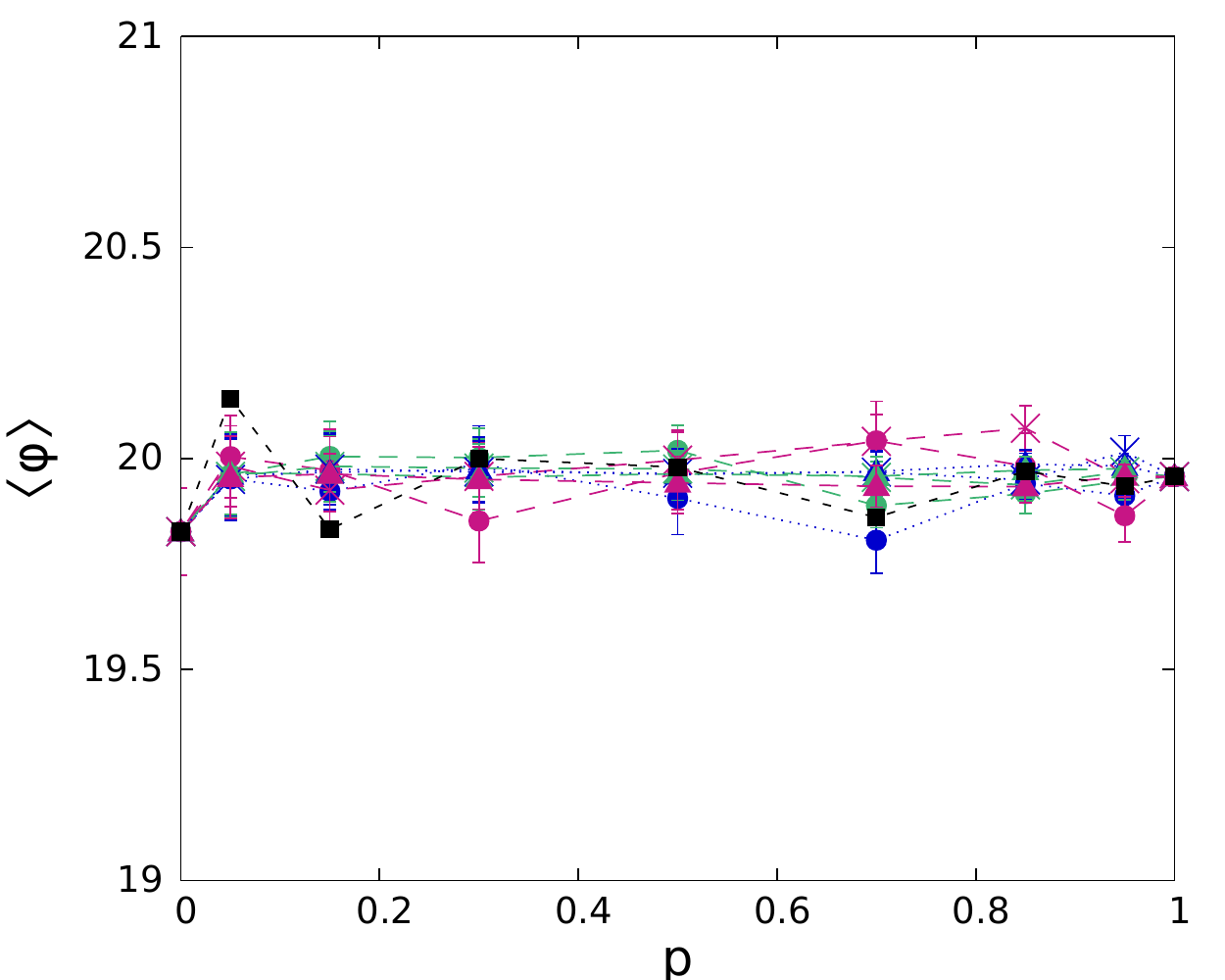}\\
\end{center}
\caption{Monte Carlo results for for the benchmark configurations: cases $1a$ (left) and $2b$ (right). Reflection probability $\langle R \rangle$ (top), transmission probability $\langle T \rangle$ (center) and scalar flux $\langle \varphi \rangle$ (bottom), as a function of $p$, for different mixing statistics $m$ and different values of ${\langle \Lambda \rangle}_{\infty}$. Black squares represent the atomic mix approximation. Blue symbols denote $m={\cal P}$, green symbols $m={\cal V}$ and red symbols $m={\cal B}$. Circles denote ${\langle \Lambda \rangle}_{\infty}=1$, crosses ${\langle \Lambda \rangle}_{\infty}=0.5$ and triangles ${\langle \Lambda \rangle}_{\infty}=0.1$. Dashed lines have been added to guide the eye.}
\label{scalars_2}
\end{figure*}

\begin{figure*}
\begin{center}
$2a$\,\,\,\,\,\,\,\,\,\,\,\,\,\,\,\,\,\,\,\,\,\,\,\,\,\,\,\,\,\,\,\,\,\,\,\,\,\,\,\,\,\,\,\,\,\,\,\,\,\,\,\,\,\,\,\,\,\,\,\,\,\,\,\,\,\,\,\,\,\,\,\,\,\,\,\,\,\,\,\,\,\,\,\,\,\,\,\,\,\,\,\,\,\,\,\,\,\,\,\,\,\,\,\,\,\,\,\,\,\,\,\,\,\,\,\,\,\,\,\,\,\,\,\,\,\,\,\,\,\,\,\,\,\,\,\,\,\,\,\,\,\,\,\,\,\,\,\,\,\,\,\,\,\,\,\,$2b$\\
\includegraphics[width=1\columnwidth]{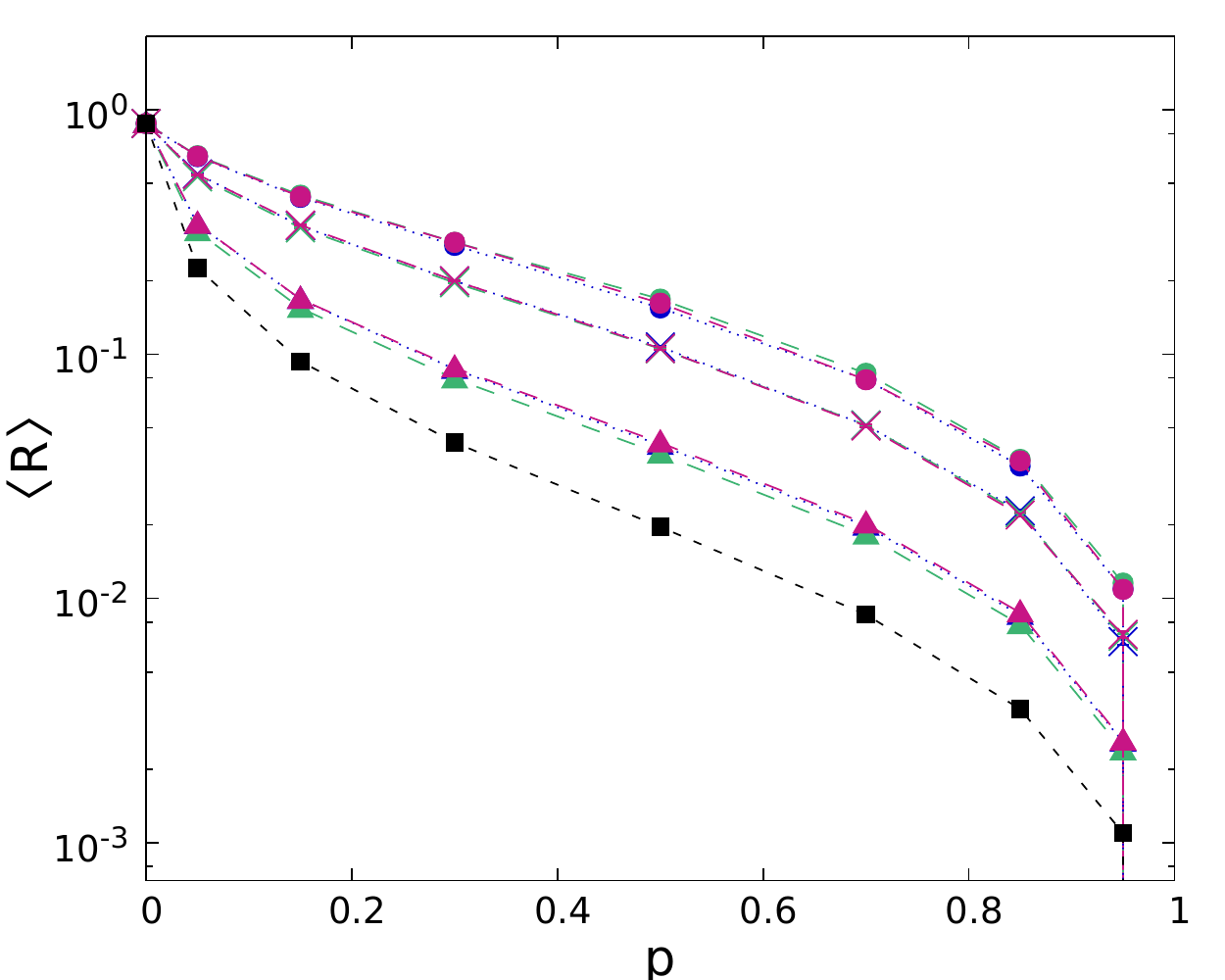}\,\,\,\,
\includegraphics[width=1\columnwidth]{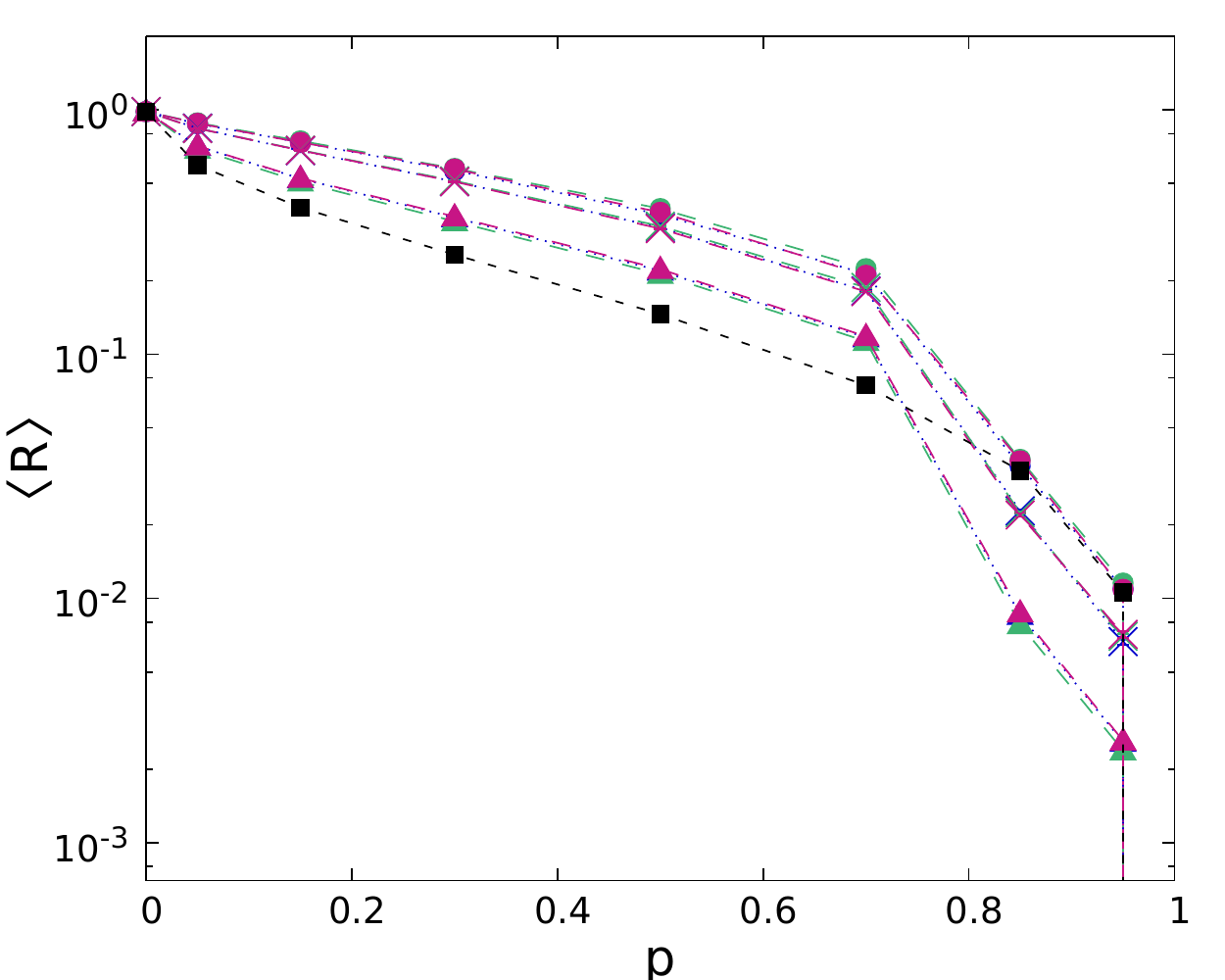}\\
\includegraphics[width=1\columnwidth]{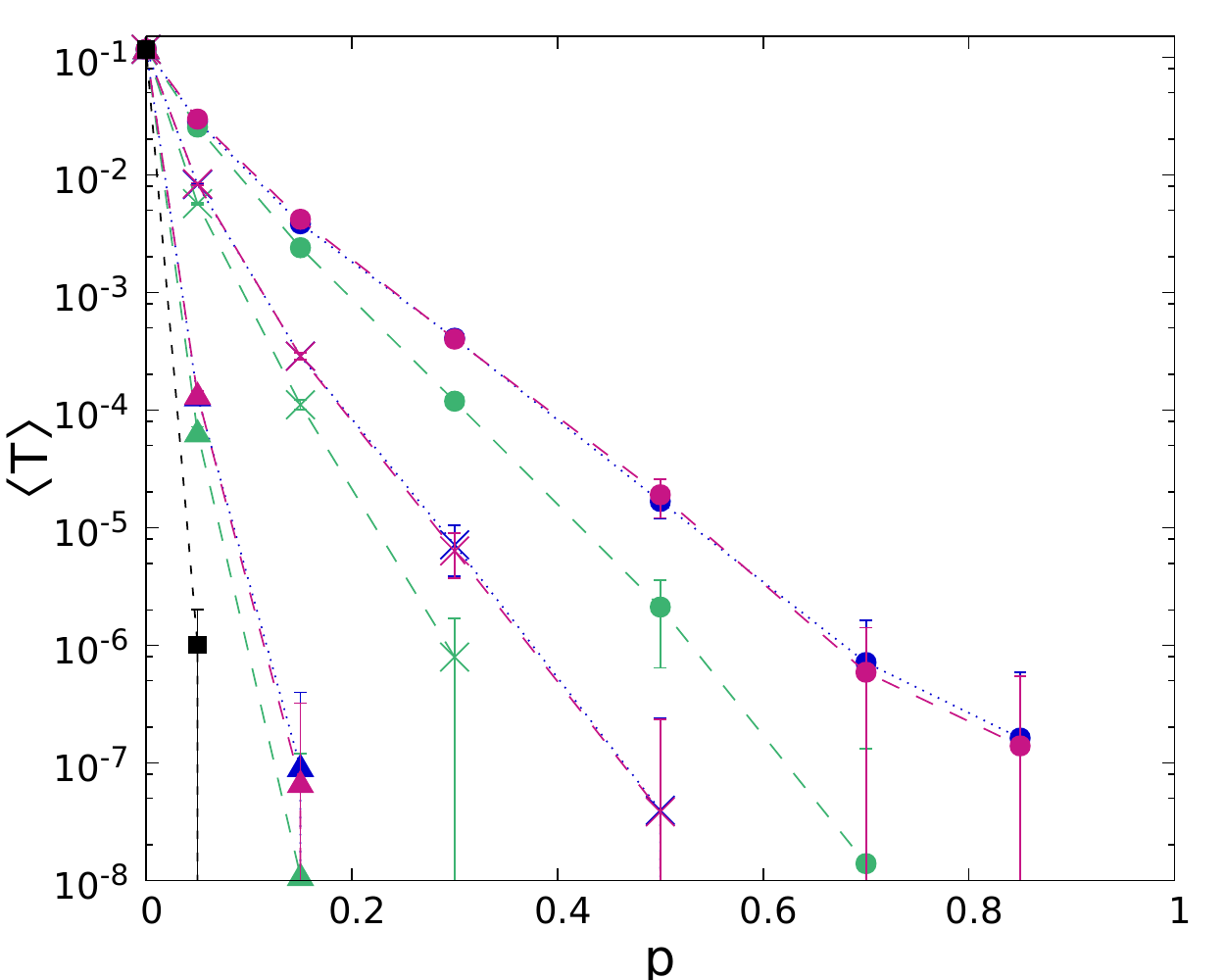}\,\,\,\,
\includegraphics[width=1\columnwidth]{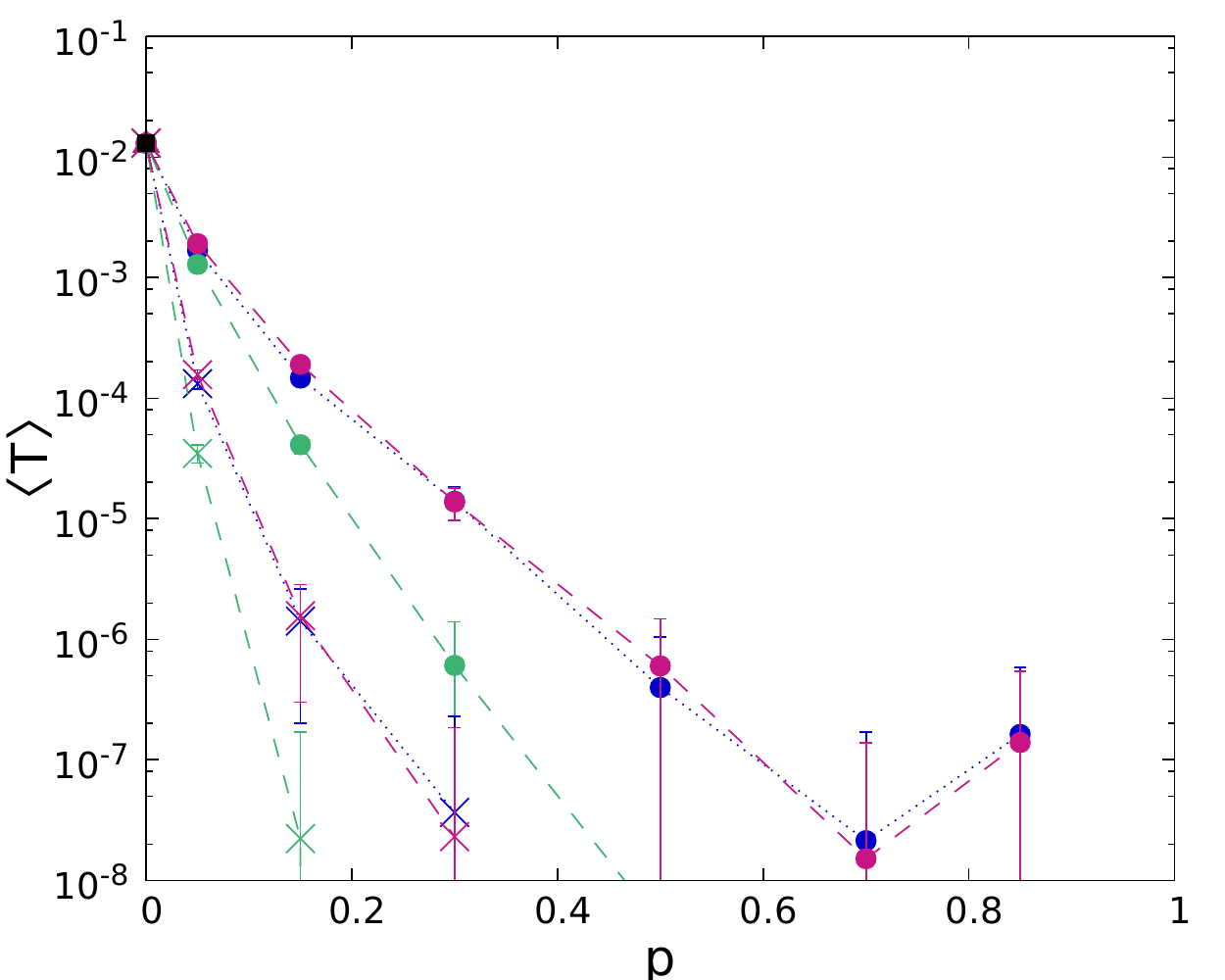}\\
\includegraphics[width=1\columnwidth]{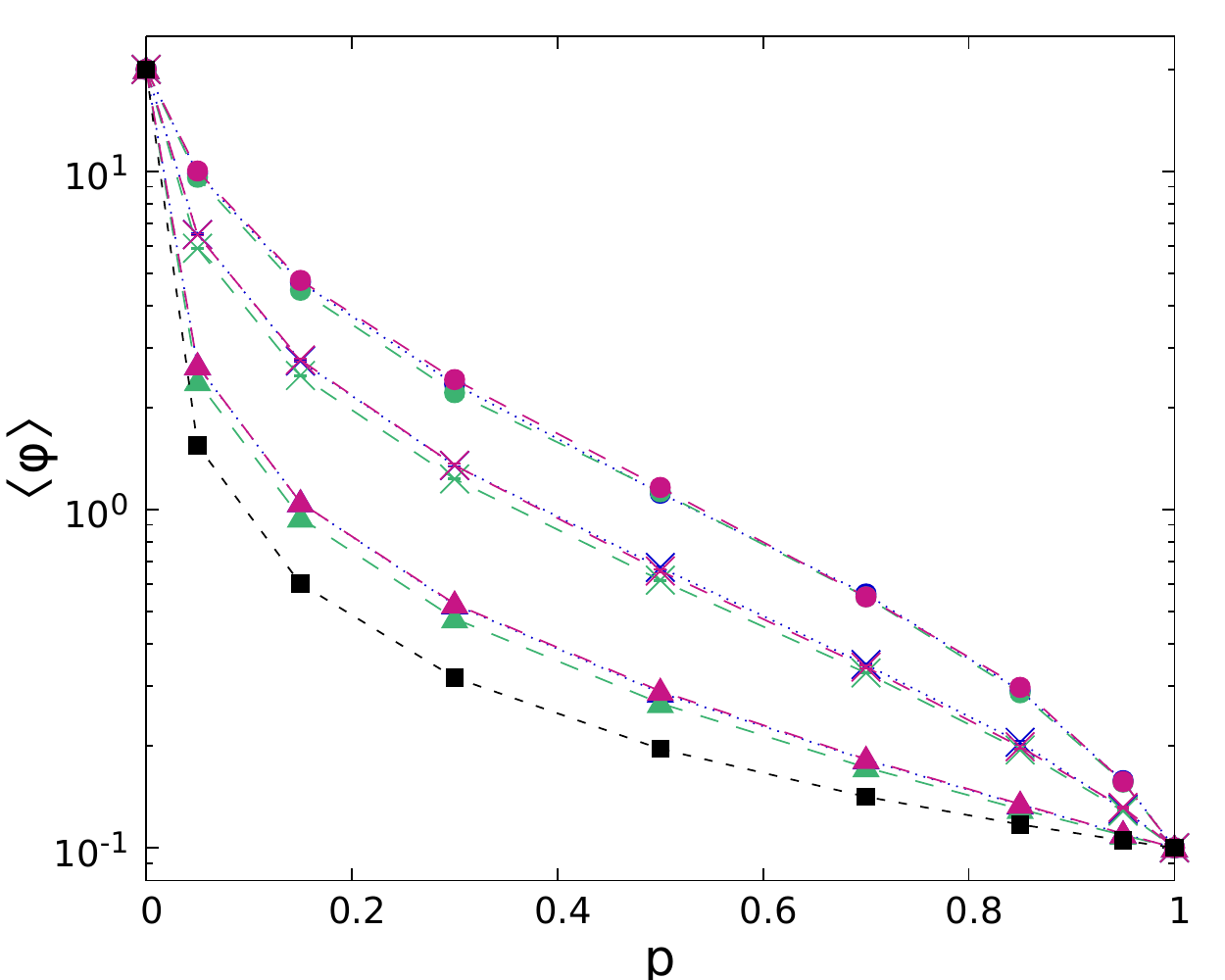}\,\,\,\,
\includegraphics[width=1\columnwidth]{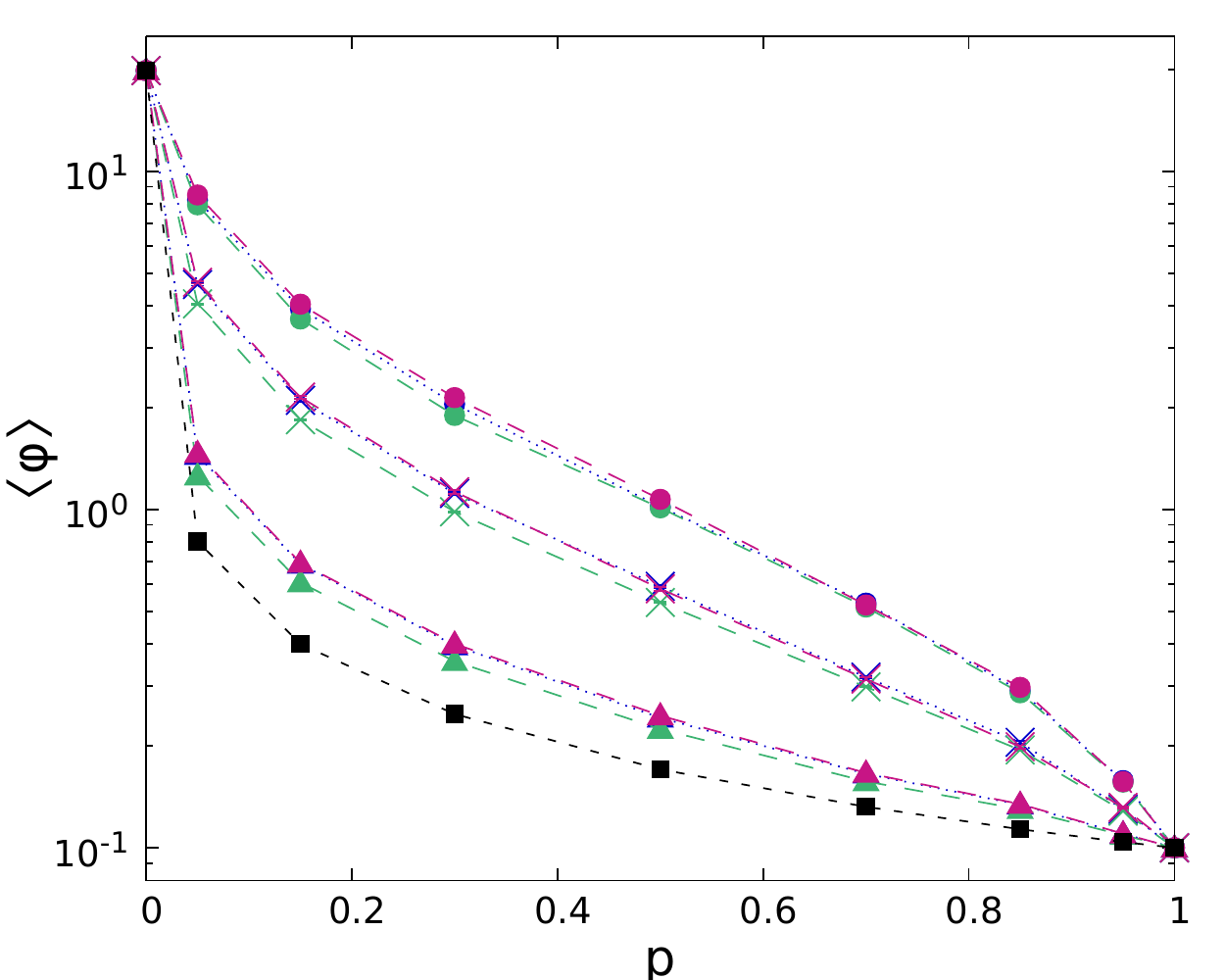}\\
\end{center}
\caption{Monte Carlo results for for the benchmark configurations: cases $2a$ (left) and $2b$ (right). Reflection probability $\langle R \rangle$ (top), transmission probability $\langle T \rangle$ (center) and scalar flux $\langle \varphi \rangle$ (bottom), as a function of $p$, for different mixing statistics $m$ and different values of ${\langle \Lambda \rangle}_{\infty}$. Black squares represent the atomic mix approximation. Blue symbols denote $m={\cal P}$, green symbols $m={\cal V}$ and red symbols $m={\cal B}$. Circles denote ${\langle \Lambda \rangle}_{\infty}=1$, crosses ${\langle \Lambda \rangle}_{\infty}=0.5$ and triangles ${\langle \Lambda \rangle}_{\infty}=0.1$. Dashed lines have been added to guide the eye.}
\label{scalars_3}
\end{figure*}

Concerning transport-related observables, the simulation results for the ensemble-averaged reflection probability $\langle R \rangle$, the transmission probability $\langle T \rangle$ and the integral flux $\langle \varphi  \rangle= \langle \int \int\varphi({\bf r}, {\boldsymbol \omega}) d {\boldsymbol \omega} d {\bf r}\rangle$ are provided in Figs.~\ref{scalars_2} and~\ref{scalars_3} for all the benchmark configurations, as a function of $p$, $m$ and ${\langle \Lambda \rangle}_{\infty}$. Atomic mix results are also provided for reference.

To begin with, we analyse the behaviour of these observables as a function of $p$. For case $1$, the transmission probability increases with the void fraction $p$. For large values of $p$, the medium is prevalently composed of voids, which enhances transmission because particle trajectories are not hindered by collisions. When $p$ decreases, the proportion of scattering material increases and so does the probability for a particle to scatter, change direction and leak from the face where the source is imposed. Symmetrically, the reflection probability decreases with $p$: this is expected on physical grounds, since for case $1$ we have $\langle T \rangle + \langle R \rangle = 1$ from mass conservation. Percolation of the void fraction appear to play no significant role for the configurations considered here: the variation of  $\langle T \rangle$ and $\langle R \rangle$ with respect to $p$ is smooth and no threshold effects at $p = p_c$ are apparent. The void fraction $p$ has no impact on the integral flux for case $1$. Actually, as stated above, $\langle \varphi \rangle = \langle \ell_V \rangle$, and from the Cauchy's formula for one-speed random walks in purely scattering domains we have
\begin{equation}
\langle \ell_V \rangle = 4\frac{V}{S_\textit{leak}}
\label{eq_cauchy_length}
\end{equation}
where $S_\textit{leak}$ is the surface area of the boundaries where leakage conditions are applied~\cite{blanco, zoia_cauchy_1, zoia_cauchy_2, zoia_cauchy_3}. This formula, which can be understood as a non-trivial generalization of the Cauchy's formula applying to the average chord lengths~\cite{blanco, zoia_cauchy_2}, holds true provided that particles enter the domain uniformly and isotropically, which is ensured here by the source that we have chosen and by symmetry considerations. Hence, the flux $\langle \varphi \rangle$ depends exclusively on the ratio of purely geometrical quantities, namely, $\langle \varphi \rangle = 4  V / S_\textit{leak}$, which for our benchmark yields $\langle \varphi \rangle = 20$.

For case $2$, reflection, transmission and integral flux all decrease with increasing absorber fraction $p$. This is also expected on physical grounds: the larger is $p$, the smaller is the survival probability of particles and the shorter is the average residence time within the box (and hence the integral flux). Although Eq.~\eqref{eq_cauchy_length} can be generalized to include also absorbing media (and even multiplication), the resulting formula will depend on the specific features of the travelling particles and will not have a universal character~\cite{zoia_cauchy_1, zoia_cauchy_2, zoia_cauchy_3}.

The impact of the average chord length on $\langle T \rangle$, $\langle R \rangle$ and $\langle \varphi \rangle$ is clearly visible in Figs.~\ref{scalars_2} and~\ref{scalars_3}. As a general consideration, for any mixing statistics, the respective observables become closer to those of the atomic mix as ${\langle \Lambda \rangle}_{\infty}$ decreases. These results suggest a convergence towards atomic mix when ${\langle \Lambda \rangle}_{\infty}$ tends to zero, i.e., for high fragmentation. However, in most cases, this convergence is not fully achieved for the range of parameters explored here. The atomic mix approximation is indeed supposed to be valid only when the chunks of different materials are optically thin, and this condition is typically not verified for our configurations (see Tabs.~\ref{tab_lambda_alpha} and~\ref{tab_compo}). Nonetheless, we notice one exception in case $2b$, for large absorber fractions in the range $0.7 < p <1$, where the relative positions of the reflection curves corresponding to tessellations are inverted with respect to those corresponding to atomic mix. In such configurations, stochastic geometries with small ${\langle \Lambda \rangle}_{\infty}$ will induce low reflection probabilities and will further enhance the discrepancy with respect to the atomic mix case. This non-trivial behaviour, which stems from finite-size and interface effects dominating the transport process, has been previously observed for the benchmark configurations analysed in~\cite{larmier_benchmark} under similar conditions, i.e., small chunks of scattering material surrounded by an absorbing medium. The threshold behaviour of $\langle R \rangle$ at $p>0.7$ might be subtly related to the percolation of the scattering material.

For cases $1a$ and $1b$, the transmission probability increases with ${\langle \Lambda \rangle}_{\infty}$: larger void chunks enhance transmission. The reflection probability is complementary to transmission and shows the opposite trend. For case $2a$ and $2b$, the transmission probability increases again with ${\langle \Lambda \rangle}_{\infty}$, and so do the reflection probability and the integral flux. Therefore, the absorption probability $\langle A \rangle = 1-\langle R\rangle-\langle T\rangle$ for case $2$ decreases with increasing ${\langle \Lambda \rangle}_{\infty}$ (see Fig.~\ref{scalars_3}).

Let us now consider the relevance of scattering cross sections $\Sigma_{\beta}^s$. For both case $1$ and $2$, the reflection probability increases when increasing $\Sigma_{\beta}^s$, whereas the transmission probability decreases. For case $2$, the resulting absorption probability $\langle A \rangle=1-\langle R \rangle -\langle T \rangle$ decreases with $\Sigma_{\beta}^s$. Moreover, larger values of $\Sigma_{\beta}^s$ enhance the discrepancies between results corresponding to different tessellations, as a function of ${\langle \Lambda \rangle}_{\infty}$: this is apparent when comparing case $1a$ to case $1b$, or case $2a$ to case $2b$. For given parameters $m$ and ${\langle \Lambda \rangle}_{\infty}$, the discrepancy with respect to atomic mix depends on both $p$ and $\Sigma_{\beta}^s$: this is maximal for large values of $p$ in case $1$ and, conversely, for small values of $p$ in case $2$.

For case $2$, where scattering is in competition with absorption, particles are all the more likely to avoid absorbing regions as the chunks of scattering materials (of linear scale ${\langle \Lambda \rangle}_{\infty}/p$) are large compared to the scattering mean free path $1/\Sigma_{\beta}^s$. If the size of scattering chunks is small compared to the scattering mean free path, particles have little or no chance of survival, because they will most often cross an absorbing region. Moreover, $\langle R\rangle$ and $\langle T\rangle$ increase with decreasing $p$ and increasing ${\langle \Lambda \rangle}_{\infty}$. When the chunks of scattering material are large, the stochastic tessellations typically contain clusters composed of scattering material spanning the geometry, forming `safe corridors' for the particles. Simulation findings suggest that a small scattering mean free path does reduce the effect of absorption, but only with respect to reflection probability: particles have a greater chance of coming back to the starting boundary when the chunks of scattering material are large compared to the scattering mean free path. On the contrary, the transmission probability is only weakly affected, which shows that for the chosen benchmark configurations of case $2$ transport is dominated by the behaviour close to the starting boundary.

Generally speaking, the impact of the mixing statistics $m$ is weaker than that of the other parameters of the benchmarks. This is not entirely surprising, since we have chosen the respective tessellation densities so to have the same ${\langle \Lambda \rangle}_{\infty}$, and transport properties will mostly depend on the average chord length of the traversed polyhedra. Such behaviour might be of utmost importance when the choice of the mixing statistics is part of the unknowns in modelling random media. For Box and Poisson tessellations, the simulation results for all physical observables are always in excellent agreement, which suggests that these mixing statistics are almost equivalent for the chosen configurations. On the contrary, results for Voronoi tessellations have a distinct behaviour, and for most cases these tessellations show systematic discrepancies with respect to those of Poisson or Box geometries, particularly in cases $2a$ and $2b$ for $\langle T \rangle$ and $\langle \varphi \rangle$. These findings are coherent with the peculiar nature of the Voronoi mixing statistics, as illustrated in the previous sections: differences in the chord length distribution and in the aspect ratio of the underlying stochastic geometries will induce small but appreciable differences in the transport-related observables.

\subsection{Integral flux distribution}

\begin{figure*}
\begin{center}
$1a$\,\,\,\,\,\,\,\,\,\,\,\,\,\,\,\,\,\,\,\,\,\,\,\,\,\,\,\,\,\,\,\,\,\,\,\,\,\,\,\,\,\,\,\,\,\,\,\,\,\,\,\,\,\,\,\,\,\,\,\,\,\,\,\,\,\,\,\,\,\,\,\,\,\,\,\,\,\,\,\,\,\,\,\,\,\,\,\,\,\,\,\,\,\,\,\,\,\,\,\,\,\,\,\,\,\,\,\,\,\,\,\,\,\,\,\,\,\,\,\,\,\,\,\,\,\,\,\,\,\,\,\,\,\,\,\,\,\,\,\,\,\,\,\,\,\,\,\,\,\,\,\,\,\,\,\,$1b$\\
\includegraphics[width=0.9\columnwidth]{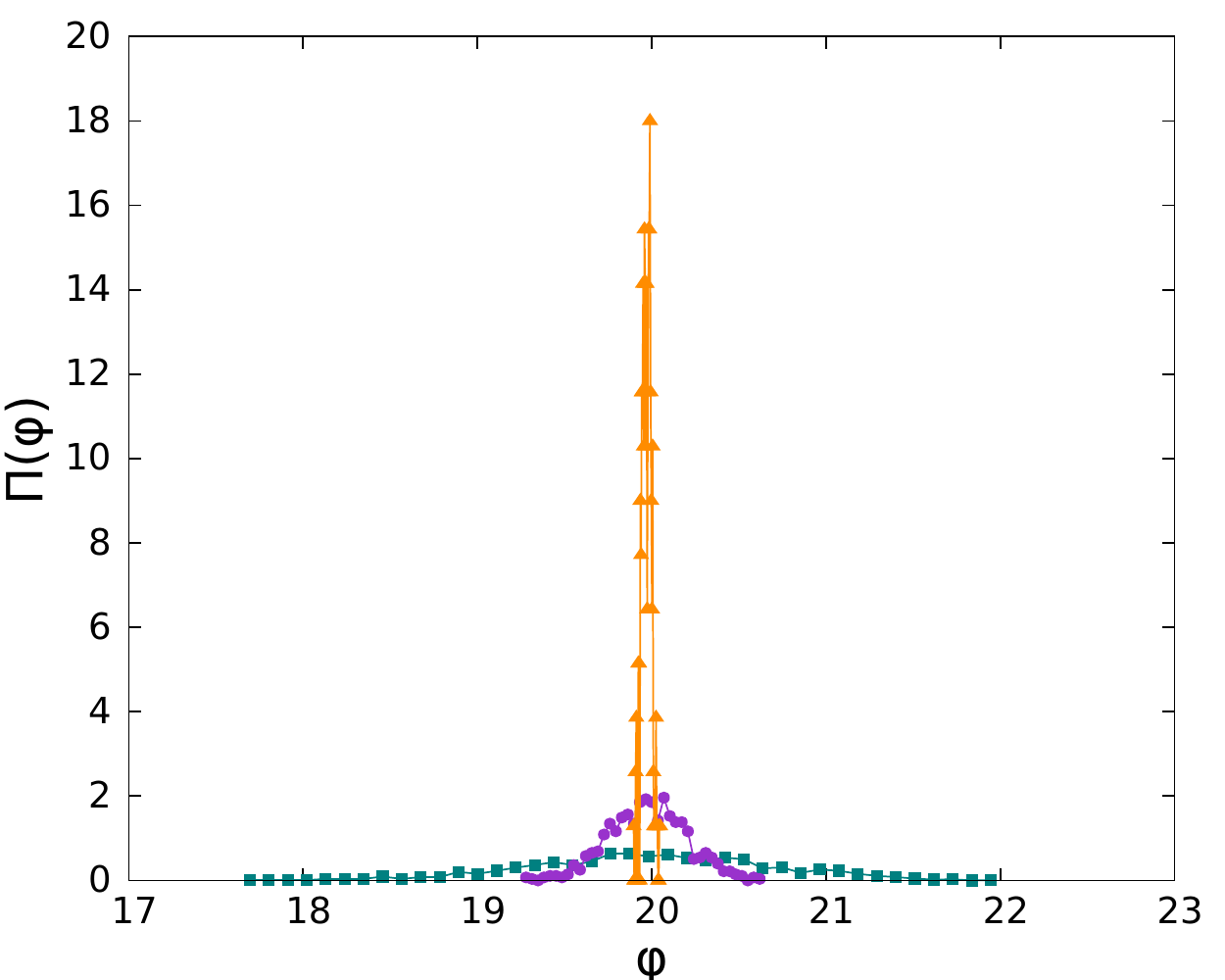}\,\,\,\,
\includegraphics[width=0.9\columnwidth]{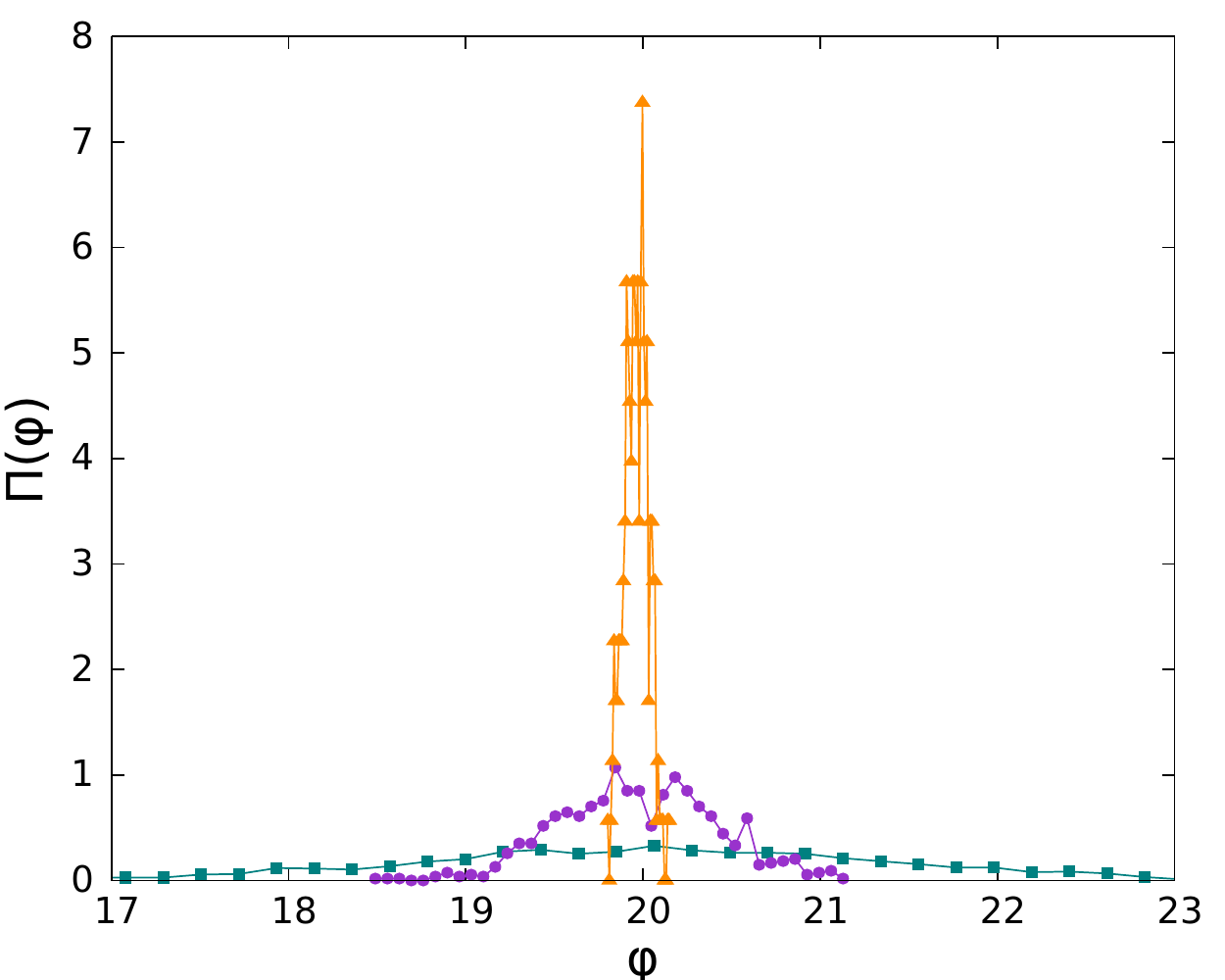}\\
\end{center}
\caption{Normalized distributions $\Pi(\varphi)$ of the scalar flux $\varphi$, for the benchmark configurations $1a$ (left) and $1b$ (right), for $m={\cal V}$ and for $p=0.5$. Dark green squares denote ${\langle \Lambda \rangle}_{\infty}=1$, violet circles ${\langle \Lambda \rangle}_{\infty}=0.5$ and orange triangles ${\langle \Lambda \rangle}_{\infty}=0.1$.}
\label{histos1}
\end{figure*}

\begin{figure*}
\begin{center}
\,\,\,\, $p=0.05$ \,\,\,\,\\
\includegraphics[width=0.9\columnwidth]{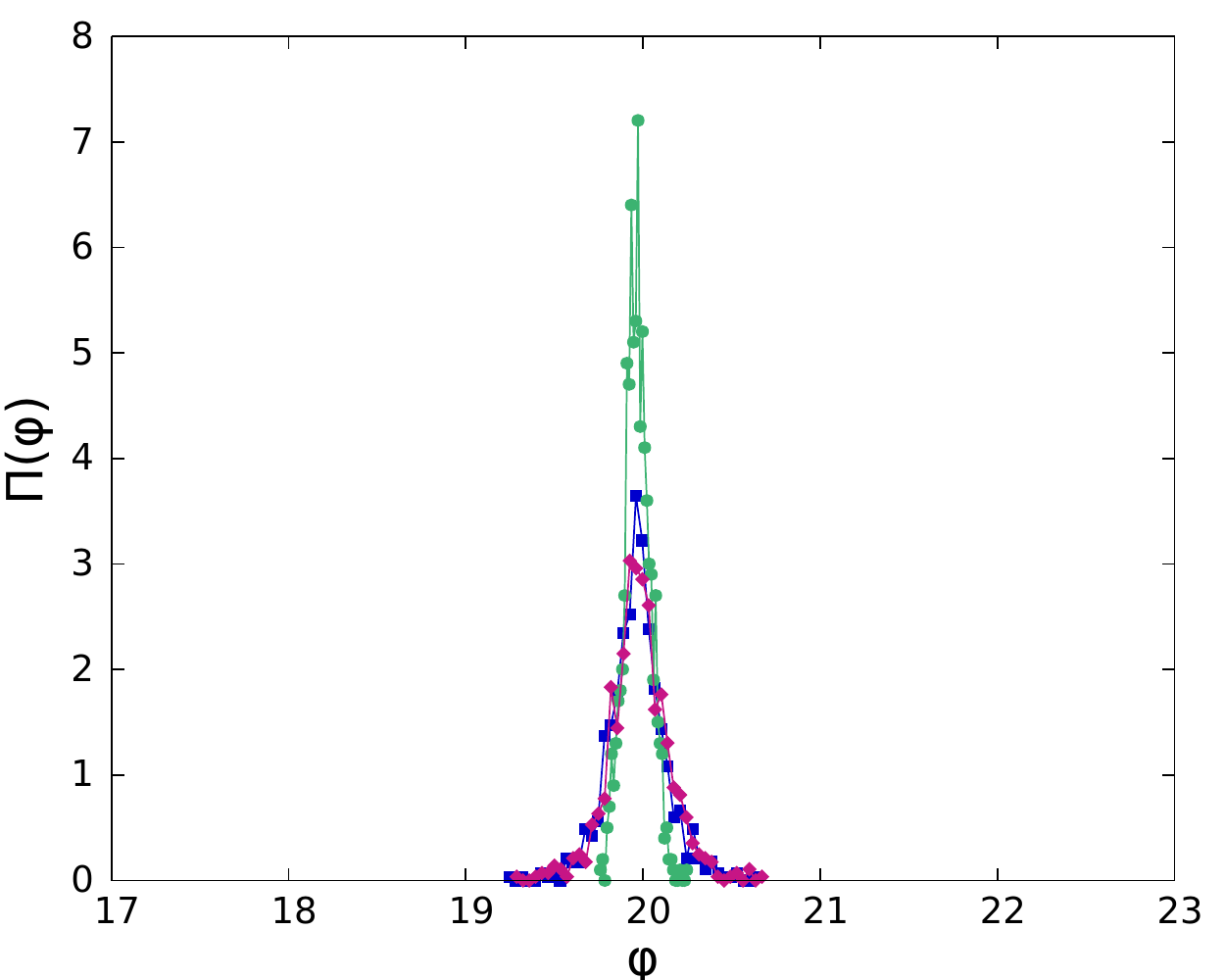}\,\,\,\,
\includegraphics[width=0.9\columnwidth]{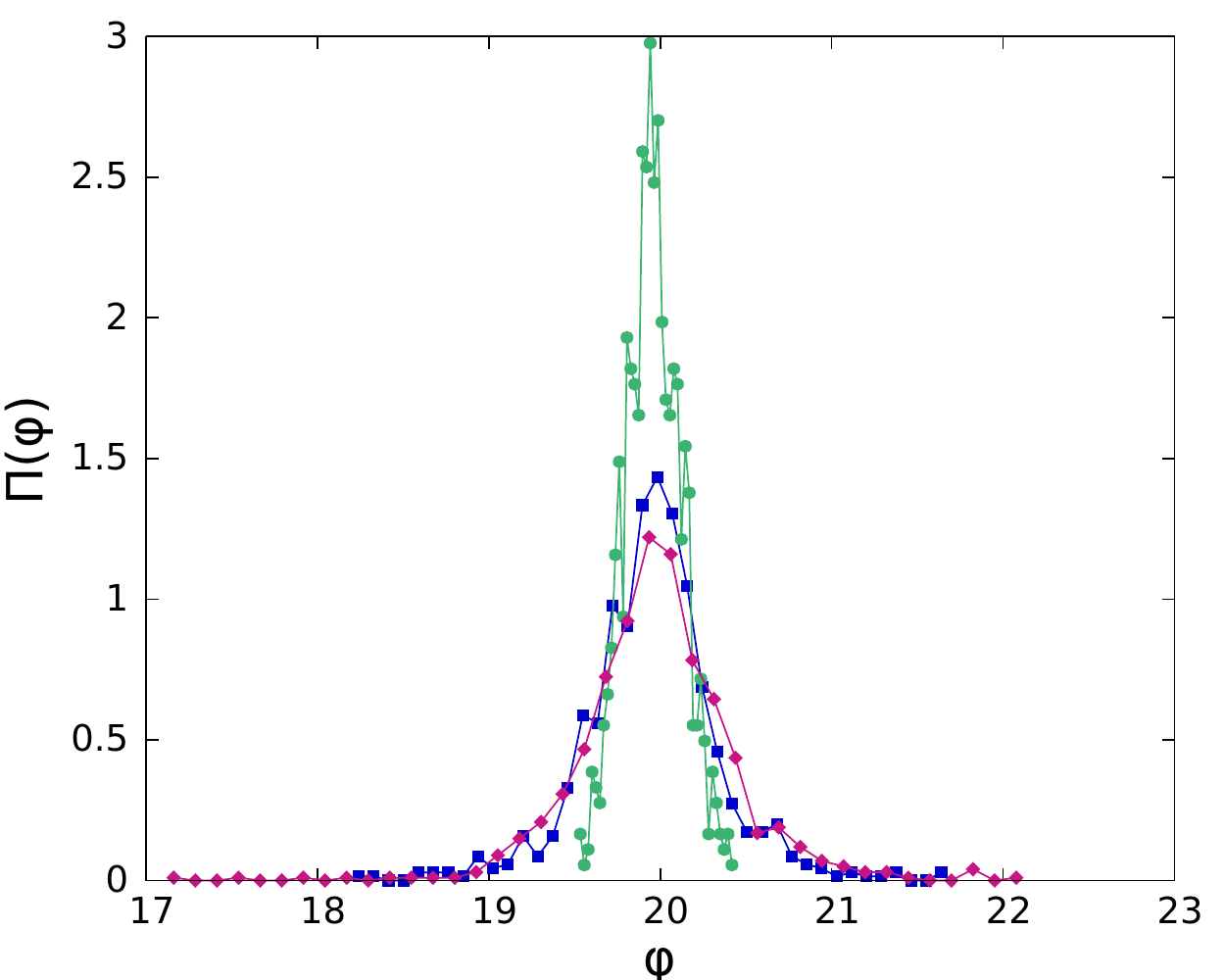}\\
\,\,\,\, $p=0.95$ \,\,\,\,\\
\includegraphics[width=0.9\columnwidth]{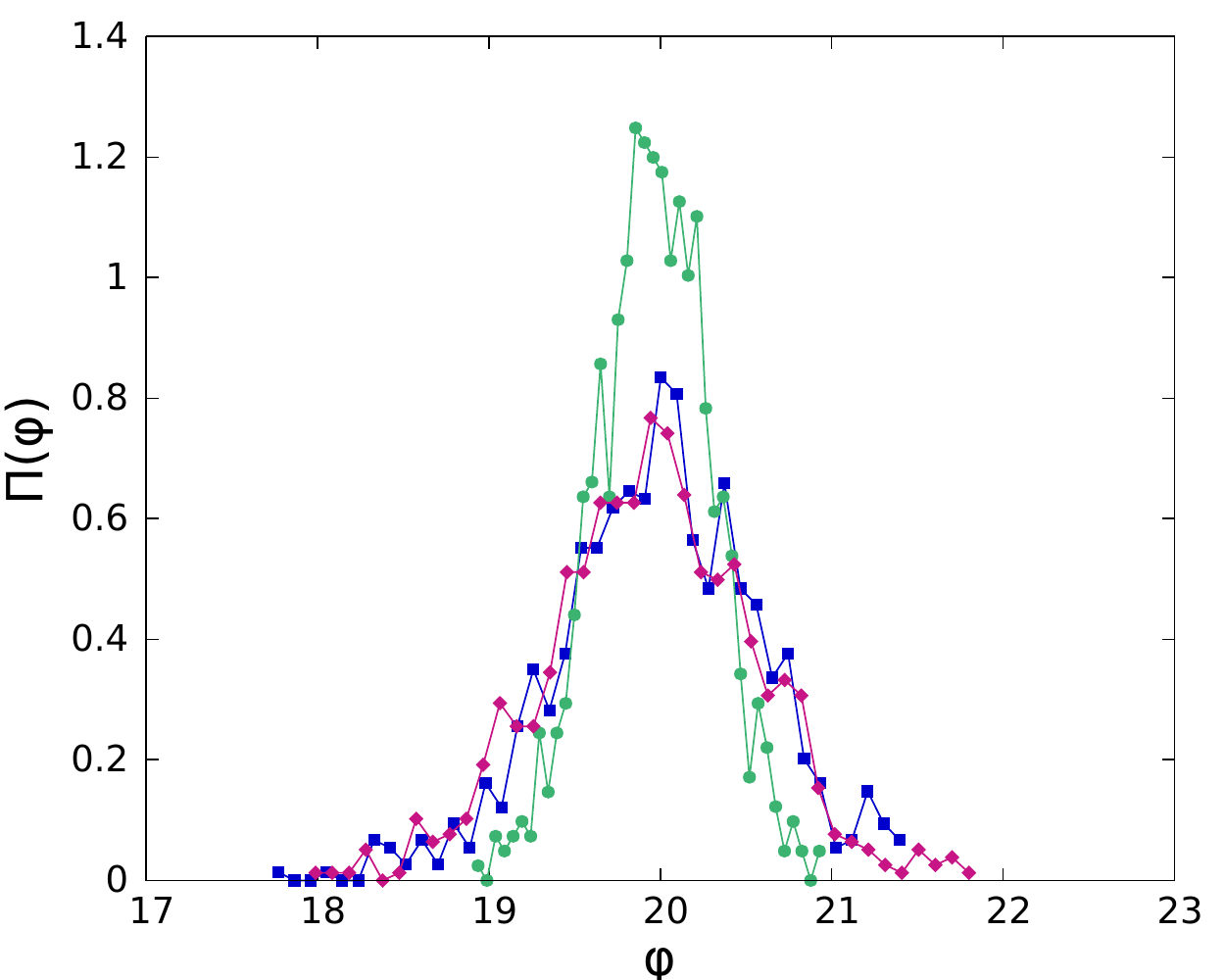}\,\,\,\,
\includegraphics[width=0.9\columnwidth]{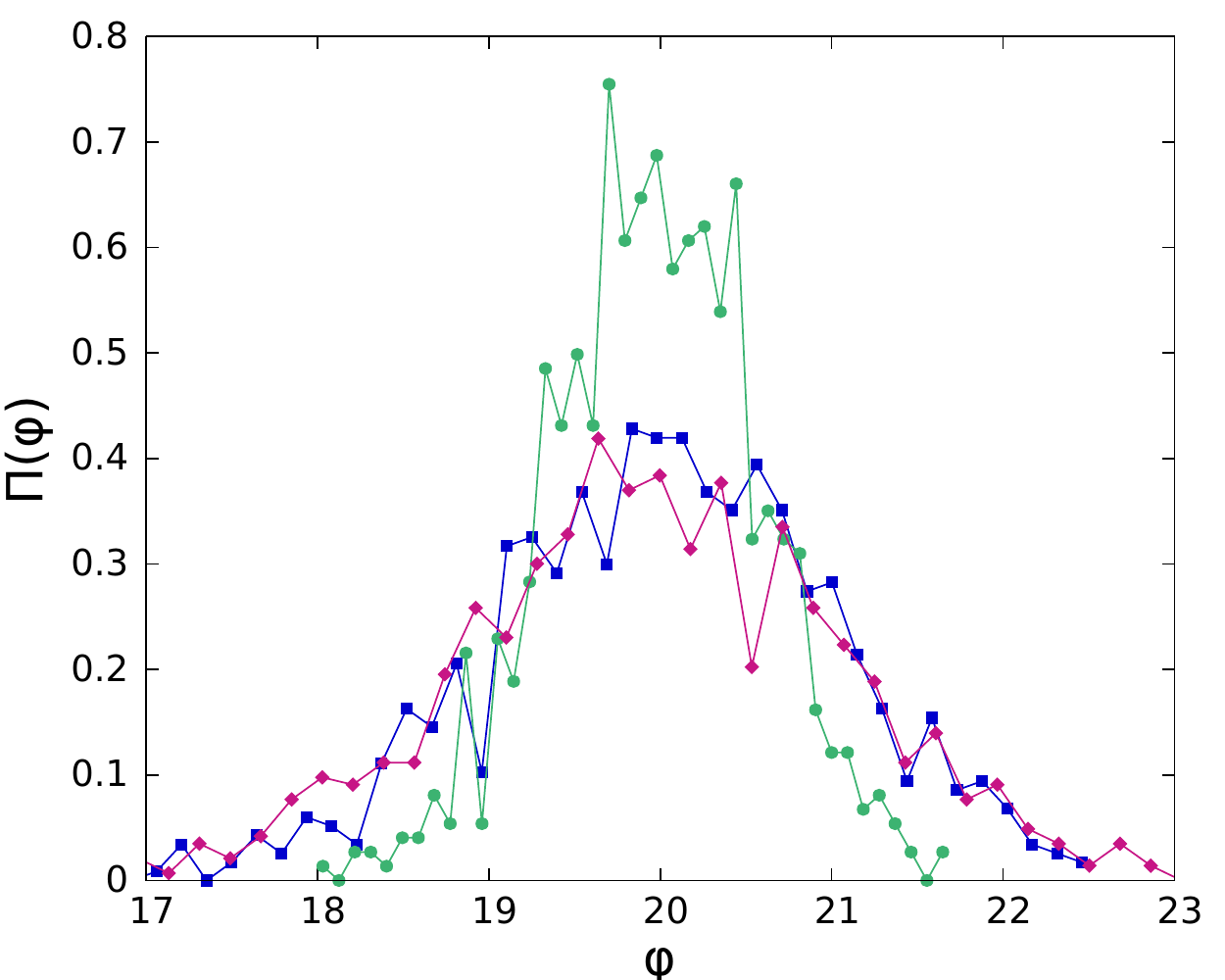}\\
\end{center}
\caption{Normalized distributions $\Pi(\varphi)$ of the scalar flux $\varphi$, for the benchmark configurations $1a$ (left) and $1b$ (right), for ${\langle \Lambda \rangle}_{\infty}=0.5$. Top: $p=0.05$; bottom $p=0.95$. Blue squares denote $m={\cal P}$, green circles $m={\cal V}$ and red diamonds $m={\cal B}$.}
\label{histos2}
\end{figure*}

In order to better assess the variability of the integral flux $\varphi$ (i.e., of the time spent by the particles in the box) around its average value, we have also computed its full distribution, based on the available realizations in the generated ensembles. The resulting normalized histograms are illustrated in Figs.~\ref{histos1} for different values of ${\langle \Lambda \rangle}_{\infty}$ and~\ref{histos2} for different values of $p$ and different mixing statistics $m$.

As a general remark, the dispersion of the integral flux decreases with decreasing ${\langle \Lambda \rangle}_{\infty}$, when the other parameters are fixed (see Fig.~\ref{histos1}, where we illustrate an example corresponding to case $1a$ and $1b$): the ensemble averages become increasingly efficient and the average values become more representative of the full distribution, which is expected on physical grounds. The same behaviour has been observed for the reflection and transmission probabilities, in any configuration, and for all mixing statistics. Moreover, numerical results show that Poisson and Box tessellations have a very close distribution for the integral flux, and that the dispersion around the average is smaller for Voronoi tessellations than for the other tessellations (see Fig.~\ref{histos2}).

For case $1$, all configurations share the same average integral flux. However, the dispersion of $\varphi$ around the average value depends on the probability $p$ and on the scattering cross section $\Sigma_{\beta}^s$ (see Fig.~\ref{histos2}), and this not universal. For this case, the impact of $p$ is clearly apparent: for small values of $p$ (chunks of void surrounded by scattering material), the distribution is rather peaked on the average value, whereas, for large values of $p$ (chunks of scattering material surrounded by void), the dispersion is more spread out. The influence of the scattering cross section $\Sigma_{\beta}^s$ is also visible when comparing cases $1a$ and $1b$: the dispersion increases with increasing $\Sigma_{\beta}^s$. 

\subsection{Spatial flux}

\begin{figure*}
\begin{center}
\,\,\,\, $p=0.95$ \,\,\,\,\\
\includegraphics[width=0.9\columnwidth]{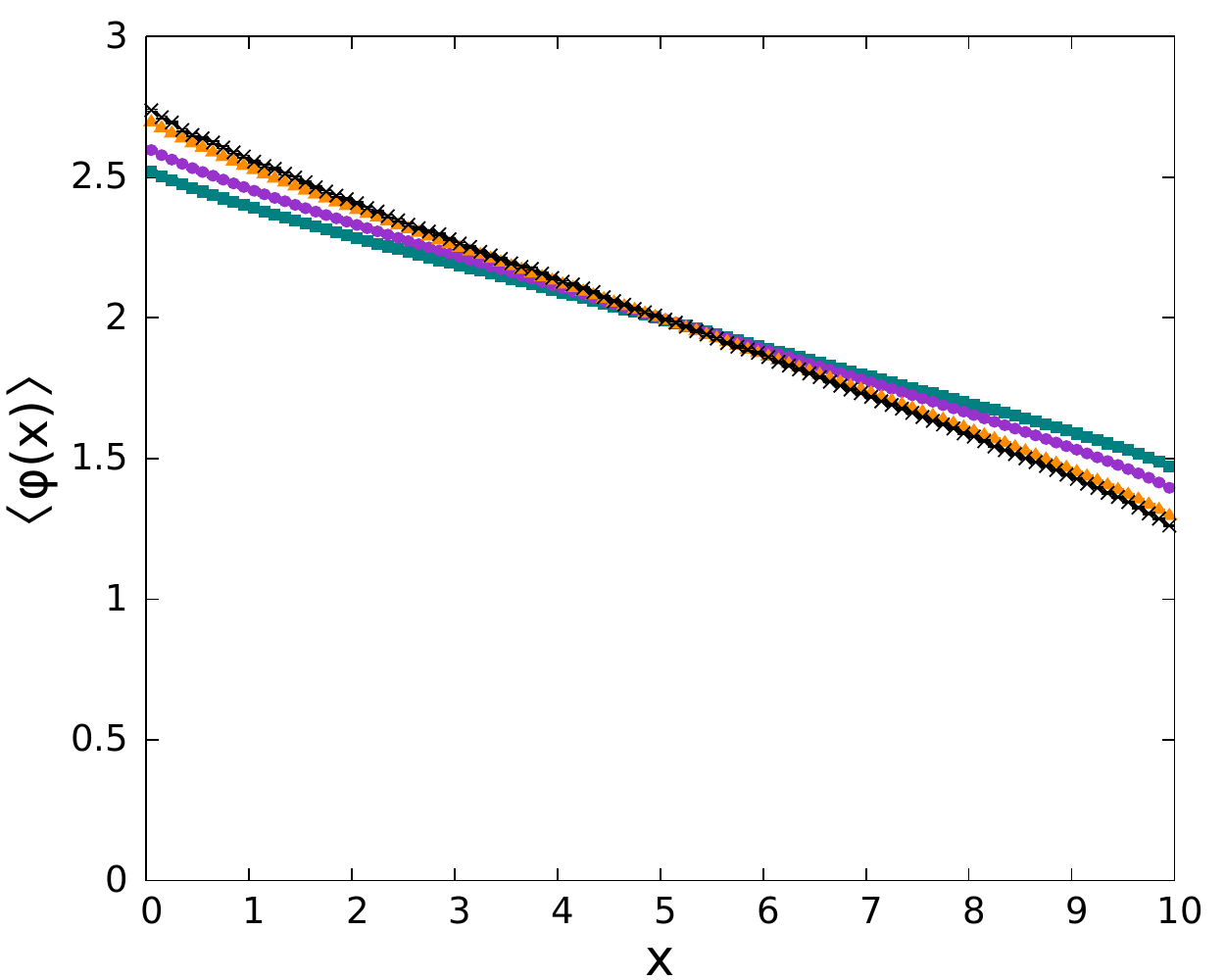}\,\,\,\,
\includegraphics[width=0.9\columnwidth]{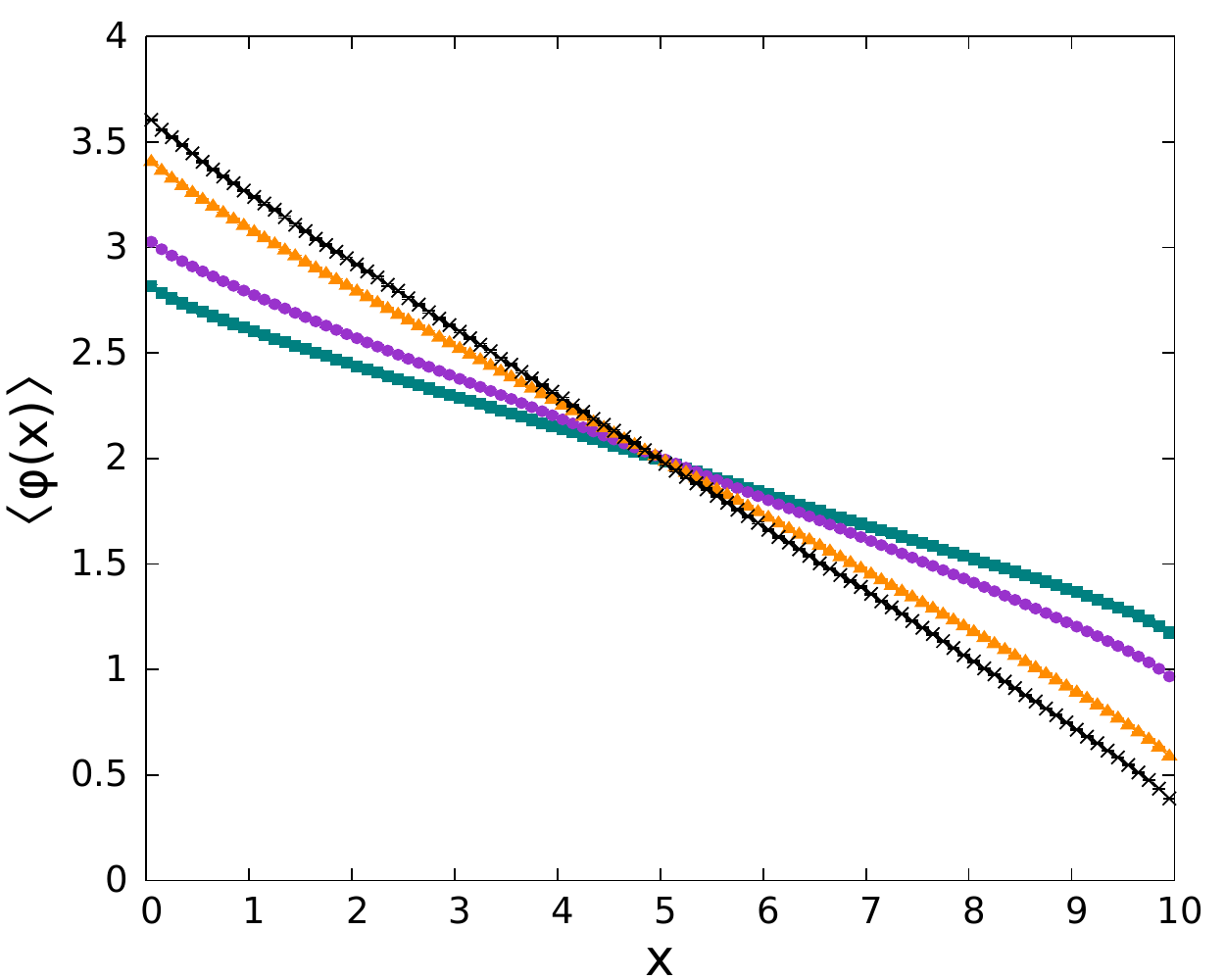}\\
\,\,\,\, $p=0.7$ \,\,\,\,\\
\includegraphics[width=0.9\columnwidth]{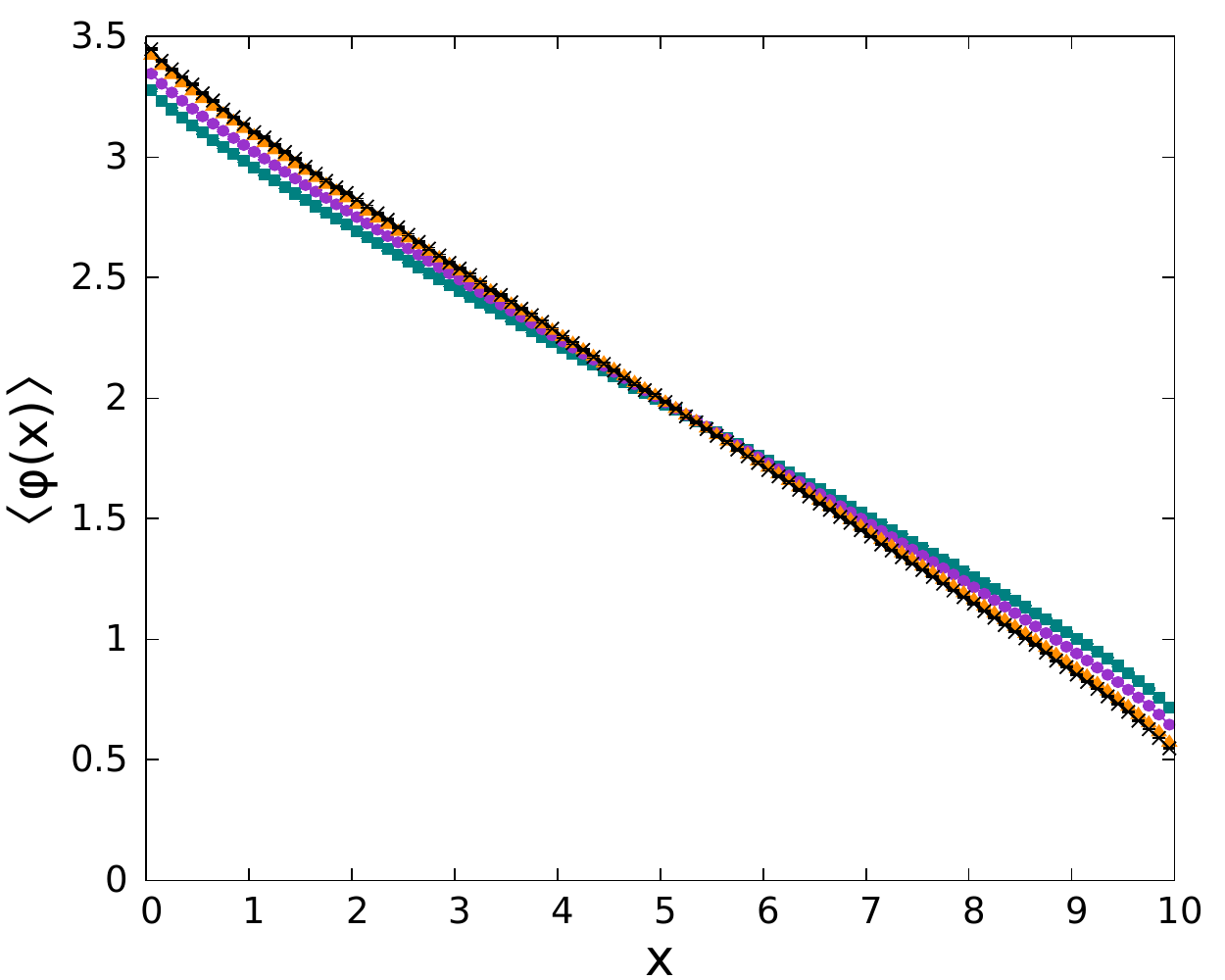}\,\,\,\,
\includegraphics[width=0.9\columnwidth]{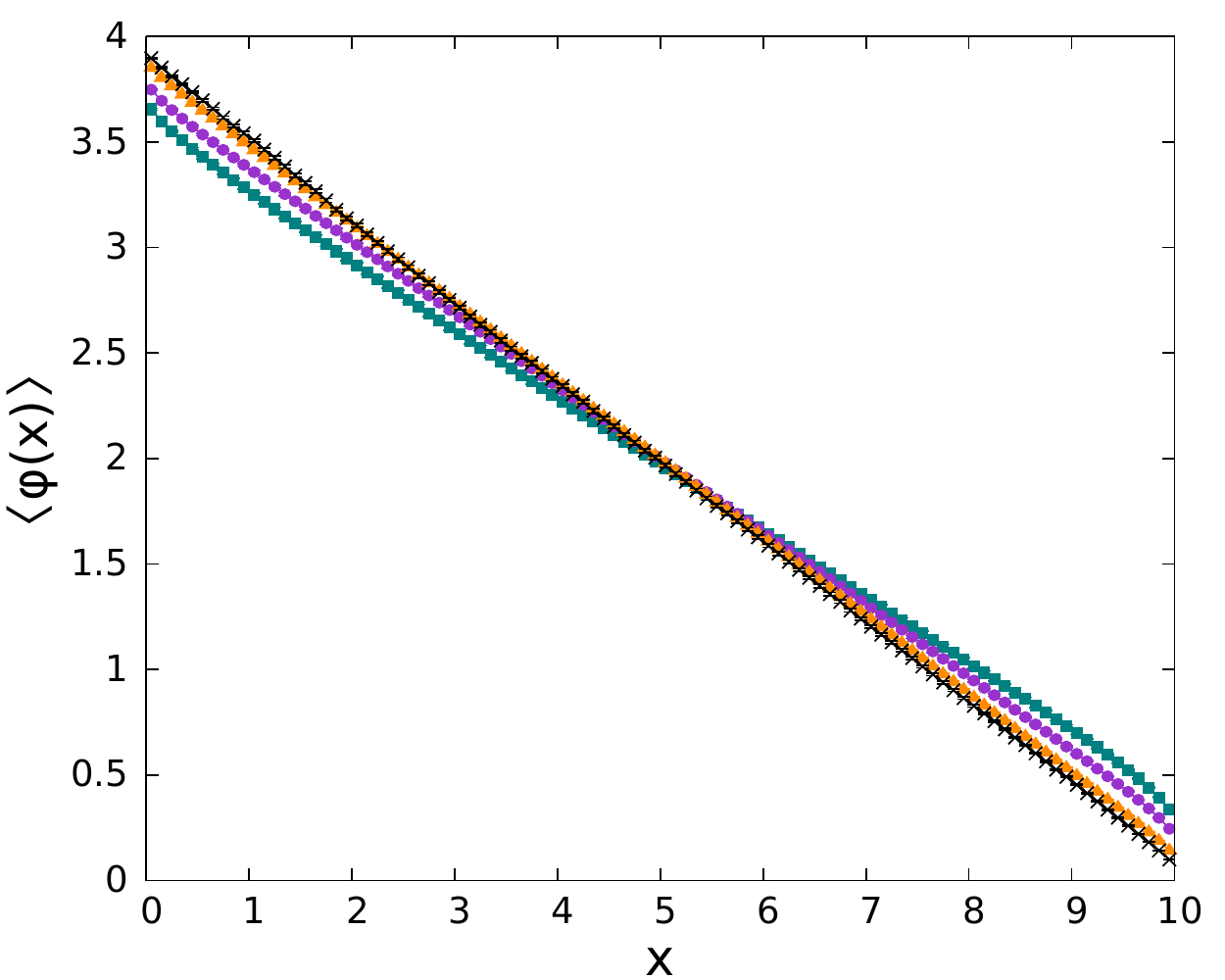}\\
\,\,\,\, $p=0.3$ \,\,\,\,\\
\includegraphics[width=0.9\columnwidth]{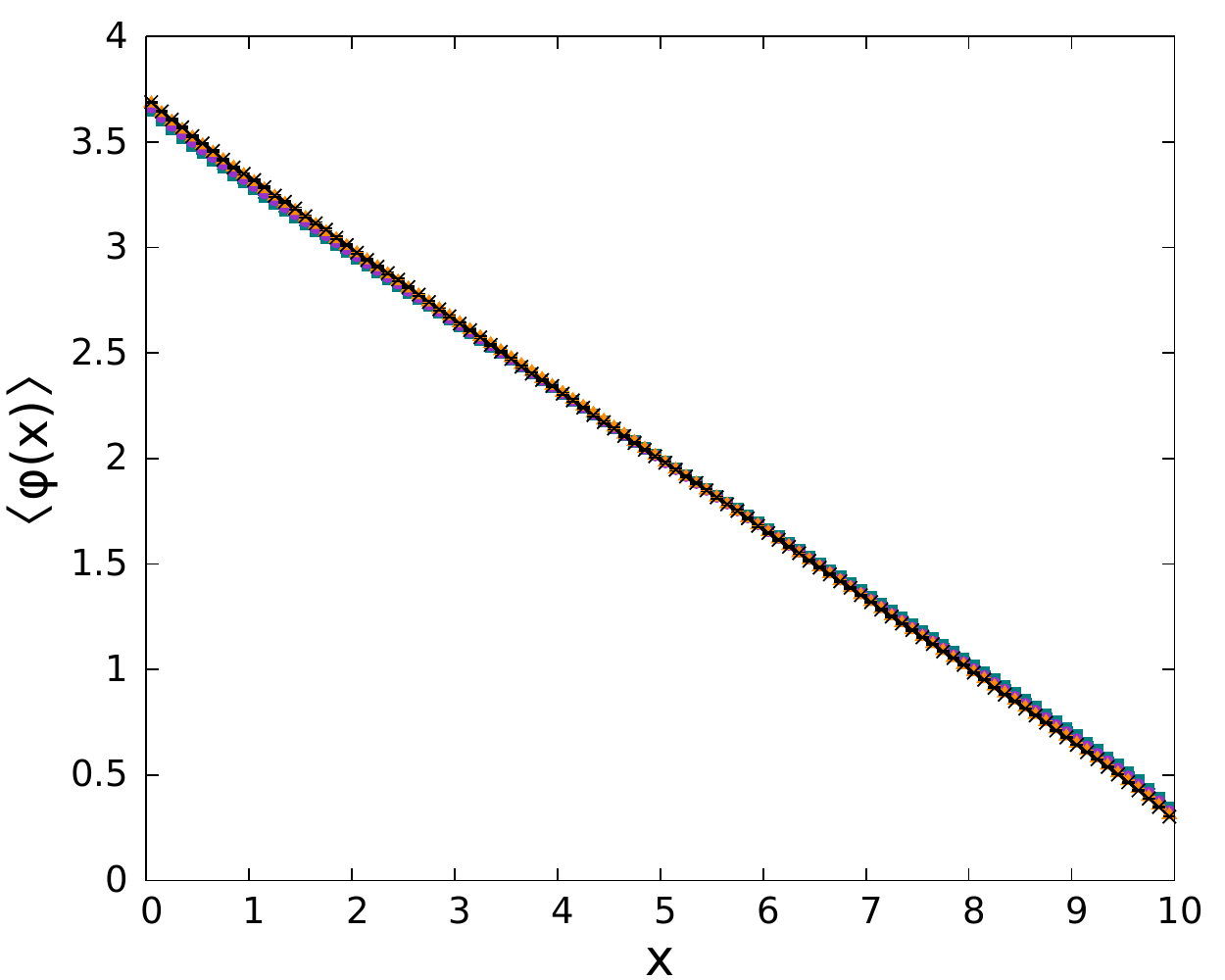}\,\,\,\,
\includegraphics[width=0.9\columnwidth]{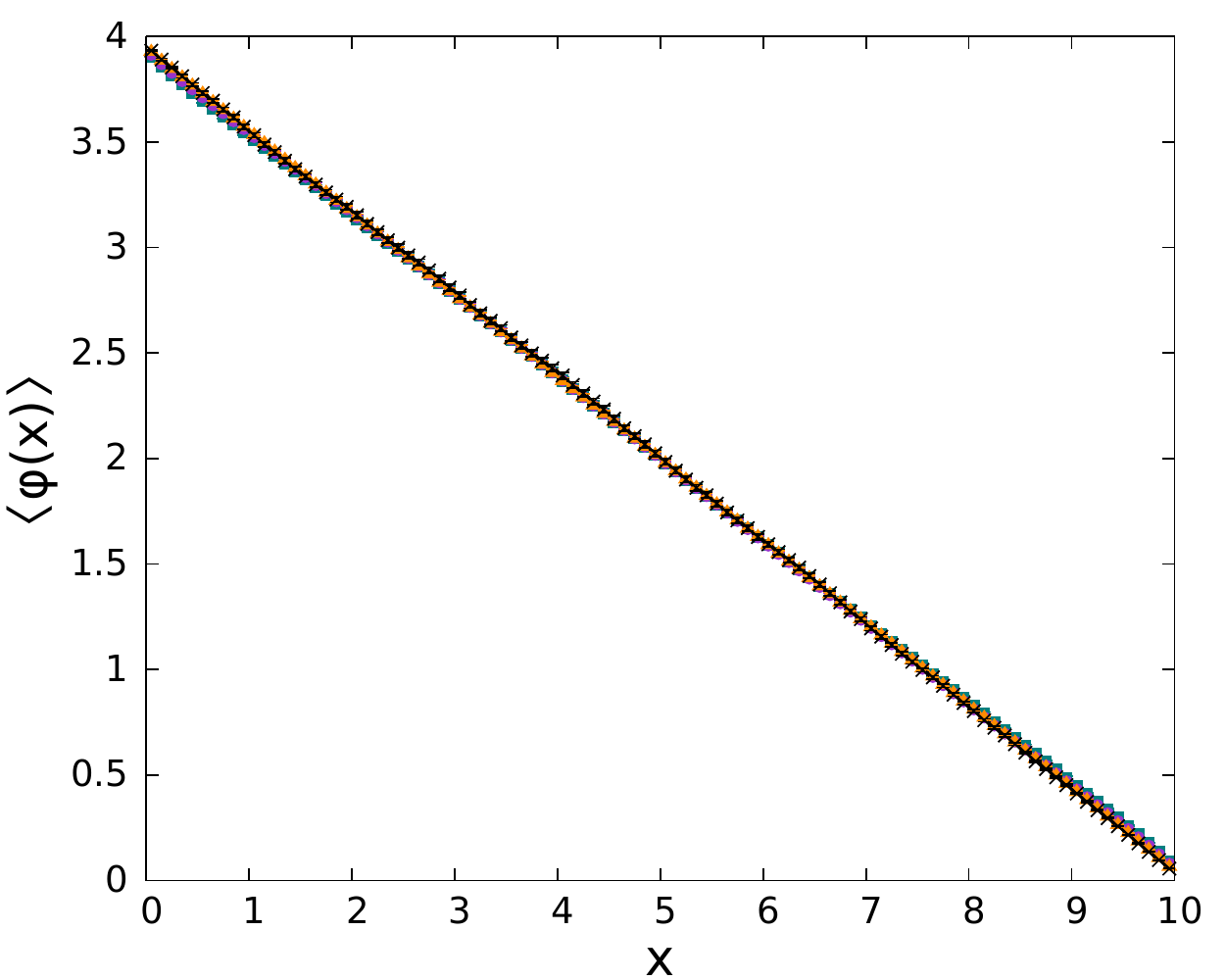}\\
\end{center}
\caption{Ensemble-averaged spatial scalar flux for Poisson tessellations ($m={\cal P}$), for the benchmark configurations: cases $1a$ (left) and $1b$ (right). Top: $p=0.95$; center$p=0.7$; bottom: $p=0.3$. Black crosses denote the atomic mix approximation, dark green squares ${\langle \Lambda \rangle}_{\infty}=1$, violet circles ${\langle \Lambda \rangle}_{\infty}=0.5$ and orange triangles ${\langle \Lambda \rangle}_{\infty}=0.1$.}
\label{density_2}
\end{figure*}

\begin{figure*}
\begin{center}
\,\,\,\, ${\langle \Lambda \rangle}_{\infty} =1$ \,\,\,\,\\
\includegraphics[width=0.9\columnwidth]{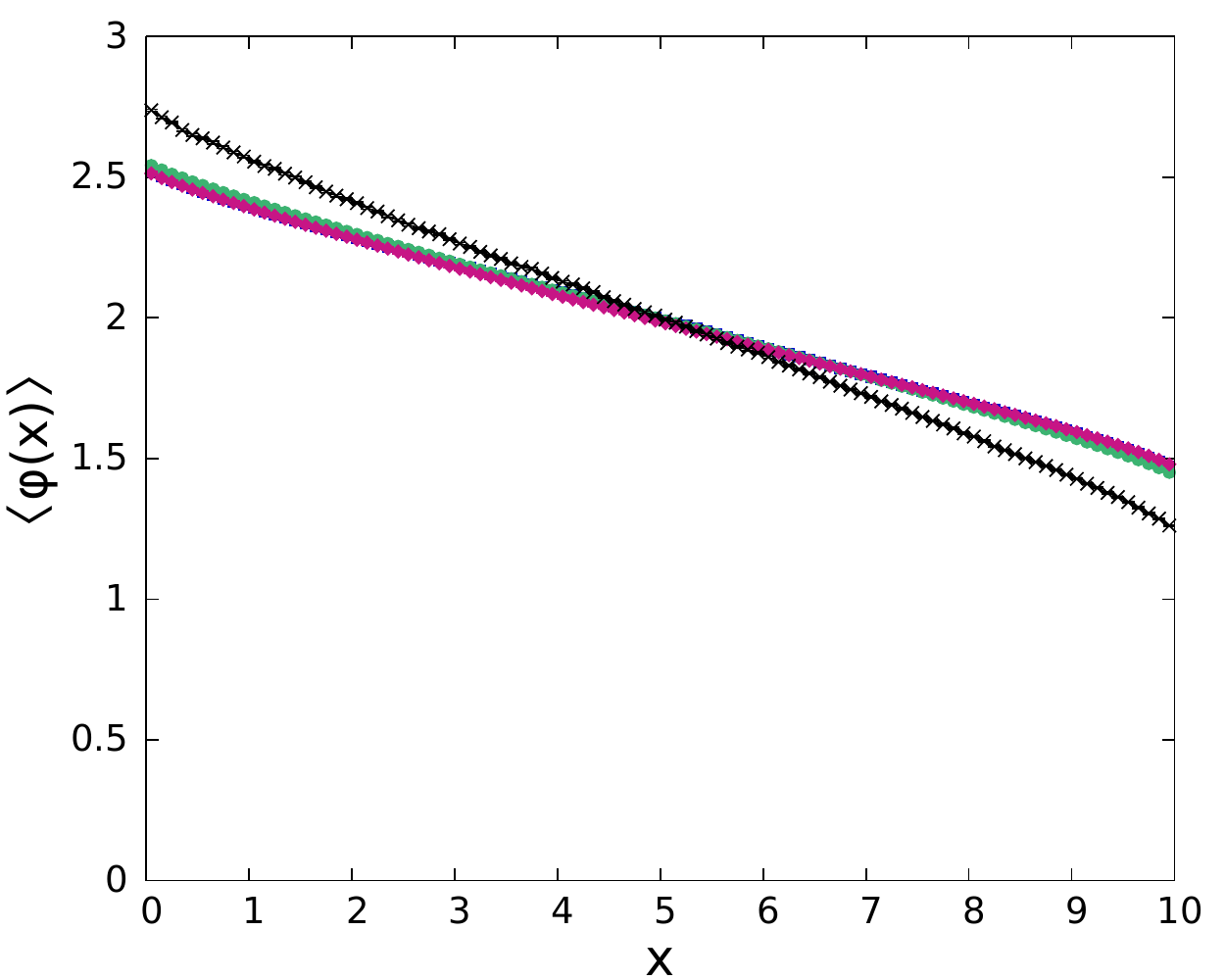}\,\,\,\,
\includegraphics[width=0.9\columnwidth]{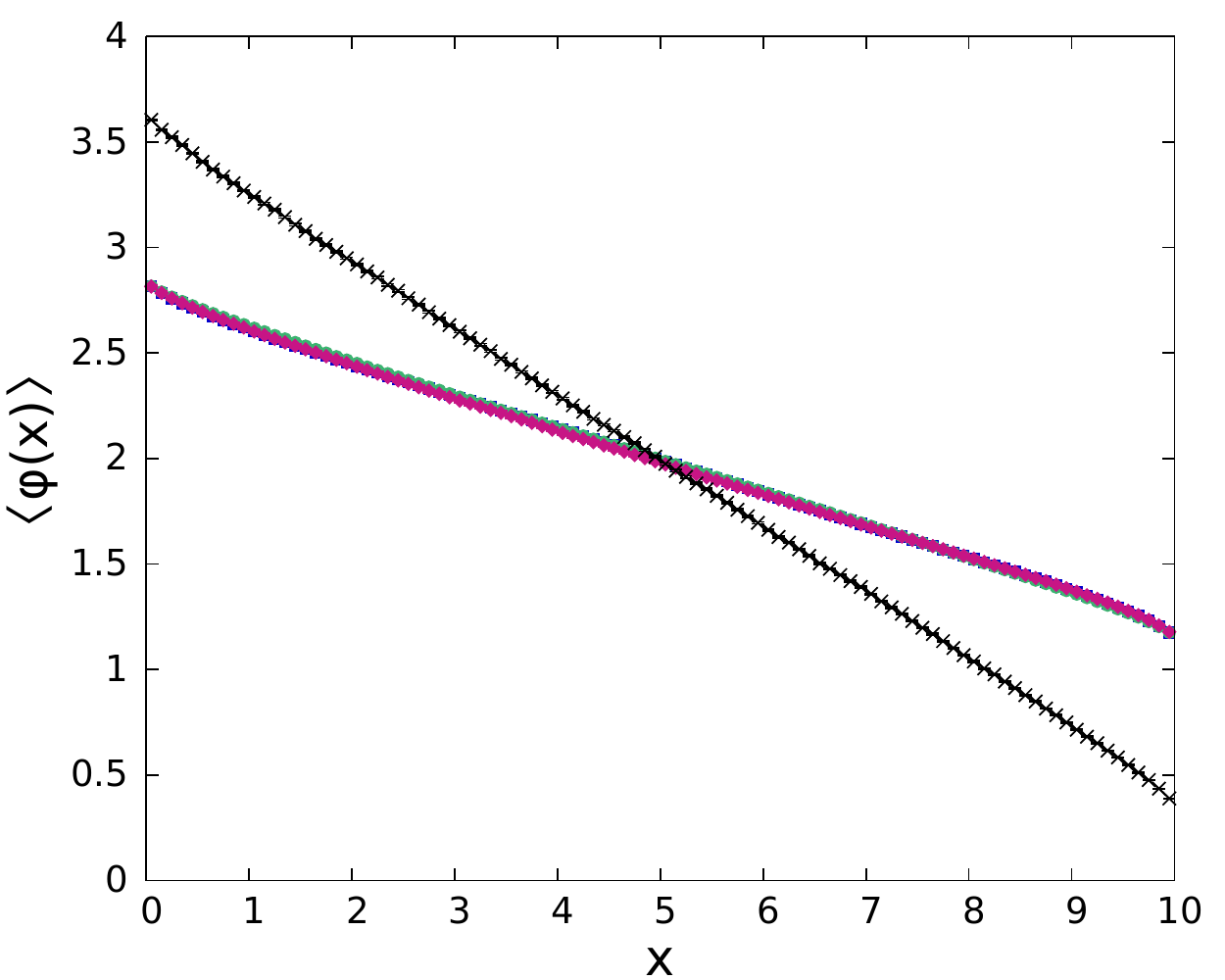}\\
\,\,\,\, ${\langle \Lambda \rangle}_{\infty} =0.5$ \,\,\,\,\\
\includegraphics[width=0.9\columnwidth]{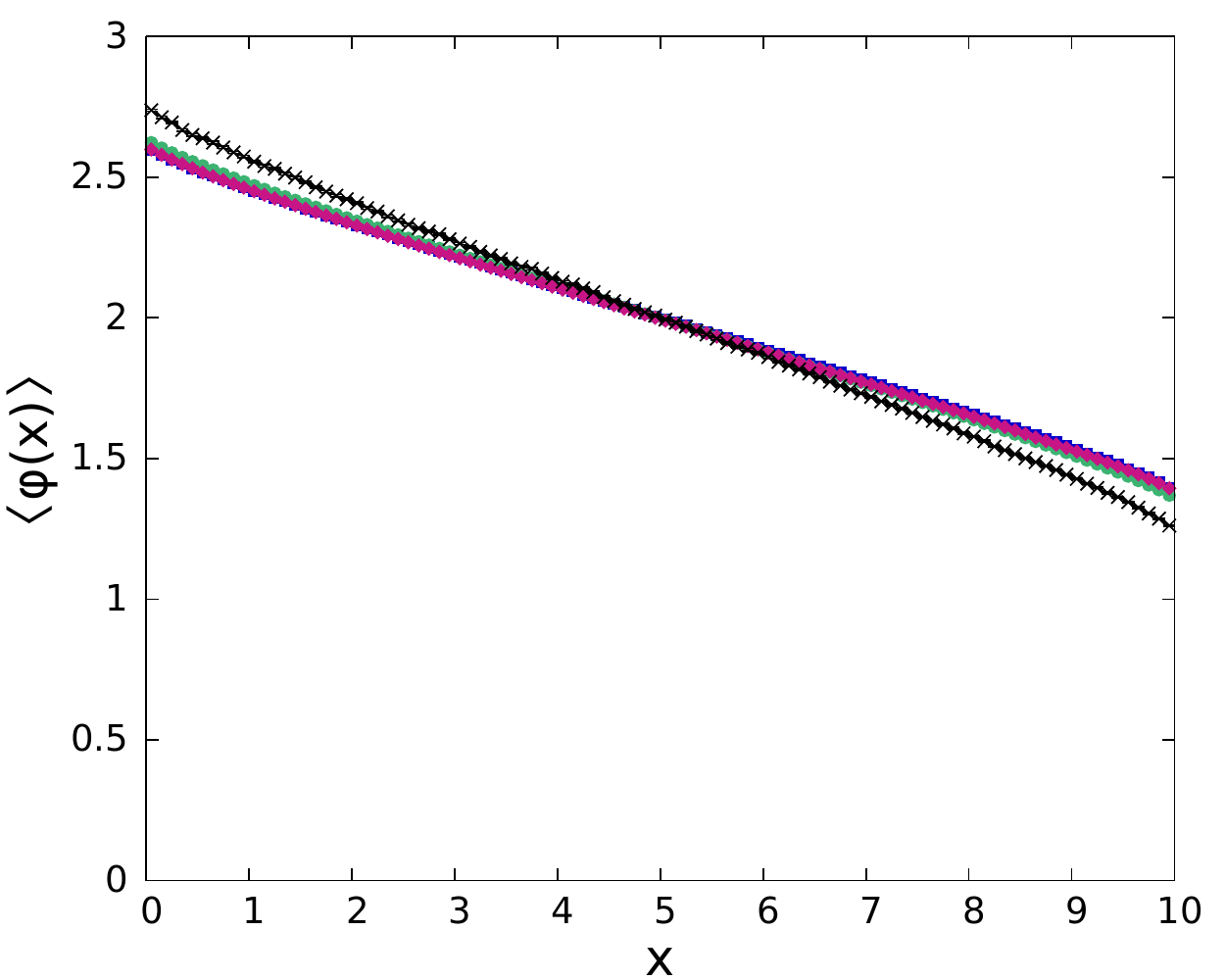}\,\,\,\,
\includegraphics[width=0.9\columnwidth]{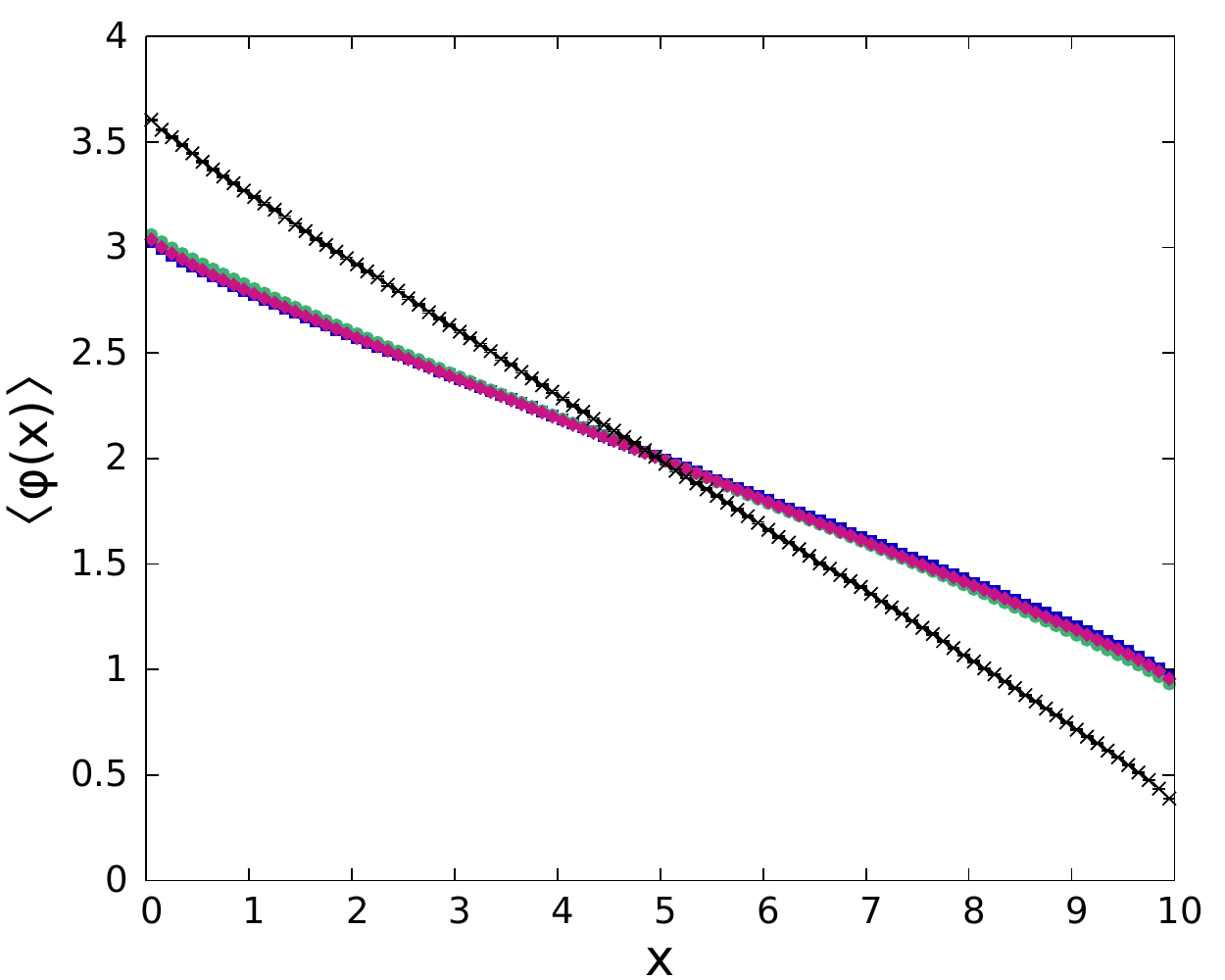}\\
\,\,\,\, ${\langle \Lambda \rangle}_{\infty} =0.1$ \,\,\,\,\\
\includegraphics[width=0.9\columnwidth]{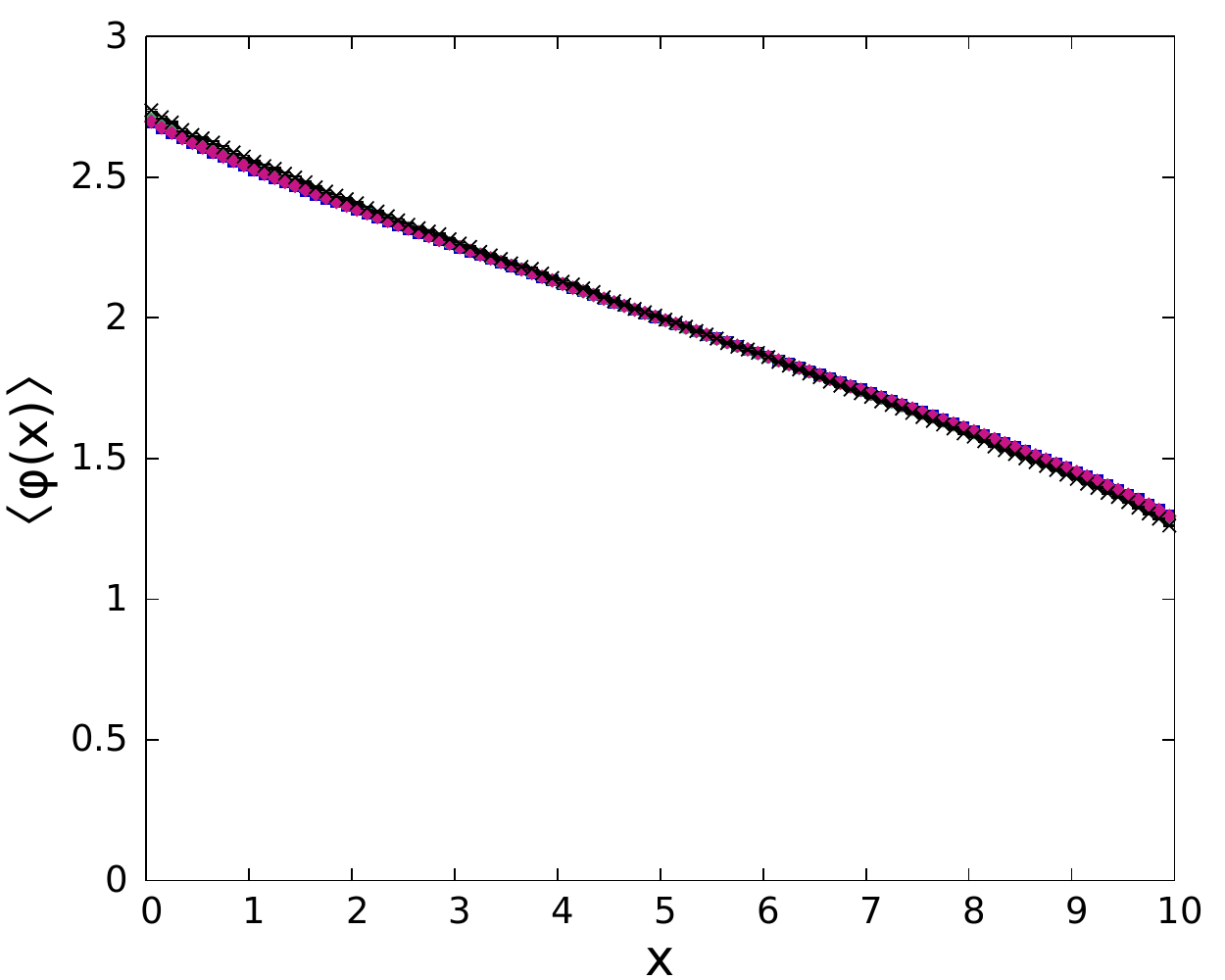}\,\,\,\,
\includegraphics[width=0.9\columnwidth]{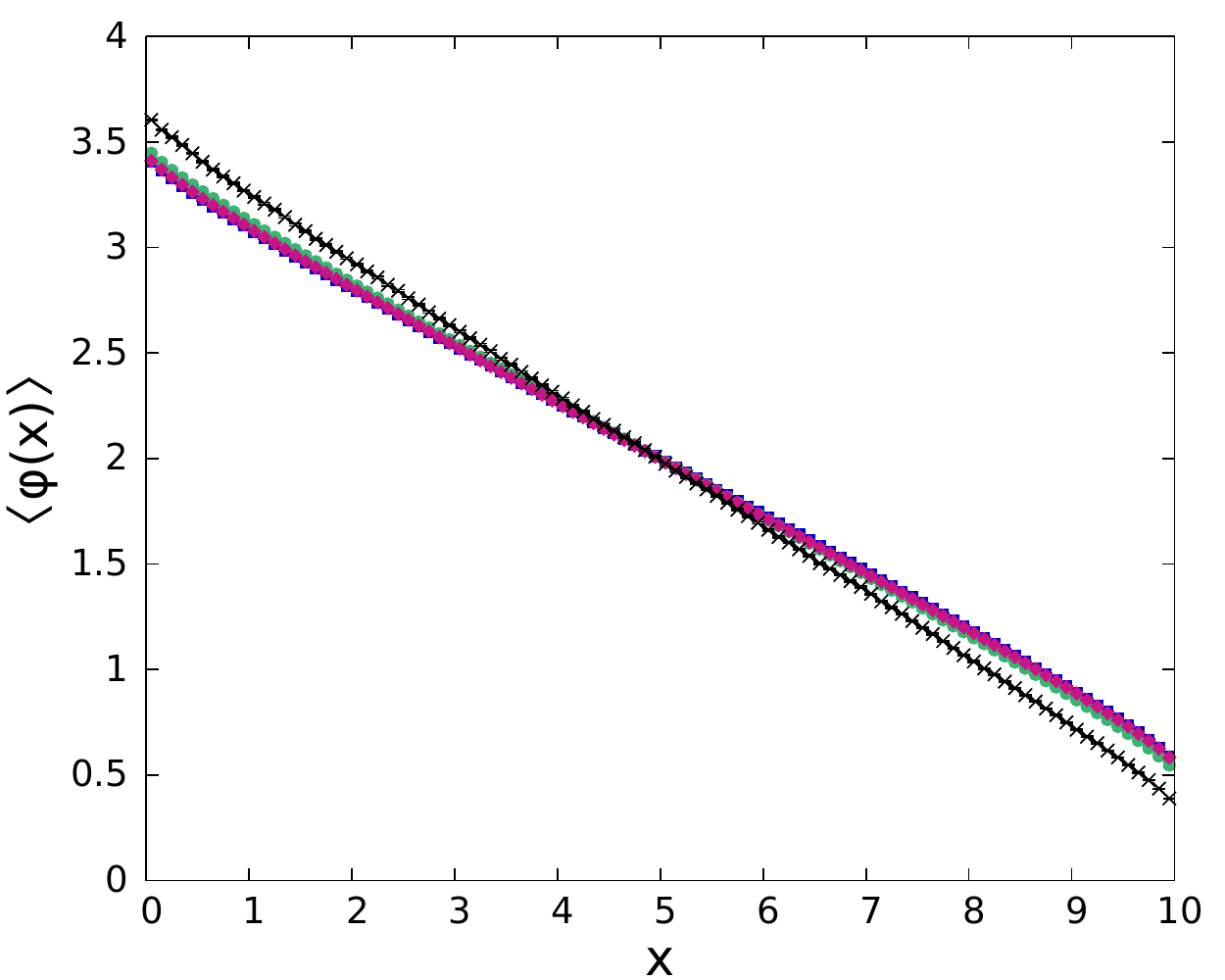}\\
\end{center}
\caption{Ensemble-averaged spatial scalar flux, for the benchmark configurations: cases $1a$ (left) and $1b$ (right), with $p=0.95$. Top: ${\langle \Lambda \rangle}_{\infty}=1$; center: ${\langle \Lambda \rangle}_{\infty}=0.5$; bottom: ${\langle \Lambda \rangle}_{\infty}=0.1$. Black crosses denote the atomic mix approximation, blue squares $m={\cal P}$, green circles $m={\cal V}$ and red diamonds $m={\cal B}$.}
\label{models_2_5}
\end{figure*}

\begin{figure*}
\begin{center}
\,\,\,\, $p=0.05$ \,\,\,\,\\
\includegraphics[width=0.9\columnwidth]{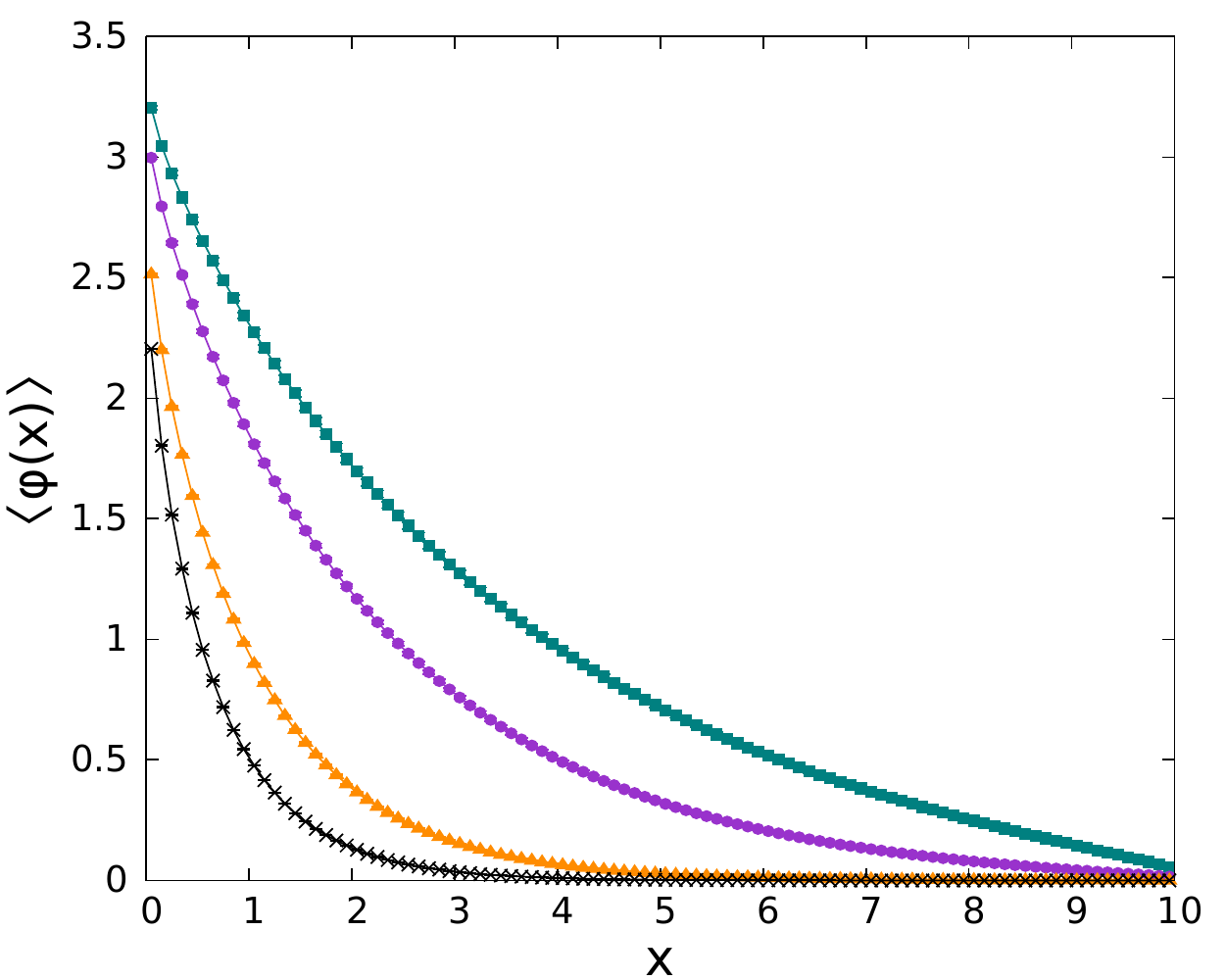}\,\,\,\,
\includegraphics[width=0.9\columnwidth]{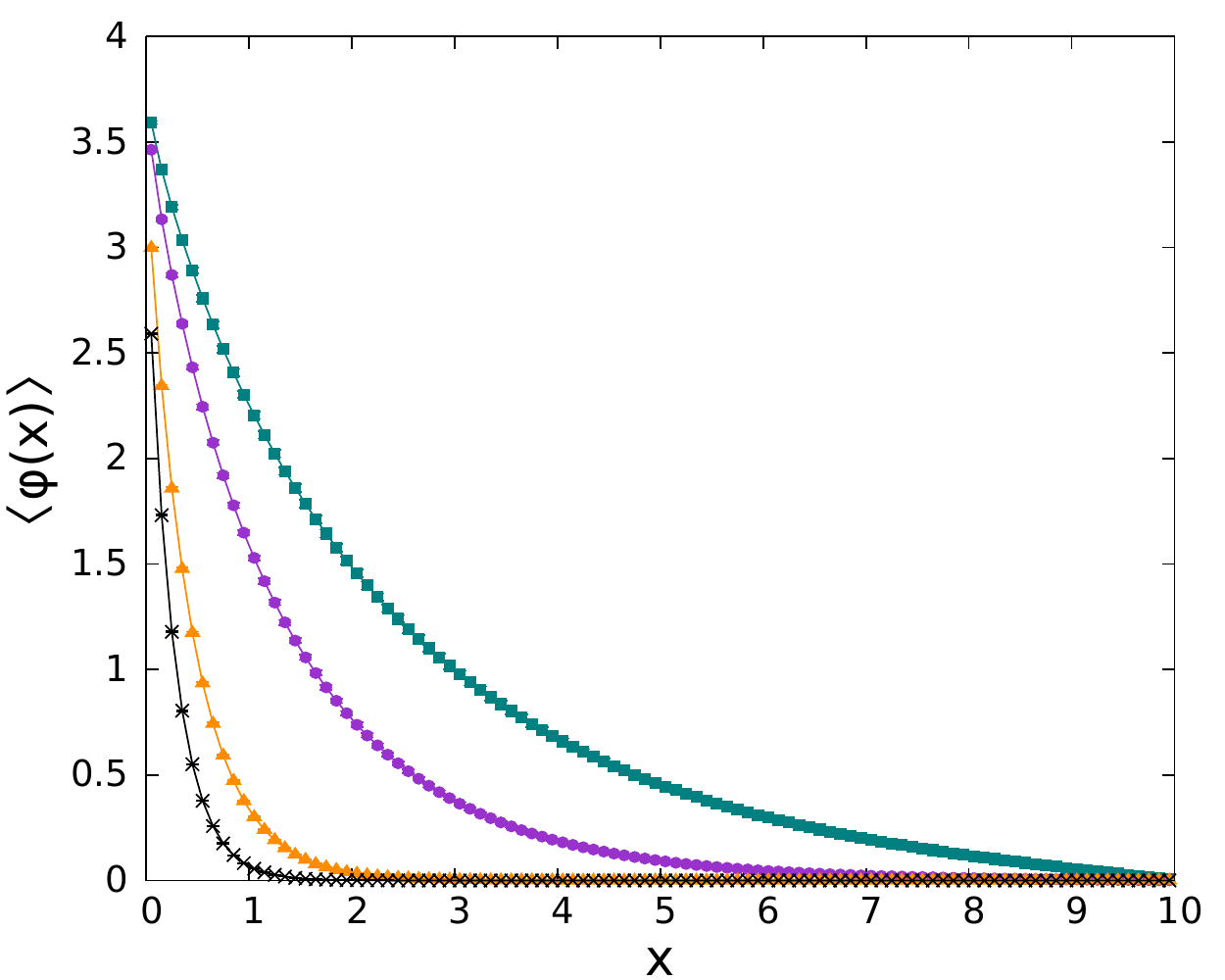}\\
\,\,\,\, $p=0.3$ \,\,\,\,\\
\includegraphics[width=0.9\columnwidth]{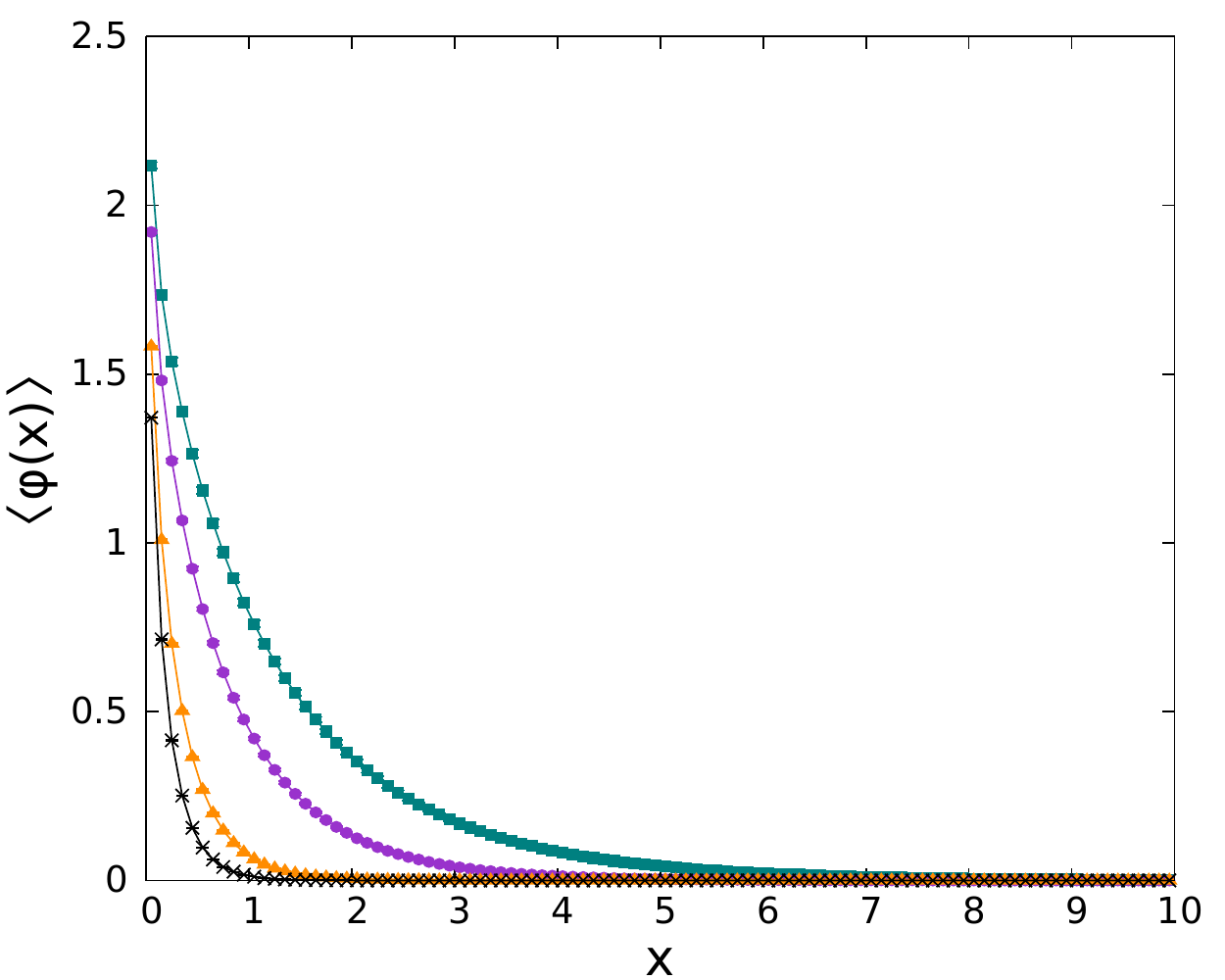}\,\,\,\,
\includegraphics[width=0.9\columnwidth]{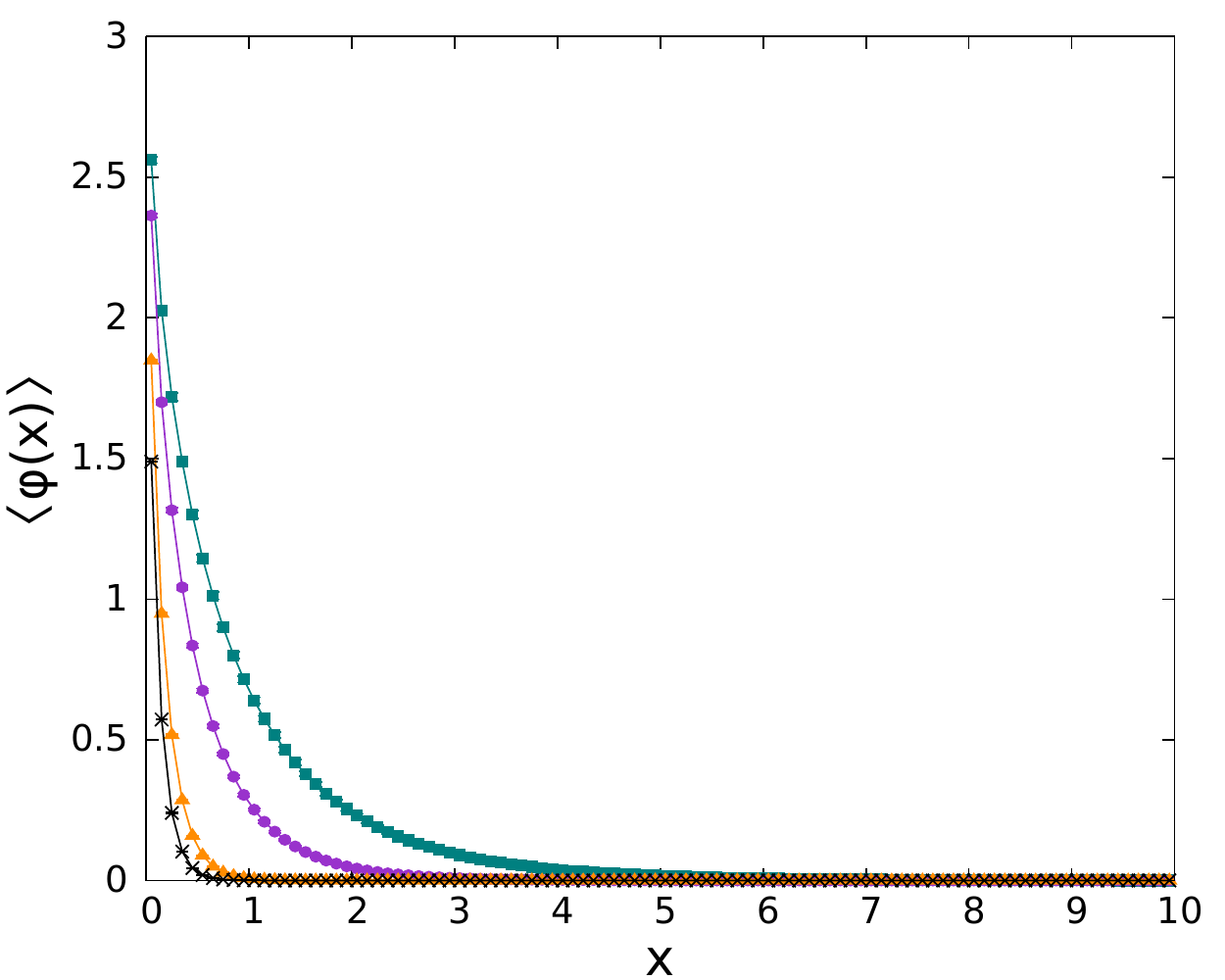}\\
\,\,\,\, $p=0.7$ \,\,\,\,\\
\includegraphics[width=0.9\columnwidth]{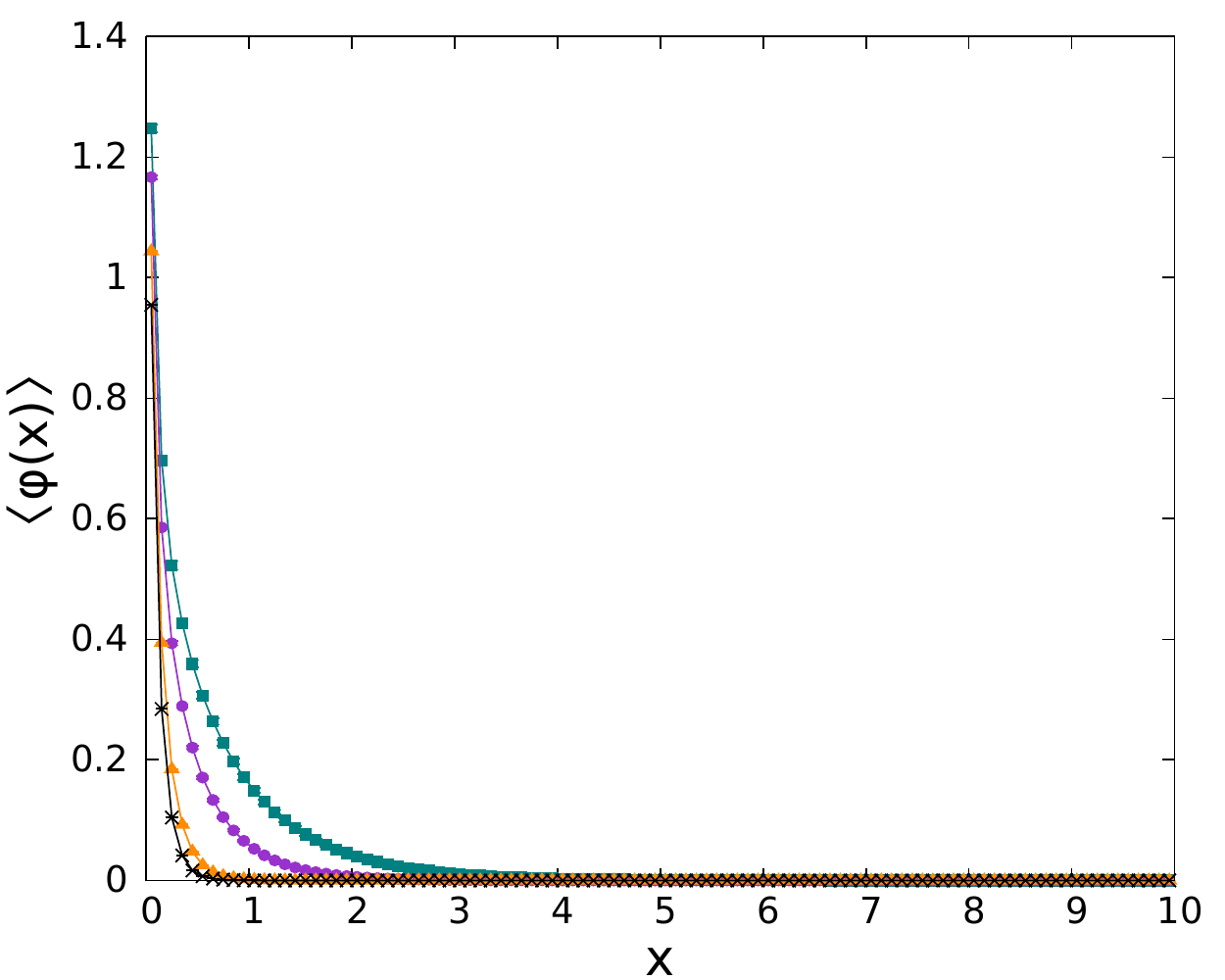}\,\,\,\,
\includegraphics[width=0.9\columnwidth]{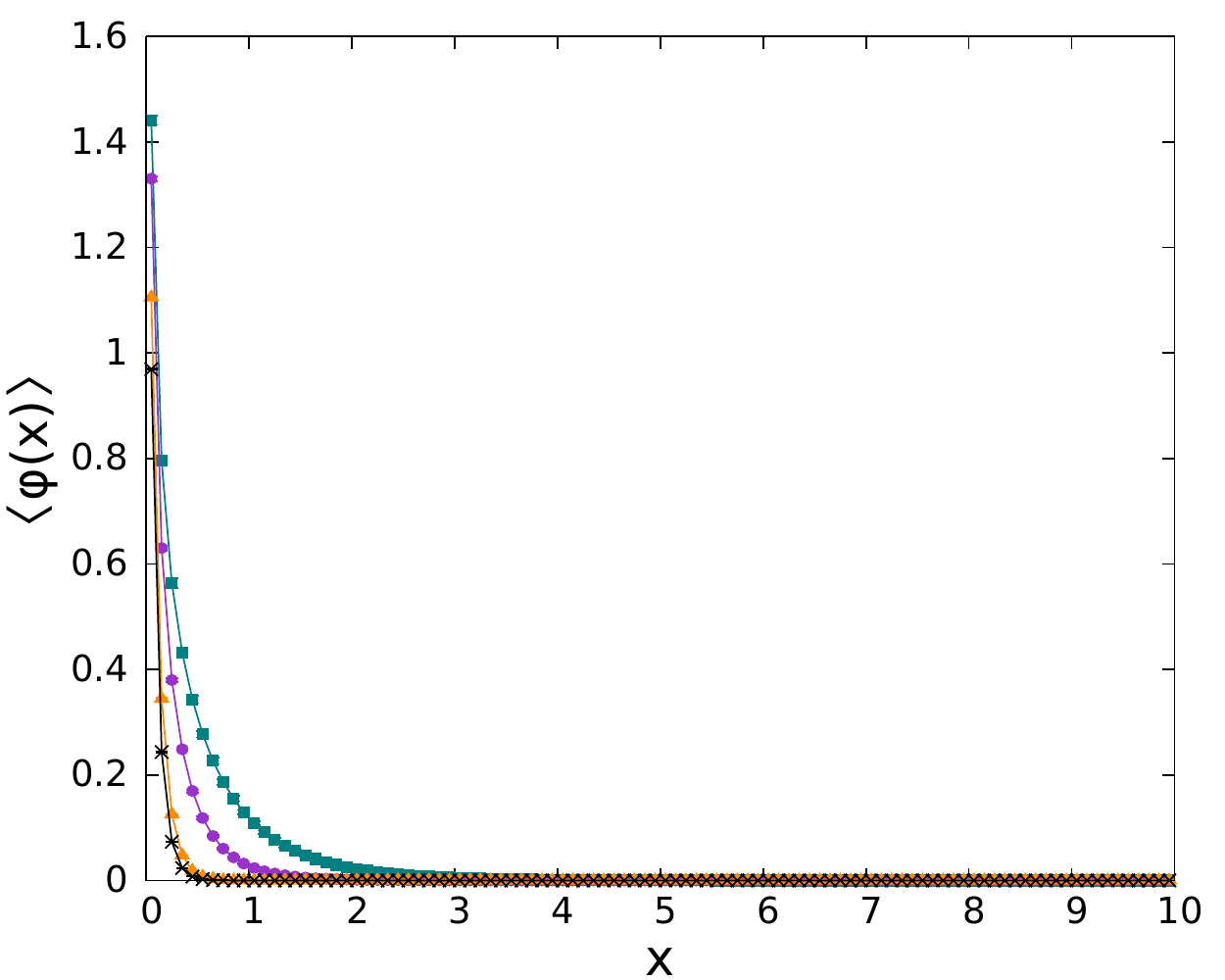}\\
\end{center}
\caption{Ensemble-averaged spatial scalar flux for Poisson tessellations ($m={\cal P}$), for the benchmark configurations: cases $2a$ (left) and $2b$ (right). Top: $p=0.05$; center: $p=0.3$; bottom: $p=0.7$. Black crosses denote the atomic mix approximation, dark green squares ${\langle \Lambda \rangle}_{\infty}=1$, violet circles ${\langle \Lambda \rangle}_{\infty}=0.5$ and orange triangles ${\langle \Lambda \rangle}_{\infty}=0.1$.}
\label{density_3}
\end{figure*}

\begin{figure*}
\begin{center}
\,\,\,\, ${\langle \Lambda \rangle}_{\infty} =1$ \,\,\,\,\\
\includegraphics[width=0.9\columnwidth]{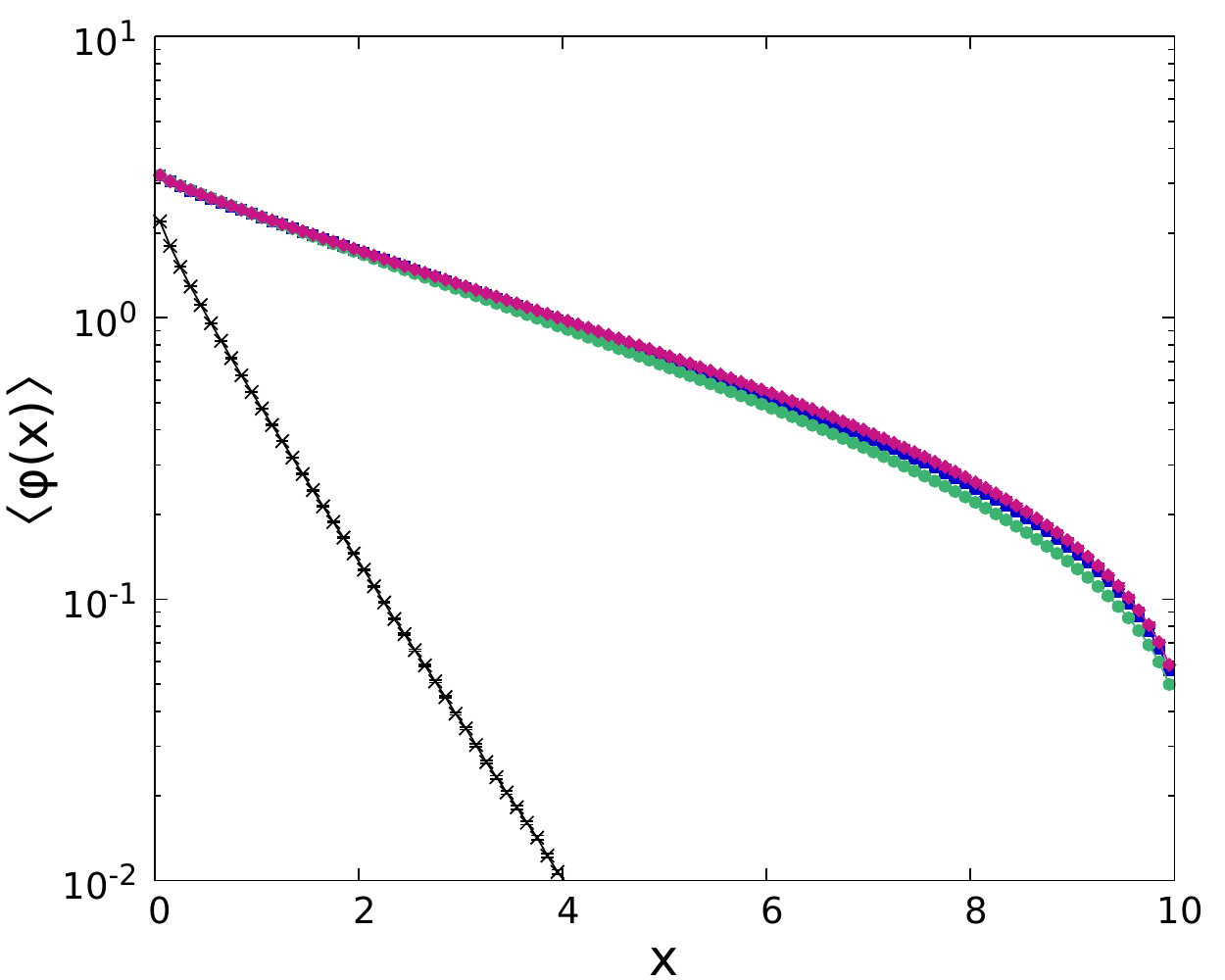}\,\,\,\,
\includegraphics[width=0.9\columnwidth]{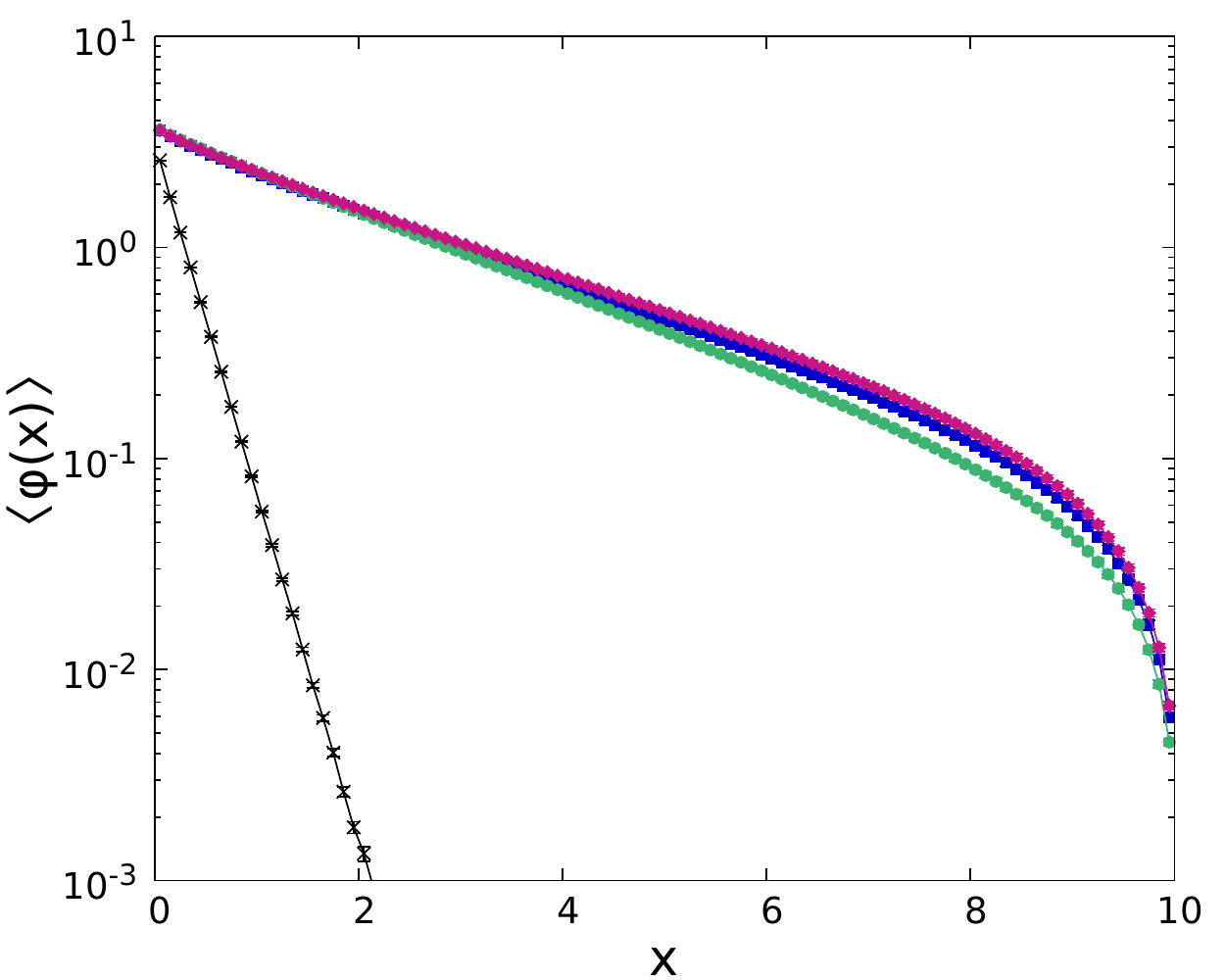}\\
\,\,\,\, ${\langle \Lambda \rangle}_{\infty} =0.5$ \,\,\,\,\\
\includegraphics[width=0.9\columnwidth]{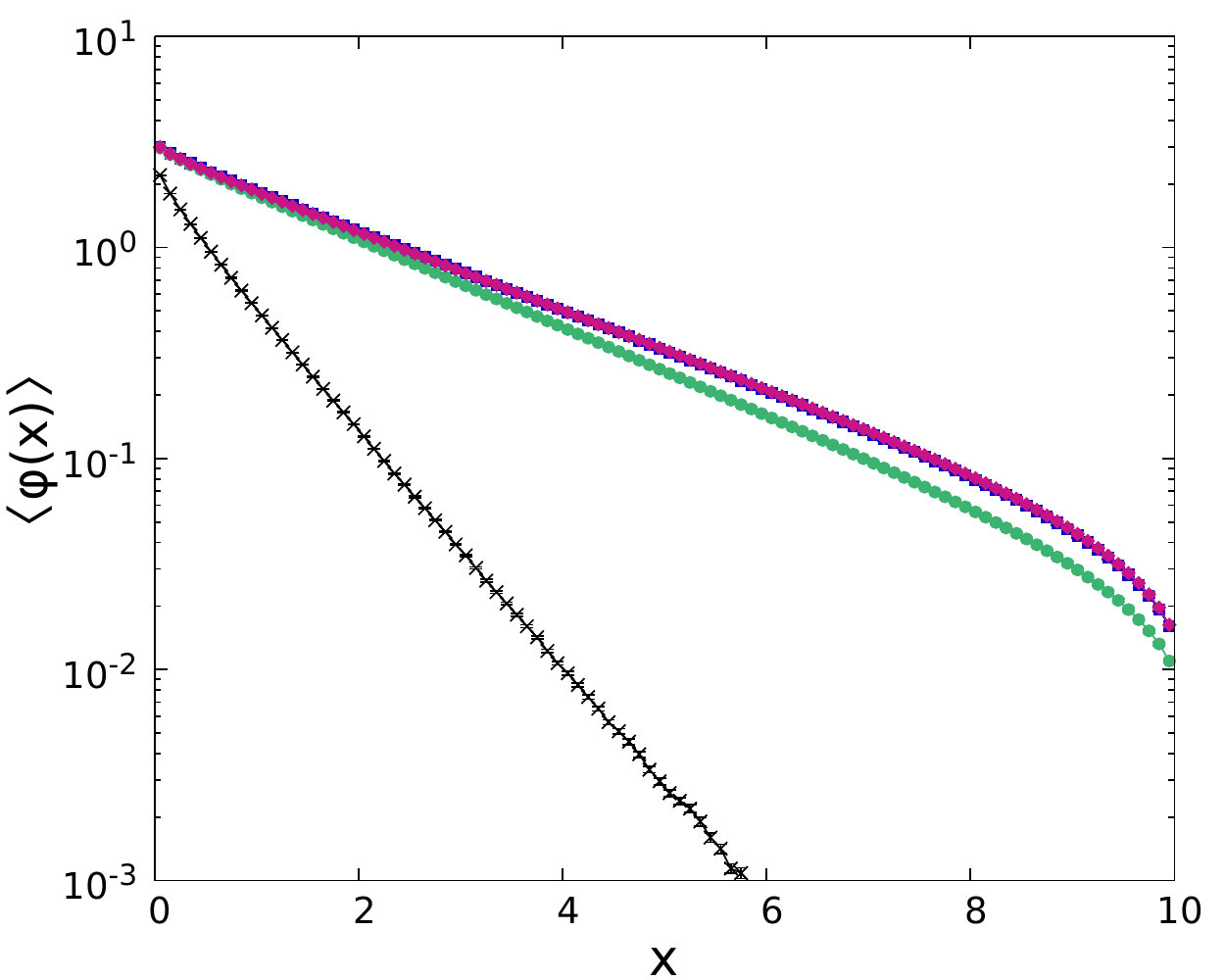}\,\,\,\,
\includegraphics[width=0.9\columnwidth]{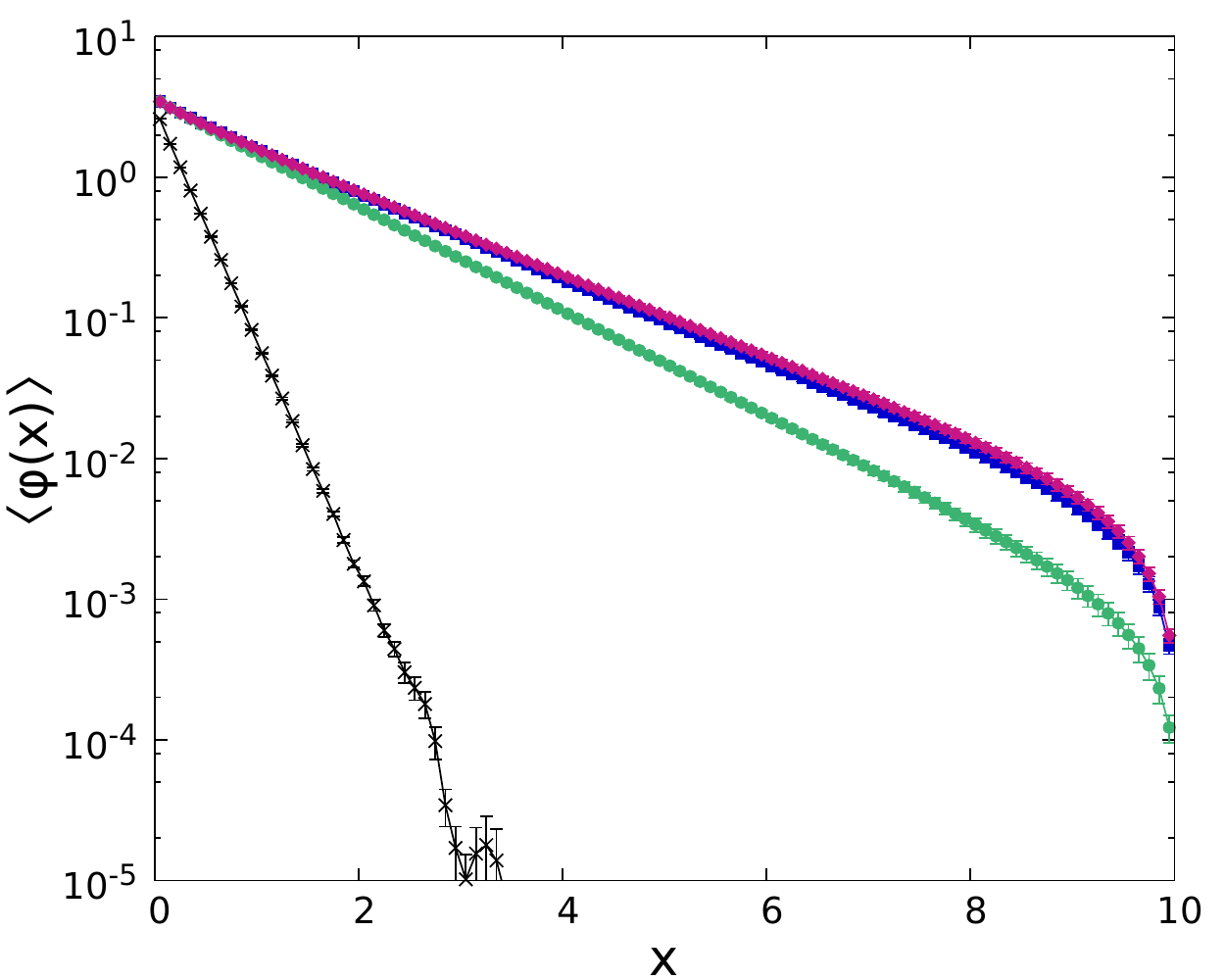}\\
\,\,\,\, ${\langle \Lambda \rangle}_{\infty} =0.1$ \,\,\,\,\\
\includegraphics[width=0.9\columnwidth]{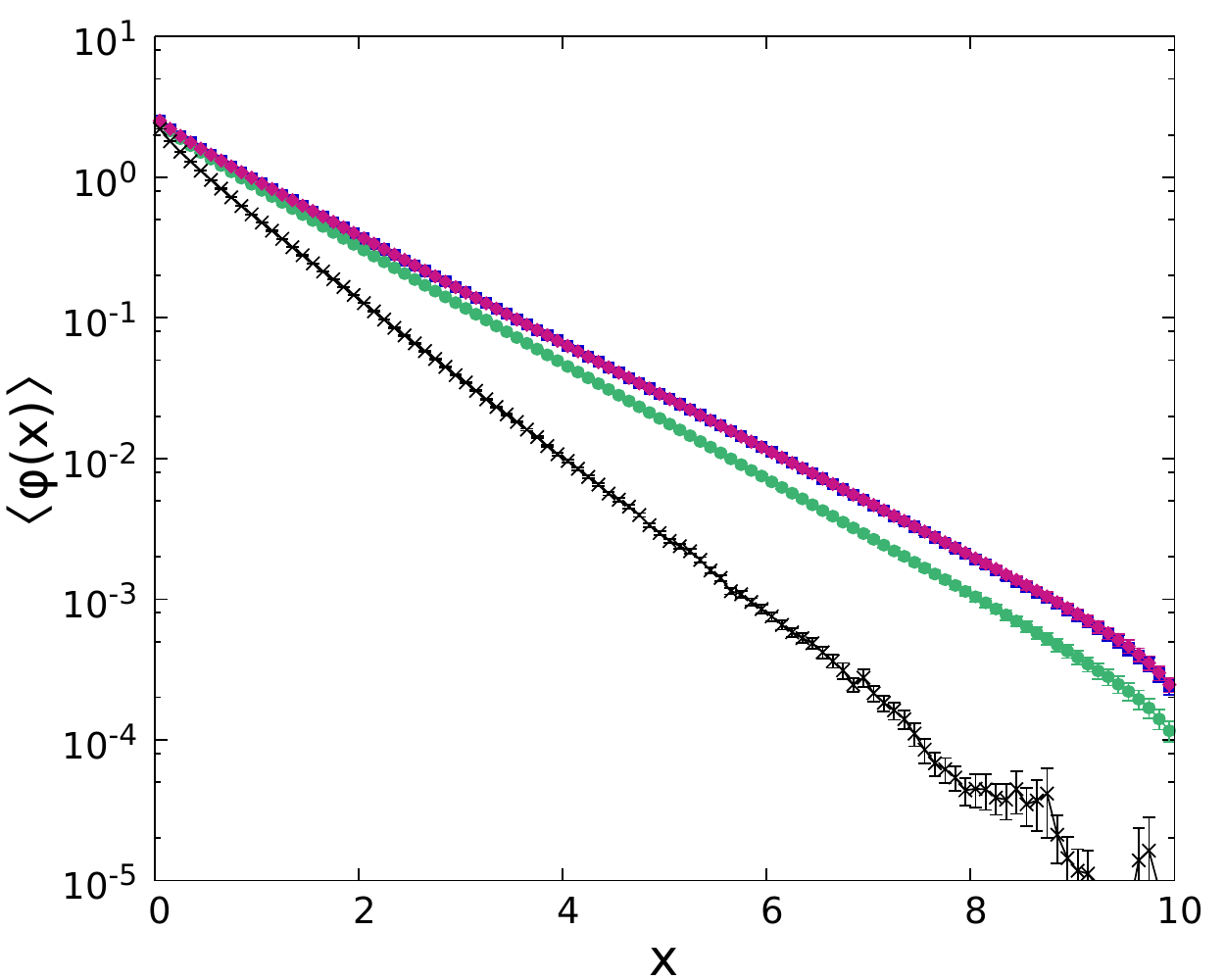}\,\,\,\,
\includegraphics[width=0.9\columnwidth]{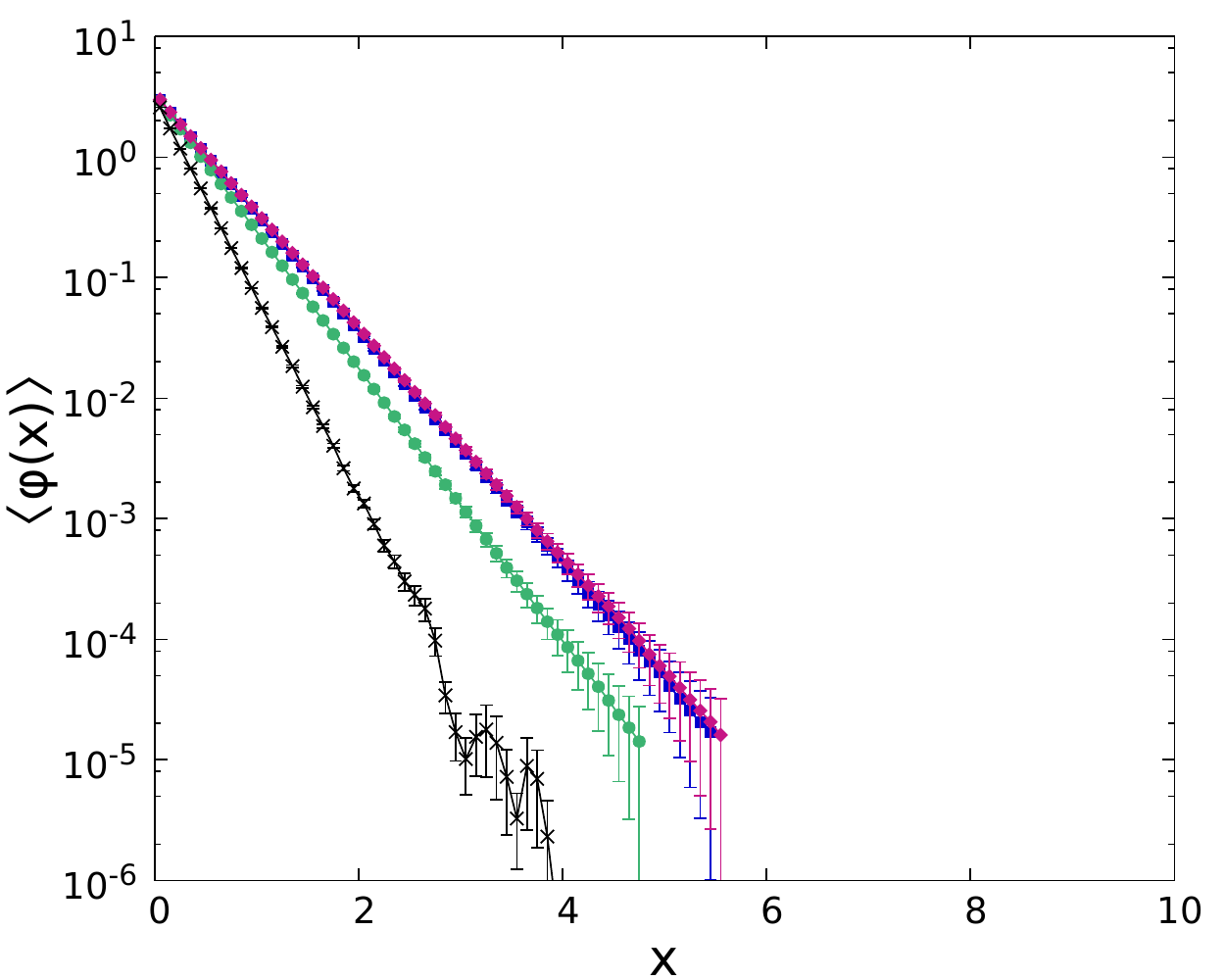}\\
\end{center}
\caption{Ensemble-averaged spatial scalar flux, for the benchmark configurations: cases $2a$ (left) and $2b$ (right), with $p=0.05$. Top: ${\langle \Lambda \rangle}_{\infty}=1$; center: ${\langle \Lambda \rangle}_{\infty}=0.5$; bottom: ${\langle \Lambda \rangle}_{\infty}=0.1$. Black crosses denote the atomic mix approximation, blue squares $m={\cal P}$, green circles $m={\cal V}$ and red diamonds $m={\cal B}$.}
\label{models_3_5}
\end{figure*}

The spatial profiles of the ensemble-averaged scalar flux $\langle \varphi(x) \rangle= \langle \int \int \int\varphi({\bf r}, {\boldsymbol \omega}) d {\boldsymbol \omega} dy dz \rangle$ along the coordinate $x$ are reported in Figs.~\ref{density_2} and~\ref{models_2_5} for case $1$ and in Figs.~\ref{density_3} and~\ref{models_3_5} for case $2$. In \tripoli{}, we estimate $\langle \varphi(x) \rangle$ by recording the flux within the spatial grid defined above and by dividing the obtained result by the volume of each mesh.

Consistently with the findings concerning the scalar observables, the impact of ${\langle \Lambda \rangle}_{\infty}$ on the spatial profile of the scalar flux is clearly visible (see Figs.~\ref{density_2} and~\ref{density_3}). The lower ${\langle \Lambda \rangle}_{\infty}$, the closer is the associated spatial flux profile to the results of the atomic mix. As for the transmission and reflection probabilities, the impact of ${\langle \Lambda \rangle}_{\infty}$ depends on the probability $p$ and on the scattering cross section $\Sigma_{\beta}^s$. The discrepancy between the atomic mix and stochastic tessellations increases with $p$ for case $1$ and decreases with $p$ for case $2$; furthermore, the discrepancy systematically increases with increasing $\Sigma_{\beta}^s$.

Mixing statistics $m$ plays also a role, but again its effect is weaker with respect to the other benchmark parameters, as illustrated in Figs.~\ref{models_2_5} and~\ref{models_3_5}. For each benchmark configuration, we observe an excellent agreement between Poisson tessellations and Box tessellations. On the contrary, spatial profiles associated with Voronoi tessellations agree with those of the other tessellations for case $1$, but show a distinct behaviour for case $2$.

\section{Conclusions}
\label{conclusions}

Benchmark solutions for particle transport in random media are mandatory in order to validate faster, albeit approximate solutions coming from closure formulas, such as the celebrated Levermore-Pomraning model, or from effective transport kernels for stochastic methods, such as the Chord Length Sampling algorithm. A common assumption when dealing with random media is that the mixing statistics obeys a Poisson (Markov) distribution~\cite{pomraning}. Most research efforts have been devoted to producing reference benchmark solutions for one-dimensional random media with slab or rod geometry: in this context, ground-breaking work was performed by Adams, Larsen and Pomraning, who proposed a series of benchmark problems in binary stochastic media with Markov mixing~\cite{benchmark_adams}. Their findings were later revisited and extended in~\cite{brantley_benchmark, brantley_conf, brantley_conf_2, vasques_suite2}. In recent works, we have examined the effects of dimension and finite-size on the physical observables of interest~\cite{larmier_benchmark, larmier}.

The impact of the mixing statistics on particles transport has received so far comparatively less attention~\cite{renewal}. In this work, we have considered the effects of varying the stochastic tessellation model on the statistical properties of the resulting random media and on the transport-related physical observables, such as the reflection and the transmission probabilities. As such, this paper is a generalization of our previous findings~\citep{larmier_benchmark, larmier}, and might be helpful for researchers interested in developing effective kernels for particle transport in disordered media. In order to single out the sensitivity of the simulation results to the various model parameters, we have proposed two benchmark configurations that are simple enough and yet retain the key physical ingredients. In the former, we have considered a binary mixture composed of diffusing materials and voids; in the latter, a binary mixture composed on diffusing and absorbing materials. The mixing statistics that have been selected for this work are the homogeneous and isotropic Poisson tessellations, the Poisson-Voronoi tessellations, and the Poisson Box tessellations.

For each benchmark configuration we have provided reference solutions by resorting to Monte Carlo simulation. A computer code that has been developed at CEA allows generating the chosen tessellations, and particle transport has been then realized by resorting to the Monte Carlo code \tripoli{}. The effects of the underlying mixing statistics, the cross sections, the material compositions and the average chord length of the stochastic geometries have been accurately and extensively assessed by varying each parameter. The distribution of the chord lengths plays an important role in characterizing the transport properties, which explains why the Voronoi tessellation, whose chord length is significantly different from the that of the other mixing statistics, shows a distinct behaviour. We conclude by remarking that Box tessellations yield almost identical results with respect to Poisson tessellations: this is a remarkable feature, in that the realizations of Box geometries are much simpler and could be perhaps adapted for deterministic transport codes.

\section*{Acknowledgements}
TRIPOLI\textsuperscript{ \textregistered} and TRIPOLI-4\textsuperscript{ \textregistered} are registered trademarks of CEA. The authors wish to thank \'Electricit\'e de France (EDF) for partial financial support.

\clearpage

\begin{table*}
\begin{center}
\begin{tabular}{ccccccccccccccccc}
\toprule
$m$ & ${\langle \Lambda \rangle}_{\infty}$ & $\langle \Lambda | L\rangle$ & ${\langle V \rangle}_{\infty}$ & $\langle V | L \rangle$ & ${\langle S \rangle}_{\infty}$ & $\langle S | L \rangle$ & ${\langle N_F \rangle}_{\infty}$ & $\langle N_F | L \rangle$ \\
 \midrule
& $1$ & $0.9247 \pm 7.10^{-4}$ & $1.9099$ & $1.90 \pm 5.10^{-2}$ & $7.63944$ & $7.5 \pm 0.1$  & $6$ & $5.9979 \pm 3.10^{-4}$ \\
${\cal P}$ & $0.5$ & $0.4755 \pm 3.10^{-4}$ & $0.238732$ & $0.243 \pm 4.10^{-3}$ & $1.90986$ & $1.93 \pm 2.10^{-2}$ & $6$ & $5.99963 \pm 6.10^{-5}$  \\
& $0.1$ & $0.09848 \pm 7.10^{-5}$ &  $0.00191$ & $0.00196 \pm 3.10^{-5}$ & $0.076394$ & $0.0777 \pm 8.10^{-4}$  & $6$ & $6.000000 \pm 2.10^{-6}$ \\
\midrule
& $1$ & $0.9554 \pm 5.10^{-4}$ & $3.081$ & $3.092 \pm 5.10^{-3}$  & $12.326$ & $12.78 \pm 2.10^{-2}$ & $15.54$ & $13.125 \pm 4.10^{-3}$   \\
${\cal V}$ & $0.5$ & $0.4880 \pm 2.10^{-4}$ & $0.3852$ & $0.3854 \pm 3.10^{-4}$ & $3.0815$ & $3.137 \pm 10^{-3}$ & $15.54$ & $14.300 \pm 10^{-3}$\\
& $10$ & $0.09952 \pm 4.10^{-5}$ &  $0.003081$ & $0.003082 \pm 4 .10^{-6}$  &  $0.12326$ & $0.12371 \pm 10^{-5}$ & $15.54$ & $15.2834 \pm 2.10^{-4}$ \\
\midrule
& $1$ & $0.9056 \pm 7.10^{-4}$ & $3.375$ & $3.18 \pm 8.10^{-2}$ & $13.5$  & $13.0 \pm 0.2$ & $6$ & $6$  \\
${\cal B}$ & $0.5$ & $0.4795 \pm 3.10^{-4}$ & $0.4219$ & $0.411 \pm 8.10^{-3}$ & $3.375$ & $3.31 \pm 4.10^{-2}$ & $6$ & $6$  \\
& $0.1$ & $0.09961 \pm 8.10^{-5}$ & $0.00338$ & $0.00325 \pm 5.10^{-5}$ & $0.135$ & $0.132 \pm 10^{-3}$ & $6$ & $6$  \\
\bottomrule
\end{tabular}
\end{center}
\caption{Statistical properties of the tessellations used for the benchmark configurations, as a function of the mixing statistics $m$ and of the average chord length ${\langle \Lambda \rangle}_{\infty}$, for a domain of linear size $L=10$. For a given observable $Q$, the Monte Carlo result $\langle Q | L \rangle$ is compared to the associated value ${\langle Q \rangle}_{\infty}$ corresponding to infinite tessellations. $\langle \Lambda | L \rangle$ is the average chord length measured by Monte Carlo ray tracing. $\langle V | L \rangle$ is the average volume of a polyhedron, $\langle S | L \rangle$ is the average total surface of a polyhedron and $\langle N_F | L \rangle$ is the average number of faces of a polyhedron.}
\label{tab_geo}
\end{table*}

\begin{table*}
\begin{center}
\begin{tabular}{cccccccccccc}
\toprule
${\langle \Lambda \rangle}_{\infty}$ &  $p$ & ${\langle \Lambda_{\alpha} \rangle}_{\infty}$ & $\langle \Lambda_{\alpha} | L \rangle ^{\cal P}$ & $\langle \Lambda_{\alpha} | L \rangle ^{\cal V}$ & $\langle \Lambda_{\alpha} | L \rangle ^{\cal B}$\\
 \midrule
& $0.05$ & $0.1053$ & $0.1032 \pm 0.0005$ & $0.1045 \pm 0.0003$ & $0.1043 \pm 0.0005$\\
& $0.15$ & $0.1176$ & $0.1163 \pm 0.0004$ & $0.1169 \pm 0.0003$ & $0.1173 \pm 0.0004$ \\
& $0.3$ & $0.1429$ & $0.1403 \pm 0.0004$ & $0.1416 \pm 0.0003$ & $0.1412 \pm 0.0004$\\
$0.1$ & $0.5$ & $0.2$ & $0.1944 \pm 0.0005$ & $0.1978 \pm 0.0004$ & $0.1973 \pm 0.0005$\\
& $0.7$ & $0.3333$ & $0.3227 \pm 0.0009$ & $0.3251 \pm 0.0008$ & $0.3257 \pm 0.0009$ \\
& $0.85$ & $0.667$ & $0.624 \pm 0.002$ & $0.631 \pm 0.002$ & $0.637 \pm 0.002$ \\
& $0.95$ & $2$ & $1.689 \pm 0.008$ & $1.691 \pm 0.008$ & $1.686 \pm 0.008$\\
\midrule
& $0.05$ & $0.526$ & $0.491 \pm 0.002$ & $0.519 \pm 0.002$ & $0.501 \pm 0.002$ \\
& $0.15$ & $0.588$ & $0.548 \pm 0.002$ & $0.575 \pm 0.001$ & $0.559 \pm 0.002$ \\
& $0.3$ & $0.714$ & $0.664 \pm 0.002$ & $0.690 \pm 0.001$ & $0.675 \pm 0.002$ \\
$0.5$ & $0.5$ & $1$ & $0.905 \pm 0.002$ & $0.943 \pm 0.002$ & $0.918 \pm 0.002$ \\
& $0.7$ & $1.667$ & $1.434 \pm 0.003$ & $1.479 \pm 0.003$ & $1.450 \pm 0.004$ \\
& $0.85$ & $3.333$ & $2.486 \pm 0.007$ & $2.536 \pm 0.006$ & $2.504 \pm 0.007$ \\
& $0.95$ & $10$ & $4.44 \pm 0.01$ & $4.53 \pm 0.01$ & $4.44 \pm 0.01$ \\
\midrule
& $0.05$ & $1.053$ & $0.932 \pm 0.005$ & $1.010 \pm 0.004$ & $0.907 \pm 0.005$\\
& $0.15$ & $1.176$ & $1.048 \pm 0.004$ & $1.122 \pm 0.003$ & $1.017 \pm 0.004$\\
& $0.3$  & $1.429$ & $1.251 \pm 0.004$ & $1.334 \pm 0.003$ & $1.228 \pm 0.004$\\
$1$ & $0.5$ & $2$ & $1.657 \pm 0.004$ & $1.772 \pm 0.004$ & $1.637 \pm 0.004$\\
& $0.7$ & $3.333$ & $2.500 \pm 0.006$ & $2.605 \pm 0.006$ & $2.459 \pm 0.006$ \\
& $0.85$ & $6.667$ & $3.769 \pm 0.009$ & $3.876 \pm 0.009$ & $3.745 \pm 0.009$ \\
& $0.95$ & $20$ & $5.38 \pm 0.01$ & $5.49 \pm 0.01$ & $5.37 \pm 0.01$ \\
\bottomrule
\end{tabular}
\end{center}
\caption{The average chord length $\langle \Lambda_\alpha | L\rangle ^m$ through clusters of composition $\alpha$, as a function of the probability $p$, of the mixing statistics $m$, and of the average chord length ${\langle \Lambda \rangle}_{\infty}$, for a domain of linear size $L=10$. The asymptotic quantity ${\langle \Lambda_{\alpha} \rangle}_{\infty}={\langle \Lambda \rangle}_{\infty}/(1-p)$, valid for infinite domains, is given for reference.}
\label{tab_lambda_alpha}
\end{table*}

\clearpage

\clearpage

\clearpage

\clearpage

\clearpage

\end{document}